\pdfoutput=1
\documentclass[11pt]{article} 

\usepackage[showcomments]{ajtex}
\DeclareRobustCommand{\DIEP}{\ensuremath{%
    \mathchoice{\includegraphics[height=2ex]{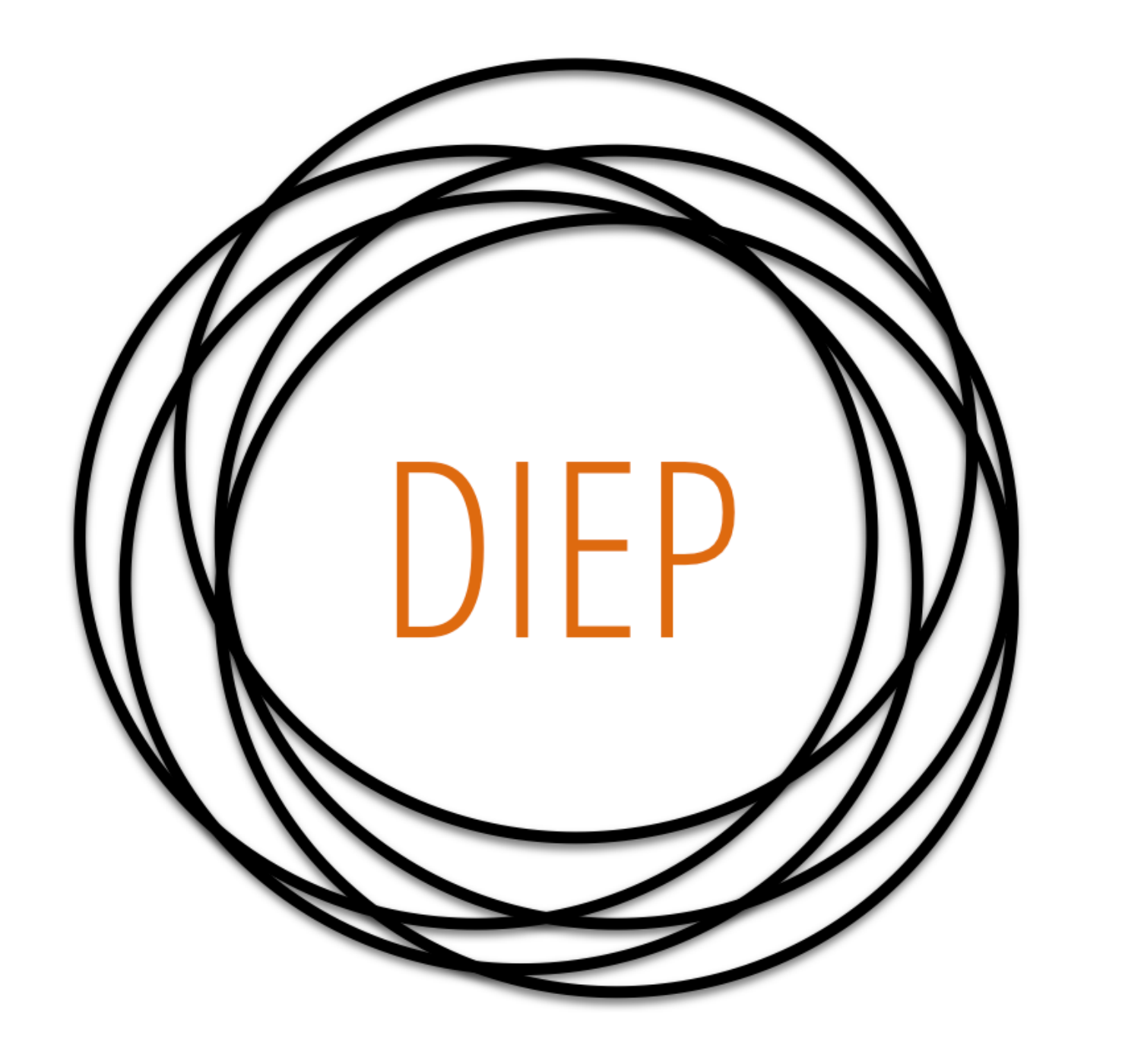}}
    {\includegraphics[height=2ex]{DIEPs.pdf}}
    {\includegraphics[height=1.5ex]{DIEPs.pdf}}
    {\includegraphics[height=1ex]{DIEPs.pdf}}
  }}
  
\title{Viscoelastic hydrodynamics and holography}

\author[a,\DIEP]{Jay Armas}\email{j.armas@uva.nl}

\author[b]{Akash Jain}\email{ajain@uvic.ca}

\affiliation[a]{Institute for Theoretical Physics, University of Amsterdam, 1090
  GL Amsterdam, The Netherlands}

\affiliation[\DIEP]{Dutch Institute for Emergent Phenomena, 1090 GL Amsterdam, The Netherlands}

\affiliation[b]{Department of Physics \& Astronomy, University of Victoria, PO
  Box 1700 STN CSC, Victoria, BC, V8W 2Y2, Canada} 

\abstract{We formulate the theory of nonlinear viscoelastic hydrodynamics of
  anisotropic crystals in terms of dynamical Goldstone scalars of spontaneously
  broken translational symmetries, under the assumption of homogeneous lattices
  and absence of plastic deformations. We reformulate classical elasticity
  effective field theory using surface calculus in which the Goldstone scalars
  naturally define the position of higher-dimensional crystal cores, covering
  both elastic and smectic crystal phases. We systematically incorporate all
  dissipative effects in viscoelastic hydrodynamics at first order in a
  long-wavelength expansion and study the resulting rheology equations. In the
  process, we find the necessary conditions for equilibrium states of
  viscoelastic materials. In the linear regime and for isotropic crystals, the
  theory includes the description of Kelvin-Voigt materials. Furthermore, we
  provide an entirely equivalent description of viscoelastic hydrodynamics as a
  novel theory of higher-form superfluids in arbitrary dimensions where the
  Goldstone scalars of partially broken generalised global symmetries play an
  essential role. An exact map between the two formulations of viscoelastic
  hydrodynamics is given. Finally, we study holographic models dual to both
  these formulations and map them one-to-one via a careful analysis of boundary
  conditions. We propose a new simple holographic model of viscoelastic
  hydrodynamics by adopting an alternative quantisation for the scalar fields. }

\newcommand\ext{\text{ext}}
\newcommand\eqb{\text{eqb}}
\newcommand\FS{\text{FS}}

\usepackage[bbgreekl]{mathbbol}

\setlist[itemize]{itemsep=0pt,topsep=0pt}

\begin{document}

\maketitle

\section{Introduction}

When undergoing deformations, most observable materials are known to exhibit
both elastic and viscous responses. Due to the coupling between fluid and
elastic behaviour, such materials are said to be \emph{viscoelastic}. Despite
being a century old subject and an active research field with multiple
technological applications \cite{gutierrez2013engineering}, the understanding of
viscoelasticity has been mostly based on phenomenological models that assume
linear strain responses such as the Kelvin-Voigt model, Maxwell model and the
Zener model as well as on some nonlinear generalisations thereof (see
e.g. \cite{christensen2013theory}). Several efforts have been made in order to
formulate viscoelasticity from general principles. In particular, the work of
Eckart \cite{PhysRev.73.373} was the key towards the geometrisation of strain
and the introduction of the notion of a dynamical material reference state.
More recently, works based on non-equilibrium thermodynamics have brought some
of these aspects to covariant form in the non-relativistic context
\cite{Azeyanagi:2009zd, Azeyanagi:2010ab} and in the relativistic context
\cite{Fukuma:2011kp, Fukuma:2011pr, Fukuma:2012ws} and recovered the stresses
and rheology of a few of the above mentioned viscoelastic models. However, while
significant, these works have not characterised the full interplay between fluid
and elastic behaviour due to defining assumptions. The aim of this paper is to
provide such a full characterisation, under relaxable conditions, in the
hydrodynamic regime.

According to Maxwell, the defining property of viscoelasticity is the capacity
for continuous media to exhibit elasticity at short time scales and fluidity at
long time scales compared to the strain relaxation time \cite{landau2013fluid,
  landau1986theory}. If the relaxation time is very large, fluid and elastic
behaviour can coexist in the hydrodynamic limit. This is the realm of (liquid)
crystal theory. We are thus interested in a long-wavelength long-distance
effective description of crystals. A crystal is characterised by a regularly
ordered lattice of points (atoms or molecules) discretely distributed over
space. More generally, the lattice cores that constitute the crystal can be
higher dimensional, such as strings and surfaces, where the atoms/molecules have
no positional ordering within the cores and can move freely like a
``liquid''. These crystals are called liquid crystals. Crystals may be present
in different phases, such as elastic (solid) phase, smectic or nematic, among
others (see e.g.\cite{chaikin_lubensky_1995, de1995physics}). In the
non-relativistic context, the hydrodynamics of (liquid) crystals has been
considered in several works \cite{PhysRevA.6.2401, JAHNIG1972129,
  chaikin_lubensky_1995} but these treatments assume isotropy, no external
currents and do not explicitly derive the constitutive relations and
stress/strain relations that couple the fluid and elastic degrees of freedom. In
this paper we focus on describing the elastic and smectic phases within a modern
framework of hydrodynamics, which includes effective field theory
\cite{Kovtun:2012rj}, offshell adiabatic analysis \cite{Haehl:2014zda,
  Haehl:2015pja}, and hydrostatic partition functions \cite{Banerjee:2012iz,
  Jensen:2012jh}.

Crystals in the elastic and smectic phases are states of matter with
spontaneously broken translational symmetries. The corresponding scalar
Goldstones $\phi^I$ with $I=1,2,\ldots, k$ associated with the broken
translation generators form the basis of classical elasticity effective field
theory (see e.g. \cite{kleinert1989gauge, BEEKMAN20171}) and are the fundamental
fields that enter the hydrodynamic description, as in \cite{PhysRevA.6.2401,
  JAHNIG1972129}. When $k=d$, the number of spatial dimensions, the theory
describes an elastic crystal with all its translation symmetries spontaneously
broken, while a generic $k\neq d$ describes a smectic crystal with only a subset
of translations broken. We note here that the scalars $\phi^I$ determine the
position of the crystal cores and, in the absence of disclinations and
dislocations, are surface forming. Thus their spacetime gradients can be used to
define an induced metric on the transverse space to the crystal cores via a
pullback map. This induced metric, when compared against the crystal intrinsic
metric defining its reference state, is a measure of strain induced in the
crystal that takes non-zero values when the Goldstone scalars acquire a
non-trivial expectation value. We use these realisations to formulate a
hydrodynamic description of anisotropic (liquid) crystals with nonlinear strains
of arbitrary strength under the assumptions of \textbf{(i)} absence of
dislocations and disclinations, \textbf{(ii)} lattice homogeneity
(i.e. invariant under $\phi^I \to \phi^I + a^I$ with $a^I$ being a constant
translation), and \textbf{(iii)} non-dynamical intrinsic crystal metric
(i.e. the crystal reference state does not change in time and thus we do not
consider plastic deformations). Within this formulation, we describe the
structure of first order elastic responses and transport properties of
viscoelastic fluids in a long-wavelength expansion. In the elastic phase of
isotropic crystals and under the assumption of linear strain responses, we
uncover 5 extra transport coefficients that have not been considered in the
literature.

Recently, it has been argued that viscoelastic hydrodynamics can be recast as a
theory of higher-form hydrodynamics making its global symmetries manifest and
avoiding the need to introduce microscopic dynamical
fields~\cite{Grozdanov:2018ewh}. This follows the recent line of research where
hydrodynamic systems with dynamical fields, such as magnetohydrodynamics with
dynamical gauge fields, are recast in terms of dual hydrodynamic systems with
higher-form symmetries \cite{Schubring:2014iwa, Grozdanov:2016tdf,
  Armas:2018ibg, Armas:2018atq, Armas:2018zbe, Glorioso:2018kcp}.  Building up
on this idea, here we provide a completely equivalent description of
viscoelastic hydrodynamics of anisotropic (liquid) crystals in arbitrary
spacetime dimensions by identifying the correct degrees of freedom of
higher-form hydrodynamics. The resulting theory describes higher-form
superfluidity in which the higher-form symmetries are partially broken (as in
the context of magnetohydrodynamics \cite{Armas:2018atq, Armas:2018zbe}). The
usefulness of this formulation resides in the fact that the hydrodynamic
description can be recast entirely in terms of symmetry principles where the
trivial conservation of the higher-form current $J^{I}=\star \df\phi^I$, with
$\star$ being the Hodge operator in $d+1$ dimensions, is a consequence of the
absence of defects (e.g. disclinations or dislocations). We provide an exact map
between the two formulations.

The lack of control over viscoelastic theories and the absence of complete
formulations have prompt a series of works where holographic methods are
employed to study putative strongly coupled viscoelastic theories and probe
regimes of elasticity and fluidity (e.g. \cite{Alberte:2017cch, Alberte:2017oqx,
  Esposito:2017qpj, Baggioli:2018bfa, Grozdanov:2018ewh, Andrade:2019zey,
  Ammon:2019wci, Ammon:2019apj, Baggioli:2019abx}). Two types of models have
been considered in the literature: gravity coupled to a set of scalar fields
$\Phi_I$ \cite{Alberte:2017oqx, Esposito:2017qpj, Andrade:2019zey} (and with
additional fields \cite{Esposito:2017qpj, Baggioli:2018bfa}) and gravity
minimally coupled to a set of higher-form gauge fields $B_{I}$
\cite{Grozdanov:2018ewh}. The former is supposed to describe the dynamics of
viscoelastic materials with spontaneously broken translation symmetries while
the latter is supposed to describe viscoelastic theories with higher-form
currents (see also~\cite{Baggioli:2018vfc, Baggioli:2019jcm}). However, the
establishment of a precise map between the two hydrodynamic formulations has
prompt us to investigate whether such a map exists at the level of holographic
models. Indeed, a careful analysis of boundary conditions has led us to propose
a simple model of viscoelasticity consisting of a set of scalars $\Phi_I$
minimally coupled to gravity but with an alternative quantisation for the scalar
fields and a double trace deformation of the boundary theory. The model is thus
the \emph{linear axion model} of \cite{Andrade:2013gsa} (see also
e.g. \cite{Bardoux:2012aw, Donos:2013eha, Davison:2014lua}) used in the context
of momentum relaxation \cite{Davison:2013jba, Blake:2015epa} but which does not
treat the fields $\Phi_I$ as background fields (as in the setting of forced
fluid dynamics \cite{Bhattacharyya:2008ji, Burikham:2016roo}). Instead, the
scalar fields are dynamical fields, as in a viscoelastic theory where the
Goldstone scalars have inherent dynamics.

Summarising, in this paper we make the following advancements and solve the
following problems/issues:
\begin{itemize}
\item We provide a complete formulation of nonlinear viscoelastic hydrodynamics
  of anisotropic (liquid) crystals in terms of Goldstone scalars of
  spontaneously broken symmetries up to first order in a long-wavelength
  expansion. We derive the Josephson equations for the Goldstone modes, akin to
  that found in the context of superfluids \cite{Jain:2016rlz}. We provide a
  classification of the response and transport of linear isotropic materials and
  recover Kelvin-Voigt materials as a special case.
  
\item We formulate the same theory of nonlinear viscoelastic hydrodynamic in
  terms of a higher-form superfluid by identifying the correct hydrodynamic
  degrees of freedom. This formulation, based on generalised global symmetries,
  provides an organisation principle and a first principle derivation of
  viscoelastic hydrodynamics that does not involve additional microscopic
  dynamical fields.
  
\item We provide holographic models for both these formulations in $D=4,5$ bulk
  spacetime dimensions as well as a map between the two models corresponding to
  each of the formulations. We identify the boundary conditions and boundary
  action necessary for obtaining holographic viscoelastic dynamics with the
  simple model of \cite{Andrade:2013gsa}. 
\end{itemize}

This paper is organised as follows. In \cref{sec:elasticity} we review the
classical elasticity effective field theory in terms of the Goldstones of broken
translational symmetries. However, we reformulate it in terms of surface
calculus, which besides aiding in understanding the appropriate degrees of
freedom of viscoelasticity, also leads to a precise covariant geometrization of
elastic strain. In \cref{sec:elastic-fluids} we formulate viscoelastic
hydrodynamics in terms of Goldstone scalars up to first order in a derivative
expansion.  We obtain the Josephson conditions and construct a hydrostatic
effective action that characterises the equilibrium viscoelastic states in the
theory. We also study the rheology equations and comment on phenomenological
viscoelastic models. In \cref{sec:higher-form} we formulate the same theory as a
novel theory of higher-form superfluidity, generalising \cite{Armas:2018ibg,
  Armas:2018atq, Armas:2018zbe} to arbitrary $d$-dimensional higher-form
currents. In this section, we also provide a detailed map between the two
formulations. \Cref{sec:holography} is devoted to the construction of
holographic models of viscoelastic dynamics and provides appropriate holographic
renormalisation procedures. It also contains a study of conformal viscoelastic
fluids. In sec.~\ref{sec:outlook} we conclude with a summary of the results
obtained in this paper together with interesting future research directions. We
also provide further details on the geometry of crystals in \cref{app:geometry},
while in \cref{app:hs-details} we give the details of hydrostatic constitutive
relations. Finally, in app.~\ref{app:comparison} we provide precise comparisons
between our different formulations and earlier ones in the literature.

\section{Broken translations and elasticity}
\label{sec:elasticity}

Utilising elements from earlier formulations (e.g. \cite{kleinert1989gauge,
  Nicolis:2013lma, BEEKMAN20171}), where Goldstones of spontaneously broken
translational symmetries play a key role, we introduce a classical effective
field theory for crystals exhibiting solid and smectic phases. As mentioned in
the introduction, crystals arrange themselves into a structured lattice of
points, strings, or surfaces (generically called lattice cores). In order to
deal with this wide range of higher-dimensional objects, we present a new
reformulation of classical elasticity effective field theory in terms of surface
calculus, which proves to be useful in later sections for tackling the
hydrodynamic regime of liquid crystals. In particular, this formulation provides
a simple and covariant notion of strain and allows us to cover solid and smectic
phases simultaneously. We begin with zero temperature considerations, moving on
to finite temperature effects towards the end of this section.

\subsection{Effective field theory of crystals}

\subsubsection{Crystal cores and strain}

We consider $(d+1)$-dimensional spacetimes where $d$ is the number of spatial
dimensions.  In the continuum limit, valid at long distances and low energies,
the worldsheets of $(d-k)$-dimensional crystal cores can be parametrised by a
set of $k$ spacetime dependent one-forms $e^I_\mu(x)$ normal to the cores, with
$I=1,2,\ldots k \leq d$. Point-like cores correspond to $k=d$, string-like to
$k=d-1$, and so on. In general, these normal one-forms have an inherent
spacetime dependent $\GL(k)$ ambiguity due to arbitrary normalisation:
$e^I_\mu \to M^I{}_{\!J}\, e^J_\mu$ with $M^I{}_{\!J} \in \GL(k)$. We keep this
redundancy unfixed for now by allowing for a local $\GL(k)$ symmetry in the
effective theory.

Given that the background spacetime is equipped with a metric $g_{\mu\nu}$
(which can be set to $\eta_{\mu\nu} = \diag(-1,1,1,\ldots)$ for crystals in flat
space), the physical distance between the cores is determined using the
\emph{crystal metric}
\begin{equation}
  \df s^2_{\text{crystal}} = h_{IJ} (e^I_\mu \df x^\mu) (e^J_\nu  \df x^\nu)~~,~~
  (h_{IJ}) = (h^{IJ})^{-1}~~,~~
  h^{IJ} = g^{\mu\nu} e^I_\mu e^J_\nu~~.
\end{equation}
The metric $h_{IJ}$ is the transverse metric to the crystal cores, obtained by
projecting the spacetime metric along the normal one-forms.  The indices
$I,J,\ldots$ can be raised/lowered using $h^{IJ}$ and $h_{IJ}$.  For later
convenience, we also define a pair of spacetime projectors transverse and along
the crystal cores by pushing forward the crystal metric
\begin{equation}
  h_{\mu\nu} = h_{IJ} e^I_\mu e^J_\nu~~,~~
  \bar h_{\mu\nu} = g_{\mu\nu} - h_{\mu\nu}~~,
\end{equation}
where $\bar h_{\mu\nu}$ is the longitudinal projector and $h_{\mu\nu}$ the
transverse projector.  On the other hand, the crystal also carries an intrinsic
\emph{reference metric} that captures the lattice structure of the crystal and
determines the ``preferred'' distance between the cores when no external factors
are at play. We define this reference metric as
\begin{equation}
  \df s^2_{\text{reference}}
  = \mathbb{h}_{IJ} (e^I_\mu \df x^\mu) (e^J_\nu  \df x^\nu)~~,
\end{equation}
where $\mathbb{h}_{IJ}$ is an arbitrary non-singular symmetric matrix. The
difference between the two metric tensors on the crystal defines the
\emph{strain tensor}
\begin{align}\label{eq:defn-strain}
  u_{IJ} = \half \lb h_{IJ} - \mathbb{h}_{IJ} \rb~~,~~
  u_{\mu\nu} = u_{IJ} e^I_\mu e^J_\nu~~,
\end{align}
which captures distortions of the crystal away from its reference
configuration. Subjecting a crystal to a strain, i.e. distorting the crystal,
causes stress depending on the physical and chemical properties of the material
that constitute the crystal. Within an effective field theory framework, we will
attempt to characterise the most generic such responses, given the symmetries of
the crystal.

\subsubsection{Crystal fields}

It is a known result in differential geometry of surfaces that a generic set of
one-forms $e^I_\mu$ does not have to be surface forming, i.e. there might not
exist a foliation of crystal core worldsheets normal to all the $e^I_\mu$. For
this to be the case, one needs to invoke the Frobenius theorem, ensuring that
there must exist a set of spacetime one-forms $a^I_{\nu J}$ such that
$\dow_{[\mu} e^I_{\nu]} = - a^I_{[\mu J} e^J_{\nu]}$. As a consequence, the
variations of the one-forms $e^I_\mu$ are not independent and we cannot use them
or the strain tensor $u_{IJ}$ directly as fundamental degrees of freedom in the
effective theory of crystals. To get around this nuisance, we assume that the
normal one-forms can locally be spanned by a set of $k$ closed one-forms, i.e.
$e^I_\mu(x) = \Lambda^I{}_{\!J}(x) \dow_\mu \phi^J(x)$, where
$\Lambda^I{}_{\!J}(x)$ is an arbitrary invertible matrix and $\phi^I(x)$ are
possibly multi-valued smooth scalar fields. This choice corresponds to the
crystal core worldsheets being level surfaces of the functions $\phi^I(x)$ and
satisfies the Frobenius condition as
$\dow_{[\mu} e^I_{\nu]} = - \Lambda^I{}_{\!K} \dow_{[\mu}
(\Lambda^{-1})^K{}_{\!J}\, e^J_{\nu]}$. The fields $\phi^I(x)$, which we refer
to as \emph{crystal fields}, describe the position of the crystal structure in
the ambient spacetime.  These crystal fields can be physically understood as
Goldstones of spontaneously broken translations.  If the crystal does not have
any topological defects such as dislocations or disclinations, the fields
$\phi^I(x)$ can be taken to be single-valued and well-behaved (see
e.g.~\cite{Beekman:2016szb}).

Recall that we had an arbitrary $\GL(k)$ renormalisation freedom in $e^I_\mu$,
which we can now fix by setting $\Lambda^I{}_{\!J}(x)$ to the identity
matrix. Consequently,
\begin{equation}
  e^I_\mu = \dow_\mu \phi^I~~, 
\end{equation}
and the physical and reference metrics of the crystal can be expressed as
\begin{equation}
  \df s^2_{\text{crystal}}
  = h_{IJ} \df\phi^I \df\phi^J~~,~~
  \df s^2_{\text{reference}}
  = \mathbb{h}_{IJ} \df\phi^I \df\phi^J~~.
\end{equation}
We should note that as such, like in any field theory, there is an arbitrary
redefinition freedom in the choice of the fundamental crystal fields
$\phi^I$. This will be useful later.

\subsubsection{Plasticity, homogeneity, and isotropy}

Generically, the reference metric $\mathbb{h}_{IJ}(x)$ of the crystal is a
dynamical field and can evolve independently with time. This physically
describes ``plastic materials'' for which the applied strain can permanently
deform the internal structure of the crystal over time. In this work, however,
we will focus on ``elastic materials'', wherein we assume that
$\mathbb{h}_{IJ}(x) = \mathbb{h}_{IJ}(\phi(x))$, i.e. the reference metric is an
intrinsic property of the crystalline structure and is not dependent on a
particular embedding of the crystal into the spacetime. The functional form of
$\mathbb{h}_{IJ}(\phi)$ is a property of the physical system under observation
and needs to be provided as input into the theory.

Furthermore, the crystals we wish to describe using this effective field theory
are homogeneous in space at macroscopic scales. Therefore, there exists a choice
of crystal fields $\phi^I$ such that the reference metric $\mathbb{h}_{IJ}$ is
constant and the theory is invariant under a constant shift
$\phi^I \to \phi^I + a^I$. In fact, for homogeneous crystals, we can utilise the
$\phi^I$ redefinition freedom to set the reference metric to be the Kronecker delta,
that is
\begin{equation} \label{eq:refchoice}
\mathbb{h}_{IJ} = \bbdelta_{IJ}~~.
\end{equation}
This leaves just a global $\SO(k)$ rotation freedom among $\phi^I$,
i.e. $\phi^I \to \Omega^I{}_{\! J}\, \phi^J$ where $\Omega^I{}_{\! J}$ is a
constant matrix valued in $\SO(k)$. As long as we properly contract the
$I,J,\ldots$ indices, we do not need to worry about this redundancy while
constructing the effective field theory.

Finally, if we wish to describe a crystal that is isotropic at macroscopic
scales (possibly due to randomly oriented crystal domains), we can impose the
aforementioned global $\SO(k)$ freedom of $\phi^I$ as an invariance of the
theory. Along with the constant shift invariance due to homogeneity, this
results in a Poincar\'e invariance on the field space. Practically, it means
that besides $e^I_\mu$ and $K^\ext_I$, the field space indices $I,J,\ldots$ in
the theory can only enter via the reference metric $\bbdelta_{IJ}$.  If,
instead, the crystal under consideration has long-range order, the parameters of
the effective theory can be arbitrary rank tensors on the field
space.\footnote{This structure only pertains to the geometric structure of the
  crystal itself. In general the atoms/molecules occupying the lattice sites can
  also carry other preferred vectors like spin or dipole moment which will need
  to be considered independently.} We provide further details on the geometry of
crystals in \cref{app:geometry}. In the bulk of this work we will assume the
crystals being described to be ``elastic'' and ``homogeneous'', while no
assumption is made regarding isotropy except in some explicit examples.


\subsection{Elasticity at zero temperature}
\label{sec:zeroTemperature}

Previously we have introduced the geometric notions required to describe
crystals but we have not yet attributed dynamics to the crystal fields. Here we
consider classical elasticity field theory at zero temperature, for which the
dynamics follows from an action principle that is written in terms of the
appropriate crystal fields (determined earlier to be $\phi^I$).

\subsubsection{Effective action}
\label{sec:0TempEffAction}

We posit that our theory of interest is described by an effective action with
functional form $S[\phi^I;g_{\mu\nu}]$, where $\phi^I$ are the dynamical
Goldstones of broken translations and $g_{\mu\nu}$ is taken to be a background
metric field. We focus on homogeneous crystals, for which the action is
invariant under a constant translation of the crystal fields
$\phi^I(x) \to \phi^I(x) + a^I$ and all the dependence on $\phi^I$ appears via
its derivatives $e^I_\mu = \dow_\mu \phi^I$.  The action can then be
parametrised as
\begin{equation} \label{eq:actelasticity}
  S[\phi^I;g_{\mu\nu}]
  = \int \df^{d+1}
  x \sqrt{-g}\, \mathcal{L}(e^I_\mu,g_{\mu\nu},\dow_\mu)~~.
\end{equation}
We define the \emph{crystal momentum currents} by varying the action with respect to
$e^I_\mu = \dow_\mu \phi^I$, that is
\begin{equation}
  \sigma^\mu_I = - \frac{1}{\sqrt{-g}} \frac{\delta S}{\delta e^I_\mu}~~.
\end{equation}
Given homogeneity, the equations of motion for $\phi^I$ simply imply the
conservation of crystal momentum currents
\begin{equation}\label{eq:defect-conservation}
  \nabla_\mu \sigma^\mu_I + K_I^\ext = 0~~,
\end{equation}
where $K^\ext_I$ is a background field, which can be understood as an external
force sourcing the crystal fields.\footnote{In order to obtain $K^\ext_I$ in
  \eqref{eq:defect-conservation} we have allowed for couplings to the external
  background field of the form $\phi^IK^\ext_I$ in \eqref{eq:actelasticity}. }
The conservation \cref{eq:defect-conservation} is not protected by any
fundamental symmetry and will in general be violated by thermal corrections as
we will see in the next section. If we further assume that all the dependence on
$e^I_\mu$ comes via $h^{IJ}$ or equivalently $u_{IJ}$, the crystal currents can
also be obtained by varying the action with respect to the strain
tensor\footnote{An exception to this comes from a dependence on the transverse
  derivatives $\bar h^{\mu\nu} e^{[I\lambda} \nabla_\nu e^{J]}_\lambda$.}
\begin{equation}
  \sigma^\mu_I = \sigma_{IJ} e^{J\mu}~~,~~
  \sigma^{IJ}
  = \frac{1}{\sqrt{-g}} \frac{\delta S}{\delta u_{IJ}}
  = - \frac{2 h^{IK} h^{JL}}{\sqrt{-g}}
  \frac{\delta S}{\delta h^{KL}}~~.
\end{equation}
Finally, we can obtain the energy-momentum tensor of the theory by varying the
action with respect to the background metric
\begin{equation}
  T^{\mu\nu} = \frac{2}{\sqrt{-g}} \frac{\delta S}{\delta g_{\mu\nu}}~~.
\end{equation}
Given that the action as constructed is invariant under background
diffeomorphisms, the energy-momentum tensor is conserved, modulo background
sources\footnote{At zero derivative order, where all the dependence on the
  metric in the Lagrangian comes via $h_{IJ}$ as well, the energy momentum
  tensor reads
  $T^{\mu\nu} = \mathcal{L}\, g^{\mu\nu} + \sigma_{IJ} e^{I\mu} e^{J\nu}$. This
  leads to the well known definition of stress tensor as the conjugate to strain
  $T^{IJ} = T^{\mu\nu} e^I_{\mu} e^J_{\nu} = \mathcal{L}\, h^{IJ} + \dow
  \mathcal{L}/\dow u_{IJ}$ (see for example section 6.3.3
  of~\cite{chaikin_lubensky_1995}).}
\begin{equation}\label{eq:EMconservation}
  \nabla_\mu T^{\mu\nu} = - K^\ext_I e^{I\nu}~~.
\end{equation}

\subsubsection{Linear isotropic materials at zero temperature}
\label{sec:LinearElasticityZeroTemp}

As an illustrative example, we consider classical elasticity theory where all
the dependence of $e^I_\mu$ comes via the strain $h^{IJ}$, and expand the
Lagrangian in a small strain expansion. In the case of homogeneous crystals, the
strain is given by $u_{IJ} = \half (h_{IJ} - \bbdelta_{IJ})$ and the most
generic such effective action for an isotropic crystal at zero-derivative order
and quadratic in strain is given by\footnote{Note that
  $\half\log\det(h_{IJ}) = \half\log\det(\bbdelta_{IJ} + 2 u_{IJ}) = h^{IJ}
  u_{IJ} + h^{I(K} h^{L)J} u_{IJ} u_{KL} + \mathcal{O}(u^2)$. So, generically,
  the $\fP$ term here can also be replaced by the term linear in strain
  $h^{IJ} u_{IJ}$ by redefining $\fB$ and $\fG$.}
\begin{equation}\label{eq:0tempLagrangian}
  \mathcal{L}
  = \half\fP\, \log\det h
  - \half C^{IJKL} u_{IJ} u_{KL} + \mathcal{O}(u_{IJ}^3,\dow)~~.
\end{equation}
Here $C^{IJKL}$ is the elasticity tensor of the crystal
\begin{equation}\label{eq:modulii}
  C^{IJKL}
  = \fB\, h^{IJ} h^{KL}
  + 2 \fG\, \lb h^{K(I} h^{J)L}
  - \frac{1}{k} h^{IJ} h^{KL} \rb~~,
\end{equation}
and the coefficients $\fB$ and $\fG$ are the bulk modulus and shear modulus of
the crystal respectively, whereas $\fP$ does not have a standard physical
interpretation in the literature. In \cref{eq:modulii}, we have chosen to
express the coefficients using $h^{IJ}$ for convenience, but we could
equivalently have used $\bbdelta_{IJ}$. This choice leads to the same physical
currents up to linear order in strain. By varying the action with respect to
$h_{IJ}$, we can read out the crystal momenta
\begin{equation}
  \sigma^\mu_I
  = \fP\, e^{\mu}_I 
    - \fB\, u^K{}_{\!\!K}\, e^{\mu}_I
    - 2 \fG \lb u_{IJ}
    - \frac{1}{k} h_{IJ} u^L{}_{\!\!L} \rb e^{J\mu}
    + \mathcal{O}(u^2,\dow)~~.
\end{equation}
On the other hand, the energy-momentum tensor is given by
\begin{align} \label{eq:EMzero}
  T^{\mu\nu}
  &= \mathcal{L}\, g^{\mu\nu}
  + \sigma^{\mu}_I e^{I\nu} \nn\\
  &= \fP \lb h^{\mu\nu} + u^\lambda{}_{\!\!\lambda} g^{\mu\nu}\rb
    - 2 \fG\, \lb u^{\mu\nu}
    - \frac{1}{k} h^{\mu\nu} u^\lambda{}_{\!\!\lambda} \rb
    - \fB\, u^\lambda{}_{\!\!\lambda}\, h^{\mu\nu}
    + \mathcal{O}(u^2,\dow)~~.
\end{align}
The bulk modulus $\fB$ and the shear modulus $\fG$ couple to the trace and
traceless parts of strain, respectively, in the energy-momentum tensor.  The
coefficient $\fP$, on the other hand, gives a constant pressure contribution
along the field directions in the energy-momentum tensor, modelling a repulsion
between lattice points. Such crystals cannot be supported without non-trivial
boundary conditions on their surface. For most phenomenological applications,
the lattice points are effectively neutral and the coefficient $\fP$ can be
dropped. However, as we will see in \cref{sec:holography}, this coefficient
appears naturally in holographic models of elasticity.

\subsection{Heating up the crystals}

So far we have focused on the effective field theory describing crystals at zero
temperature. However, for the phenomenological applications that we have in
mind, we need to take into account the effects of finite temperature.  In this
section we discuss crystals in thermodynamic equilibrium using the Matsubara
formalism of finite temperature field theory and introduce the equilibrium
effective action. Towards the end we motivate the hydrodynamic formulation of
crystals seen as small dynamical perturbations around thermodynamic equilibrium,
which we later elaborate in \cref{sec:elastic-fluids}.

\subsubsection{Equilibrium effective action}

The fundamental entity of interest at finite temperature is the thermal
partition function written in a given statistical ensemble. However, our
understanding of a complete partition function describing arbitrary
non-equilibrium thermal processes in a quantum field theory is still very
limited. Nevertheless, if we focus on just equilibrium (time-independent states)
in the theory, the grand canonical partition function can be computed using the
Matsubara imaginary time formalism
\begin{equation}
  \mathcal{Z}^\eqb
  = \int \mathcal{D}\phi^I \exp \lb - S^\eqb \rb~~.
\end{equation}
Here $S^\eqb$ is the equilibrium effective action of the theory that is,
naively, obtained by Wick rotating the Lorentzian action $S$. It should be noted
that defining the partition function above requires us to pick a preferred time
coordinate with respect to which the equilibrium is defined and with respect to
which the Wick rotation is to be performed. Consequently, in an effective field
theory approach, the equilibrium effective action $S^\eqb$ can contain many new
terms dependent on the preferred timelike vector that have no analogue in the
original zero temperature effective action $S$. To make this precise, let us
define $K^\mu = \delta^\mu_t/T_0$ to be the preferred timelike vector, with
$T_0$ being the inverse radius of the Euclidean time circle interpreted as the
global temperature of the thermal state under consideration. The requirement of
equilibrium implies that the Lie derivative of the constituent fields
$g_{\mu\nu}$ and $\phi^I$ along $K^\mu$ is zero, leading to
\begin{equation}\label{eq:K-variations}
  \delta_\scK g_{\mu\nu} = 2 \nabla_{(\mu} K_{\nu)}
  = \frac{1}{T_0} \dow_t g_{\mu\nu} = 0~~,~~
  \delta_\scK \phi^I = K^\mu e^I_\mu
  = \frac{1}{T_0} \dow_t \phi^I = 0~~.
\end{equation}

The equilibrium effective action and the resulting thermal partition function
for a crystal can be schematically represented as
\begin{align}\label{eq:eqbPFSchematic}
  S^\eqb[\phi^I;g_{\mu\nu}]
  &= \int_\Sigma \df \sigma_\mu\, (N^\mu)_\eqb(K^\mu,e^I_\mu,g_{\mu\nu},\dow_\mu) \nn\\
  &= \frac{1}{T_0}\int \df^d x \sqrt{-g}\,
    \mathcal{L}^\eqb(K^\mu,e^I_\mu,g_{\mu\nu},\dow_\mu)~~, \nn\\
  \mathcal{Z}^\eqb[g_{\mu\nu}]
  &= \int \mathcal{D}\phi^I \exp \lb - S^\eqb [\phi^I;g_{\mu\nu}] \rb~~.
\end{align}
The integral in the first line is performed over a constant-time Cauchy slice
$\Sigma$ with the respective differential volume-element denoted by
$\df \sigma_\mu$. The \emph{free-energy current} $(N^\mu)_\eqb$ is conserved
\begin{equation}
  \nabla_\mu (N^\mu)_\eqb = 0~~,
\end{equation}
rendering the effective action independent of the choice of Cauchy slice.  The
finite temperature action, for instance, can have dependence on the scalar
$K^2 = K^\mu K^\nu g_{\mu\nu} = g_{tt}/T_0^2$, which has no analogue in the zero
temperature effective action. This scalar is related to the local observable
temperature $T_\eqb$ in the field theory (as opposed to the global thermodynamic
temperature $T_0$) as
\begin{equation}\label{eq:eqbTemp}
  T_\eqb = \frac{T_0}{\sqrt{- g_{tt}}} = \frac{1}{\sqrt{- K^2}}~~.
\end{equation}

Once the effective action is at hand, we can work out the finite
temperature version of the $\phi^I$ equations of motion
\eqref{eq:defect-conservation} with the crystal momentum currents
\begin{equation}
  (\sigma^\mu_I)_\eqb = - \frac{1}{\sqrt{-g}} \frac{\delta S^\eqb}{\delta e^I_\mu}~~.
\end{equation}
We can also read out the energy-momentum tensor of the theory in thermal
equilibrium to be
\begin{equation}
  (T^{\mu\nu})_\eqb = \frac{2}{\sqrt{-g}} \frac{\delta S^\eqb}{\delta g_{\mu\nu}}~~,
\end{equation}
which satisfies the conservation equation \eqref{eq:EMconservation} owing to the
background diffeomorphism invariance of the equilibrium effective action.

\subsubsection{Linear isotropic materials at thermal equilibrium}

Focusing on the model of linear elasticity from
\cref{sec:LinearElasticityZeroTemp}, it is possible to heat it up to finite temperature while
keeping it in equilibrium. At zero-derivative order, the equilibrium effective
action has a form similar to \cref{eq:0tempLagrangian}, except that here we also
need to take into account the dependence on $T_\eqb$. We find
\begin{align}
  \mathcal{L}^\eqb
  &= P_{\text{f}}(T_\eqb)
  + \half \fP(T_\eqb)\, \log\det h
  - \half C^{IJKL}\, u_{IJ} u_{KL} + \mathcal{O}(u_{IJ}^3)
  + \mathcal{O}(\dow)~~,
\end{align}
where
\begin{equation}
  C^{IJKL}
  = \fB(T_\eqb)\, h^{IJ} h^{KL}
  + 2 \fG(T_\eqb)\, \lb h^{K(I} h^{J)L}
  - \frac{1}{k} h^{IJ} h^{KL} \rb~~.
\end{equation}
Here $P_{\text f}$ is interpreted as the thermodynamic pressure of the crystal,
which is purely a finite temperature effect. Note that at finite temperature,
the elastic modulii $\fB$ and $\fG$ of the crystal as well as the crystal
pressure $\fP$ are functions of the local temperature.  Varying the resulting
effective action, we can read out
\begin{align}\label{eq:linear-isotropic-consti-eqb}
  (\sigma^\mu_I)_\eqb
  &=
    \fP\, e^{\mu}_I
    - \fB\, u^\lambda{}_{\!\!\lambda}\, e^{\mu}_I
    - 2 \fG \lb u_{IJ}
    - \frac{1}{k} h_{IJ} u^\lambda{}_{\!\!\lambda} \rb e^{J\mu}
    + \mathcal{O}(u^2,\dow)~~, \nn\\
  (T^{\mu\nu})_\eqb
  &=
    T^3_{\eqb} \lb \dow_{T_\eqb} P_{\text f}
    + \dow_{T_\eqb} \fP\, u^\lambda{}_{\!\!\lambda} \rb
    K^\mu K^\nu
    + P_{\text{f}}\, g^{\mu\nu} \nn\\
  &\qquad
    + \fP \lb h^{\mu\nu} + u^\lambda{}_{\!\!\lambda} g^{\mu\nu} \rb
    - 2 \fG \lb u^{\mu\nu}
    - \frac{1}{k} h^{\mu\nu} u^\lambda{}_{\!\!\lambda} \rb
    - \fB\, u^\lambda{}_{\!\!\lambda}\, h^{\mu\nu}
    + \mathcal{O}(u^2,\dow)~~,
\end{align}
which enter \eqref{eq:defect-conservation} and \eqref{eq:EMconservation} to give
the $\phi^I$ equations of motion and energy-momentum conservation in thermal
equilibrium respectively. These can be directly compared to their zero
temperature counterparts in \cref{sec:LinearElasticityZeroTemp}. The crystal
momentum currents remain similar in form except for the temperature dependence
of the coefficients, while the energy-momentum tensor has a few novel terms. The
first of these terms corresponds to the thermodynamic energy density while the
second to the thermodynamic pressure, as promised.

\subsubsection{Leaving equilibrium -- hydrodynamics}

Although generic non-equilibrium processes in a thermal field theory are not
accessible with the machinery at hand, we can leave equilibrium perturbatively
using the framework of hydrodynamics. The basic premise of hydrodynamics is that
we can describe slight departures from thermal equilibrium by replacing the
isometry $K^\mu$ with a slowly varying dynamical field $\beta^\mu$. The
time-evolution of these fields is governed by the energy-momentum conservation
\eqref{eq:EMconservation}, which in the out of equilibrium context is not trivially satisfied as a mathematical
identity. It is customary to isolate the normalisation piece
and re-express $\beta^\mu$ as
\begin{equation}
  \beta^\mu = \frac{u^\mu}{T} ~~ \text{such that} \quad u^\mu u_\mu = -1~~.
\end{equation}
Here $u^\mu$ is the fluid velocity and $T$ is the fluid temperature out of
equilibrium. Note that in equilibrium, obtained by setting $\beta^\mu = K^\mu$,
the temperature $T$ reverts back to its equilibrium value $T_\eqb$ in
\cref{eq:eqbTemp}, while the fluid velocity $u^\mu$ is just a unit vector along
$\delta^\mu_t$ describing a fluid at rest.

Out of equilibrium, we no longer have the luxury to derive the $\phi^I$ equation
of motion or the conserved energy-momentum tensor using an effective
action. Instead we assume the existence of these as the starting
point of hydrodynamics. To wit
\begin{equation}\label{eq:RawEOM}
  K_I + K_I^\ext = 0~~,~~
  \nabla_\mu T^{\mu\nu} = - K^\ext_I e^{I\nu}~~.
\end{equation}
Note that we have replaced $\nabla_\mu \sigma^\mu_I$ from
\cref{eq:defect-conservation} by an arbitrary operator $K_I$ out of equilibrium,
making contact with our previous comment that there is no fundamental symmetry
at play to enforce the $\phi^I$ equations of motion to take the form of a
conservation law. As such, $K_I$ and $T^{\mu\nu}$ can be arbitrary operators made
out of the constituent fields in the theory. However, the existence of a
partition function in thermal equilibrium implies that the theory must admit a
free energy current $N^\mu$ which reduces to the conserved free-energy
current $(N^\mu)_\eqb$ in equilibrium (upon setting $\beta^\mu =
K^\mu$). Performing a time-dependent deformation of the equilibrium effective
action in \cref{eq:eqbPFSchematic}, it is not hard to convince oneself that
\begin{equation}\label{eq:RawAdiabaticity}
  \nabla_\mu N^\mu = \half T^{\mu\nu} \delta_\scB g_{\mu\nu} + K_I \delta_\scB
  \phi^I
  + \Delta~~,
\end{equation}
where $\Delta$ is at least quadratic in $\delta_\scB$. Here $\delta_\scB$ is
defined similar to \cref{eq:K-variations} and denotes Lie derivatives along
$\beta^\mu$. This is commonly referred to as the \emph{adiabaticity equation} and
determines the allowed terms in $K_I$ and $T^{\mu\nu}$ in agreement with the
equilibrium partition function.

Generally, hydrodynamic systems are required to satisfy a slightly stronger
constraint: the second law of thermodynamics. It is possible to define an entropy
current $S^\mu = N^\mu - T^{\mu\nu} \beta_\nu$, which upon using
\cref{eq:RawAdiabaticity} and \cref{eq:RawEOM} satisfies
\begin{equation}
  \nabla_\mu S^\mu = \Delta~~.
\end{equation}
The second law requires that the divergence of the entropy current should be
locally positive semi-definite, forcing $\Delta$ in \cref{eq:RawAdiabaticity} to
be a positive semi-definite quadratic form.

Having motivated hydrodynamics of crystals from the viewpoint of thermal field
theories, in the next section we will reintroduce hydrodynamics as its own
framework based on the second law of thermodynamics. We will revisit the
hydrodynamic elements discussed here and work out the equations governing a
crystal in the hydrodynamic regime up to first order in derivatives in agreement
with the second law.

\section{Viscoelastic hydrodynamics}
\label{sec:elastic-fluids}

In this section we formulate viscoelastic hydrodynamics as a theory of viscous
fluids with broken translation invariance, analogous to \cite{PhysRevA.6.2401,
  JAHNIG1972129}.  This is done by introducing the set of crystal (scalar)
fields, one for each spatial dimension along the crystal, as in the previous
section, which can be seen as Goldstones of broken momenta. Contrary to
previously studied cases of forced fluid dynamics \cite{Bhattacharyya:2008ji}
and models of momentum relaxation \cite{Blake:2015epa} where the scalar fields
are background fields, these Goldstone fields are dynamical. Their dynamics is
governed by a Josephson-type condition similar to that encountered in the context
of superfluids. We formulate viscoelastic fluids with one-derivative corrections
in arbitrary dimensions and study carefully the case of isotropic crystals with
linear responses in strain.  Attention is given to the resulting rheology
equations and a linearised fluctuation analysis is carried out, identifying
dispersion relations for phonons and sound modes.

\subsection{The setup}
\label{sec:conventional-setup}

As discussed in detail in the previous section, a crystal can be characterised
by a set of normal one-forms $e^I_\mu$, with $I = 1,2,\ldots k$.  In the
hydrodynamic regime, the dynamics of a viscoelastic crystal is governed by the
conservation of energy-momentum tensor
\begin{subequations}\label{eq:convetional-EOM}
  \begin{equation} \label{eq:elastic} \nabla_\mu T^{\mu\nu} = - K^\ext_I
    e^{I\nu}~~,
  \end{equation}
  along with the ``no topological defect'' constraint that requires the normal
  one-forms describing the crystal to be closed
  \begin{equation}\label{eq:no-defects}
    \partial_{[\mu}e^{I}_{\nu]}=0~~,
  \end{equation}
  and the crystal evolution equations
  \begin{equation}\label{eq:phiI-EOM-formal}
    K_I + K_I^\ext = 0 ~~.
  \end{equation}
\end{subequations}
The conservation of the energy-momentum tensor $T^{\mu\nu}$ is being sourced by
the external sources $K_I^\ext$ coupled to $e^I_\mu$. It governs the time
evolution of a set of hydrodynamic fields: fluid velocity $u^\mu$ (normalised
such that $u^\mu u_\mu=-1$) and fluid temperature $T$. The
constraint in \cref{eq:no-defects} can be identically solved by introducing a
set of crystal fields $\phi^I$ such that $e^I_\mu = \dow_\mu
\phi^I$. Physically, the crystal fields can be understood as Goldstones of
broken momentum generators.\footnote{A closely related formulation of
  viscoelastic fluids is found in~\cite{Fukuma:2011pr}, which models
  viscoelasticity as a sigma model given by $d+1$ scalar fields, seen as
  coordinates on an internal worldsheet. We provide a discussion of the
  similarities and distinctions between the two formulations in
  \cref{app:sakatani}.} The dynamics of these crystal fields themselves is
governed by \cref{eq:phiI-EOM-formal}, where $K_I$ is an effective macroscopic
operator composed of the field content of the theory. A priori, we do not have
any knowledge of the form of this operator. However, much like $T^{\mu\nu}$,
within the hydrodynamic derivative expansion, constitutive relations for $K_I$
can be fixed using the second law of thermodynamics~\cite{Jain:2016rlz}.
\Cref{eq:phiI-EOM-formal} is the finite temperature counterpart of
\cref{eq:defect-conservation} but since effective actions for viscoelastic
fluids describing dissipative dynamics have not yet been constructed, there is
no first principle derivation of $K_I$.\footnote{A nice parallel can be made
  with the theory of magnetohydrodynamics where the crystal fields $\phi^I$
  are replaced by the photon $A_\mu$ and the normal one-forms $e^I_\mu$ by the
  field strength $F_{\mu\nu}$. The three equations in \eqref{eq:convetional-EOM}
  find their respective analogues in energy-momentum conservation
  $\nabla_\mu T^{\mu\nu} = - F^{\nu\rho} J^\ext_\rho$, Bianchi identity
  $\dow_{[\mu} F_{\nu\rho]} = 0$, and Maxwell's equations
  $J^\mu + J^\mu_\ext = 0$. See~\cite{Armas:2018atq,Armas:2018zbe} for more
  details.}

Hydrodynamics is an effective theory where the most generic
constitutive relations for $T^{\mu\nu}$ and $K_I$ are obtained order-by-order in a derivative expansion
in terms of the constituent fields $u^\mu$, $T$, $\phi^I$ and background field
$g_{\mu\nu}$. These constitutive relations are required to
satisfy the second law of thermodynamics, which states that there must exist an
entropy current $S^\mu$, whose divergence is positive semi-definite in an
arbitrary $\phi^I$-offshell configuration. To wit
\begin{equation}\label{eq:elastic2ndLaw}
  \nabla_\mu S^\mu + \beta_\nu \lb \nabla_\mu T^{\mu\nu} - K_I e^{I\nu} \rb
  = \Delta \geq 0~~,
\end{equation}
where $\beta^\mu$ is an arbitrary multiplier that can be chosen to be $u^\mu/T$
using the inherent redefinition freedom in the hydrodynamic fields. A more
helpful version of the second law is obtained by defining a free energy current
\begin{equation}
  N^\mu_{\text{elastic}} = S^\mu + \frac{1}{T}T^{\mu\nu} u_\nu~~,
\end{equation}
which converts \cref{eq:elastic2ndLaw} into the adiabaticity equation
\begin{equation}\label{eq:adiabaticity-elastic}
  \nabla_\mu N^\mu_{\text{elastic}}
  = \half T^{\mu\nu} \delta_\scB g_{\mu\nu}
  + K_I \delta_\scB \phi^I  + \Delta~~,~~
  \Delta \geq 0~~,
\end{equation}
where we have denoted the Lie derivatives of $g_{\mu\nu}$ and $\phi^I$ along
$\beta^\mu$ as
\begin{equation}
  \delta_{\scB} g_{\mu\nu} = 2 \nabla_{(\mu} \beta_{\nu)}~~,~~
  \delta_\scB \phi^I = \beta^\mu \dow_\mu \phi^I = \beta^\mu e^I_\mu~~.
\end{equation}
To obtain the hydrodynamic constitutive relations allowed by the second law of
thermodynamics, we need to find the most generic expressions for $T^{\mu\nu}$
and $K_I$, within a derivative expansion, which satisfy
\cref{eq:adiabaticity-elastic} for some $N^\mu_{\text{elastic}}$ and $\Delta$.
It is thus required to establish a derivative counting scheme.  Following usual
hydrodynamic treatments, we consider $u^\mu$, $T$, and $g_{\mu\nu}$ to be
$\mathcal{O}(1)$ in the derivative expansion. On the other hand, we treat the
derivatives of the scalars $\phi^I$ as $\mathcal{O}(1)$, formally pushing the
scalars themselves to $\mathcal{O}(\dow^{-1})$. We also treat the sources
$K_I^{\ext}$ coupled to the scalars to be $\mathcal{O}(\dow)$. This counting
scheme is reminiscent of the one employed in the context of superfluids, and
guarantees that the crystal cores composing the lattice, which are responsible
for the elastic behaviour, appear at ideal order in the constitutive
relations. Thus, we will be describing viscoelastic fluids with arbitrary
strains, avoiding working in the restrictive regime of small strains as in
\cite{Fukuma:2011pr}.

Similar to the case of magnetohydrodynamics with dynamical gauge fields, not all
terms in the adiabaticity equation \bref{eq:adiabaticity-elastic} appear at the
same derivative order. In particular, $\delta_\scB \phi^I$ is $\mathcal{O}(1)$
while $\delta_\scB g_{\mu\nu}$ is $\mathcal{O}(\dow)$. This leads to order
mixing in the constitutive relations, that is, the same transport coefficients
can appear across derivative orders, forcing the analysis of the constitutive
relations to consider multiple derivative orders simultaneously - an expression
of one of the fallbacks of hydrodynamic formulations with dynamical fields.  In
sec.~\ref{sec:higher-form}, we show that this problem can be avoided by working
instead with formulations in terms of higher-form symmetries.

\subsection{Ideal viscoelastic fluids}

Given the establishment of a derivative counting scheme, we can use the
adiabaticity equation \bref{eq:adiabaticity-elastic} in order to find the
constitutive relations of a viscoelastic fluid at ideal order. It is possible to
infer that at leading order in derivatives, the adiabaticity equation has the
solution
\begin{equation} \label{eq:josephson-consti}
  K_I = - T \sigma_{IJ} \delta_\scB \phi^J + \mathcal{O}(\dow)~~,~~
  T^{\mu\nu} = N^\mu_{\text{elastic}} = \mathcal{O}(1)~~,~~
  \Delta = T \sigma_{IJ} \delta_\scB \phi^I \delta_\scB \phi^J + \mathcal{O}(\dow)~~.
\end{equation}
The coefficient matrix $\sigma_{IJ}$ can be arbitrary except that its
eigenvalues are constrained to be positive semi-definite.\footnote{The symbol
  $\sigma$ has been used to draw a parallel with the respective term in
  magnetohydrodynamics, where higher-form fluids find another useful
  application~\cite{Armas:2018atq,Armas:2018zbe}. There, the non-hydrodynamic
  field is the electromagnetic photon $A_\mu$ with the respective equation of
  motion given schematically as
  $J^\mu = \ldots - T \sigma P^{\mu\nu} \delta_\scB A_\nu + \ldots = -
  J^\mu_\ext$.}  Noting that $K^I_\ext = \mathcal{O}(\dow)$, the $\phi_I$
equation of motion \bref{eq:phiI-EOM-formal} requires that
\begin{equation}\label{phi-EOM}
  u^\mu \dow_\mu \phi^I = \mathcal{O}(\dow)~~. 
\end{equation}
This is the equivalent of the Josephson equation for superfluids and implies
that the crystal fields are stationary at ideal order in
derivatives.\footnote{Note that, unlike superfluids, we do not have a chemical
  potential whose redefinition freedom could be used to absorb the plausible
  derivative corrections in \cref{phi-EOM}. Technically, the fluid velocity
  itself serves as a chemical potential along spontaneously broken
  translations. To see this, one can expand the Goldstones along a reference
  position as $\phi^I = \delta^I_i x^i + \delta\phi^I$ and note that
  \cref{phi-EOM} becomes
  $u^0 \dow_0 \delta\phi^I = - \delta^I_i u^i - u^i \dow_i \delta\phi^I +
  \mathcal{O}(\dow)$. One can in principle absorb the derivative corrections
  into the redefinitions of $u^i$, but such redefinitions will be incompatible
  with the manifest Lorentz covariance of the theory.} In practice, this
equation algebraically determines the time-derivatives of the crystal fields.
It is useful to define the independent spatial derivatives of the crystal fields
as
\begin{equation} \label{eq:spatiald}
  P^{I\mu} = P^{\mu\nu} \dow_\nu \phi^I~~,
\end{equation}
where $P^{\mu\nu} = g^{\mu\nu} + u^\mu u^\nu$ is the projector orthogonal to the
fluid velocity. The spatial derivatives \eqref{eq:spatiald} capture all the
onshell independent information contained in $\phi_I$.

In order to proceed further, we consider \cref{eq:adiabaticity-elastic} at one-derivative
order, i.e. $N^\mu_{\text{elastic}}$ and $T^{\mu\nu}$ appear at ideal order in derivatives
order while $K_I$ only appears at one-derivative order. The most generic
constitutive relations are characterised by a free-energy current of the form
\begin{equation} \label{eq:free_elastic}
  N^\mu_{\text{elastic}} = P(T,h^{IJ}; \mathbb{h}_{IJ})\, \beta^\mu + \mathcal{O}(\dow)~~,
\end{equation}
where the fluid pressure $P(T,h^{IJ}; \mathbb{h}_{IJ})$ is an arbitrary function
of all the zero-derivative scalar fields in the theory, namely, the temperature
$T$ and the crystal metric $h^{IJ} = g^{\mu\nu} e^I_\mu e^J_\nu$.
In particular we have allowed for an independent dependence on each component of
$h^{IJ}$.  Additionally, $\mathbb{h}_{IJ}$ labels the reference state of the
material but has no inherent dynamics so hereafter we omit it for simplicity.
Introducing \eqref{eq:free_elastic} in the adiabaticity equation
\bref{eq:adiabaticity-elastic} and noting that
$\nabla_\mu(P \beta^\mu) = \delta_\scB P + \half P g^{\mu\nu} \delta_\scB
g_{\mu\nu}$ along with
\begin{equation}
  \delta_\scB T = \frac{T}{2} u^\mu u^\nu \delta_\scB g_{\mu\nu}~~,~~
  \delta_\scB h^{IJ} = 
  - e^{I\mu} e^{J\nu} \delta_\scB g_{\mu\nu}
  + 2 e^{(I\mu} \nabla_\mu \delta_\scB \phi^{J)}~~,
\end{equation}
we find the ideal viscoelastic fluid constitutive relations
\begin{align} \label{eq:const_elastic}
  T^{\mu\nu}
  &= (\epsilon + P) u^\mu u^\nu + P g^{\mu\nu}
  - r_{IJ} e^{I\mu} e^{J\nu} + \mathcal{O}(\dow)~~, \nn\\
  K_I
  &= - T \sigma_{IJ} \delta_\scB \phi^J
    - \nabla_\mu \lb r_{IJ} e^{J\mu} \rb
    + \mathcal{O}(\dow)~~, \nn\\
  N^\mu_{\text{elastic}}
  &= \frac{1}{T} P u^\mu - r_{IJ} e^{I\mu} \delta_\scB \phi^J + \mathcal{O}(\dow)~~,
\end{align}
with $\Delta$ remaining the same as \cref{eq:josephson-consti}. In writing \eqref{eq:const_elastic}, we have
defined the thermodynamic relations
\begin{equation} \label{eq:elastic_thermo}
  \df P = s \df T + \half r_{IJ} \df h^{IJ}~~,~~
  \epsilon + P = sT~~.
\end{equation}
Thus, we can identify $P$ as the thermodynamic pressure, $\epsilon$ as the energy
density, $s$ as the entropy density, and $r_{IJ}$ as the thermodynamic
stress that models elastic responses. The $\phi^I$ equation of motion now becomes 
\begin{equation}
  u^\mu \dow_\mu \phi^I
  = (\sigma^{-1})^{IJ} \lB K^{\text{ext}}_J - \nabla_\mu \lb r_{JK} e^{K\mu} \rb
  \rB
  + \mathcal{O}(\dow)~~.
\end{equation}
The constitutive relations \eqref{eq:const_elastic} are quite general at this
point but we will specialise to the case of an isotropic viscoelastic fluid
later in \cref{sec:isotropic} leading to more familiar expressions. It is worth
noticing that the same transport coefficient $P(T,h^{IJ})$ that is introduced at
zero-derivative order in $N^\mu_{\text{elastic}}$, appears at zero-derivative
order in $T^{\mu\nu}$ but at one-derivative order in $K_I$ (via thermodynamic
relations). However, as we will see in the next subsection, both $T^{\mu\nu}$
and $K_I$ get further corrections at one-derivative order. Hence, the
constitutive relations for $K_I$ mix different derivative orders. This is the
manifestation of order-mixing that we alluded to above.

For later use, it is helpful to explicitly write the energy-momentum
conservation equations \cref{eq:elastic}, given the constitutive relations \eqref{eq:const_elastic}. In particular, we find
\begin{gather}
  T \nabla_\mu (s u^\mu)
  =
  (\sigma^{-1})^{IJ}
  \lB K_I^\ext - \nabla_\mu \lb r_{IJ} e^{J\mu} \rb \rB
  \lB K^\ext_J - \nabla_\mu \lb r_{JK} e^{K\mu} \rb
  \rB
  + \mathcal{O}(\dow^2)~~, \nn\\
  sT\, P^{K\nu} \lb \frac1T \dow_\nu T + u^\mu \nabla_\mu u_\nu \rb
  + \lB K_I^\ext - \nabla_\mu \lb r_{IJ} e^{I\mu} \rb \rB
  P^{K\nu} e^{I}_\nu
  = \mathcal{O}(\dow^2)~~,
\end{gather}
or equivalently
\begin{gather}
  \delta_\scB (Ts)
  + \frac{Ts}{2} g^{\mu\nu} \delta_\scB g_{\mu\nu}
  =
  T \sigma_{IJ} \delta_\scB \phi^I \delta_\scB \phi^J
  + \mathcal{O}(\dow^2)~~, \nn\\
  P^{K\nu}  \lb sT u^\mu \delta_\scB g_{\mu\nu}
  + \sigma_{IJ} e^{I}_\nu \delta_\scB \phi^J \rb
  = \mathcal{O}(\dow^2)~~.
  \label{eq:first-order-eom}
\end{gather}
Formally, these equations can be used to eliminate
$u^\mu \delta_\scB g_{\mu\nu}$ at one-derivative order in favour of
$P^{I\mu}P^{J\nu} \delta_\scB g_{\mu\nu}$ and $\delta_\scB \phi^I$.

\subsection{One derivative corrections}
\label{sec:oned}

The philosophy implemented for ideal viscoelastic fluids can also be extended to include 
one-derivative corrections to the constitutive relations. For simplicity,
we focus on the elastic phase of crystals (as opposed to
liquid crystals) for which $k=d$. The derivative corrections can naturally be
classified into hydrostatic and non-hydrostatic constitutive relations: those
that do not vanish when promoting $\beta^\mu = u^\mu/T$ to an isometry and those that do
vanish, respectively (see~\cite{Jain:2016rlz}).

In order to characterise the hydrostatic sector, we need all the one-derivative
hydrostatic scalars that will make up the respective hydrostatic free-energy
current. For this purpose, we list all the hydrostatic one-derivative structures
\begin{equation}
  P^{\mu\nu}\dow_\nu T~~,~~
  2 P^{\mu\rho} P^{\nu\sigma} \dow_{[\rho} u_{\sigma]}~~,~~
  P^{\rho(\mu} P^{\nu)\sigma} \nabla_\rho e^I_\sigma~~.
\end{equation}
The presence of the vectors $u^\mu$ and $e^{I\mu}$ in the theory completely
breaks the Poincar\'e invariance, so we can convert all of these into
independent scalars\footnote{Note that
  $2 e^{(I\rho} e^{J)\sigma} \nabla_\rho e^K_\sigma = 2 e^{(I\rho} \nabla_\rho
  h^{J)K} - e^{K\sigma} \nabla_\sigma h^{IJ}$.}
\begin{equation}
  \frac{1}{T} e^{I\mu} \dow_\mu T~~,~~
  2T e^{I\mu} e^{J\nu} \dow_{[\mu} u_{\nu]}~~,~~
  e^{I\mu} \dow_\mu h^{JK}~~.
\end{equation}
When $k\neq d$, this is no longer true and the counting of independent scalars
needs to be more carefully implemented.  Supplementing with arbitrary
transport coefficients $f^1_I, f^2_{[IJ]}, f^3_{I(JK)}$ as functions of $T$ and $h^{IJ}$, we construct the
hydrostatic free energy density at first order in derivatives as
\begin{equation}\label{eq:N-conventional}
  \mathcal{N}
  = P
  + f^1_I \frac{1}{T} e^{I\mu} \dow_\mu T
  + 2T f^2_{[IJ]} e^{I\mu} e^{J\nu} \dow_{[\mu} u_{\nu]}
  + f^3_{I(JK)} e^{I\mu} \dow_\mu h^{JK}
  + \mathcal{O}(\dow^2)~~,
\end{equation}
with $N^\mu_{\text{elastic,hs}} = \mathcal{N}\, \beta^\mu$. Noting that
$\nabla_\mu (\mathcal{N} \beta^\mu) = \delta_\scB \mathcal{N} + \half
\mathcal{N} g^{\mu\nu} \delta_\scB g_{\mu\nu}$ and using the adiabaticity
equation \bref{eq:adiabaticity-elastic}, we can read off the respective modified
hydrostatic constitutive relations (see app.~\ref{app:hs-details} for details). 
The free energy density is defined up to total derivative terms.
Hence, it is possible to use the total derivative term
$\nabla_\mu e^{I\mu}$ to eliminate the trace part of
$f^3_{I(JK)} \sim f^{3\mathrm T}_I h_{JK}$ and take $f^3_{I(JK)}$ to be
traceless in the $JK$ indices without loss of generality.

In the non-hydrostatic sector, the constitutive relations are
the most generic expressions that involve $\delta_\scB g_{\mu\nu}$ and
$\delta_\scB \phi^I$. At one-derivative order, the contribution to the respective free energy
density happens to be zero, while the actual constitutive relations are 
\begin{equation}
  \begin{pmatrix}
    T^{\text{nhs}}_{IJ} \\ K^{\text{nhs}}_I
  \end{pmatrix}
  =
  - T \begin{pmatrix}
    \eta_{IJKL}
    & \chi_{IJK} \\
    \chi'_{IKL}
    & \sigma_{IK}
  \end{pmatrix}
  \begin{pmatrix}
    \half P^{K\mu} P^{L\nu} \delta_\scB g_{\mu\nu} \\
    \delta_\scB \phi^K
  \end{pmatrix}~~.
\end{equation}
We have defined
$T^{\mu\nu}_{\text{nhs}} = P^{I\mu} P^{J\nu} T^{\text{nhs}}_{IJ}$ and have used
the first order conservation equations \bref{eq:first-order-eom} to eliminate
$u^\mu \delta_\scB g_{\mu\nu}$ as well as to set the Landau frame condition
$T^{\mu\nu}_{\text{nhs}} u_\nu = 0$. The associated quadratic form is given as
\begin{equation}
  T \Delta
  =
  \begin{pmatrix}
    \half P^{I\mu} P^{J\nu} \delta_\scB g_{\mu\nu} \\
    \delta_\scB \phi^I
  \end{pmatrix}^{\mathrm T}
  \begin{pmatrix}
    \eta_{IJKL}
    & \half(\chi_{IJK} + \chi'_{IJK}) \\
    \half(\chi_{IKL} + \chi'_{IKL})
    & \sigma_{IK}
  \end{pmatrix}
  \begin{pmatrix}
    \half P^{K\mu} P^{L\nu} \delta_\scB g_{\mu\nu} \\
    \delta_\scB \phi^K
  \end{pmatrix}~~.
\end{equation}
The second law \eqref{eq:adiabaticity-elastic} requires that all the eigenvalues of the coefficient matrix are
non-negative.

To summarise, the constitutive relations of a viscoelastic fluid,
including the most generic one derivative corrections, are given by
\begin{align}\label{eq:net-EM}
  T^{\mu\nu}
  &= (\epsilon + P) u^\mu u^\nu + P g^{\mu\nu}
    - r_{IJ} e^{I\mu} e^{J\nu}
    + T^{\mu\nu}_{f_1}
    + T^{\mu\nu}_{f_2}
    + T^{\mu\nu}_{f_3} \nn\\
  &\qquad
    - P^{I\mu} P^{J\nu} \eta_{IJKL} P^{K\rho} P^{L\sigma} \nabla_{(\rho}
    u_{\sigma)}
    - P^{I\mu} P^{J\nu} \chi_{IJK} u^\rho \dow_\rho \phi^K
    + \mathcal{O}(\dow^2)~~,
\end{align}
where the contributions $T^{\mu\nu}_{f_i}$ are given in app.~\eqref{eq:consti-hs-conventional}. The
$\phi^I$ equations of motion modify to
\begin{equation}
  u^\mu \dow_\mu \phi^I
  = (\sigma^{-1})^{IJ}\lB
  K_J^\ext
  - \nabla_\mu \lb r_{JK} e^{K\mu} \rb
  - \chi'_{JKL} P^{K\mu} P^{L\nu} \nabla_{(\mu} u_{\nu)} \rB
  + \mathcal{O}(\dow^2)~~,
  \label{dBphiI-map}
\end{equation}
which is now correct up to two derivative terms. These constitutive relations
describe the dynamics of a viscoelastic fluid fully non-linearly in strain. In 
the next subsection we focus on the linear regime.

\subsection{Linear isotropic materials}
\label{sec:isotropic}

For concreteness, we study the constitutive relations of an isotropic
viscoelastic fluid.  In this case, all the $I,J,\ldots$ indices appear due to
the crystal metric $h_{IJ}$ and the reference metric $\mathbb{h}_{IJ}$. For
simplicity, we work linearly in strain
$u_{IJ} = \half (h_{IJ} - \mathbb{h}_{IJ})$, though the formalism introduced
previously is sufficient to handle any possible non-linearities.

\subsubsection{Constitutive relations}

Firstly, we note that we cannot construct an odd-rank tensor or an antisymmetric
2-tensor (in field space) using just $h_{IJ}$ and $\mathbb{h}_{IJ}$. Therefore
we are forced to set
\begin{equation}
  f^1_I = f^2_{[IJ]} = f^3_{I(JK)} = \chi_{IJK} = \chi'_{IJK} = 0~~,
\end{equation}
and hence the hydrostatic sector \eqref{eq:N-conventional} at one-derivative order is rendered trivial. The ideal
order pressure $P$ can be expanded up to quadratic terms in strain as
\begin{equation}
  P(T,h^{IJ}) =  P_{\text{f}}(T)
  + \half \fP(T)\, \log\det h
  - \half C^{IJKL} u_{IJ} u_{KL} + \mathcal{O}(u^3)~~,
\end{equation}
where
\begin{equation}
  C^{IJKL}
  = \fB(T)\, h^{IJ} h^{KL}
  + 2 \fG(T)\, \lb h^{K(I} h^{J)L}
  - \frac{1}{k} h^{IJ} h^{KL} \rb~~.
\end{equation}
This should be contrasted with the zero-temperature Lagrangian density in
\cref{eq:0tempLagrangian}.  We have expanded the pressure up to quadratic terms
because their derivatives can generically contribute to the constitutive
relations with terms linear in strain via thermodynamics (see
\bref{eq:elastic_thermo}). Thus
\begin{gather}
\label{eq:thermo-expansions-linear}
  P = P_{\text{f}} + \fP\, u^\lambda{}_{\!\!\lambda} + \mathcal{O}(u^2)~~,~~
  \epsilon = sT - P_{\text{f}} - \fP\, u^\lambda{}_{\!\!\lambda} +
  \mathcal{O}(u^2)~~,~~
  s = \dow_T P_{\text{f}} + \dow_T\fP\,u^\lambda{}_{\!\!\lambda} + \mathcal{O}(u^2)~~, \nn\\
  r_{IJ}
  = -\fP\, h_{IJ}
    + \fB\, u^\lambda{}_{\!\!\lambda}\, h_{IJ}
    + 2 \fG \lb u_{IJ}
    - \frac{1}{k} h_{IJ} u^\lambda{}_{\!\!\lambda} \rb
    + \mathcal{O}(u^2)~~.
  \end{gather}
In the non-hydrostatic sector, we can expand
the coefficients $\eta_{IJKL}$ and $\sigma_{IJ}$ linearly in strain and obtain
\begin{align}
  \eta_{IJKL}
  &= 2\lb \eta + \eta^u_1\, u^\lambda{}_{\!\!\lambda} \rb
    \lb h_{IK} h_{JL} - \frac{1}{k} h_{IJ} h_{KL} \rb
  + \lb \zeta + \zeta^u_1\, u^\lambda{}_{\!\!\lambda} \rb h_{IJ} h_{KL} \nn\\
  &\qquad + 2\eta^u_2\lb h_{IK} u_{JL}
    - \frac{1}{k} h_{IJ} u_{KL}
    - \frac{1}{k} u_{IJ} h_{KL}
    + \frac{1}{k^2} \, u^\lambda{}_{\!\!\lambda} h_{IJ} h_{KL} \rb \nn\\
  &\qquad
    + 2 \zeta^u_2\lb h_{IJ} u_{\langle KL\rangle} + u_{\langle IJ\rangle} h_{KL} \rb
    + 2 \bar\zeta^u\lb h_{IJ} u_{\langle KL\rangle} - u_{\langle IJ\rangle} h_{KL}
 \rb
     + \mathcal{O}(u^2)~~, \nn\\
  \sigma_{IJ}
  &= \lb \sigma + \sigma^u_1\, u^\lambda{}_{\!\!\lambda} \rb h_{IJ}
  + \sigma^u_2\, u_{IJ} + \mathcal{O}(u^2)~~,
\end{align}
together with the associated quadratic form 
\begin{align}
  T\Delta
  &=
  \half \begin{pmatrix}
    \sigma_{\mu\nu} \\ u^\lambda{}_{\!\!\lambda} \sigma_{\mu\nu} \\
    u_{\langle\mu|\sigma} \sigma_{\nu\rangle}{}^{\sigma}
  \end{pmatrix}^{\mathrm T}
  \begin{pmatrix}
    \eta & \half\eta^u_1 & \half \eta^u_2 \\
    \half\eta^u_1 & \ldots & \ldots \\
    \half \eta^u_2 & \ldots & \ldots
  \end{pmatrix}
  \begin{pmatrix}
    \sigma^{\mu\nu} \\ u^\lambda{}_{\!\!\lambda} \sigma^{\mu\nu} \\
    u^{\langle\mu}{}_\sigma \sigma^{\nu\rangle\sigma}
  \end{pmatrix}
  + \begin{pmatrix}
    \Theta \\ u^\lambda{}_{\!\!\lambda} \Theta \\
    u_{\mu\nu} \sigma^{\mu\nu}
  \end{pmatrix}^{\mathrm T}
  \begin{pmatrix}
    \zeta & \half\zeta^u_1 & \zeta^u_2 \\
    \half\zeta^u_1 & \ldots & \ldots \\
    \zeta^u_2 & \ldots & \ldots
  \end{pmatrix}
  \begin{pmatrix}
    \Theta \\ u^\lambda{}_{\!\!\lambda} \Theta \\
    u_{\mu\nu} \sigma^{\mu\nu}
  \end{pmatrix} \nn\\
  &\qquad
    + \begin{pmatrix}
      h_{\mu\nu} u^\nu \\ u^\lambda{}_{\!\!\lambda} h_{\mu\nu} u^\nu \\ u_{\mu\nu} u^\nu
    \end{pmatrix}
  \begin{pmatrix}
    \sigma & \half \sigma^u_1 & \half\sigma^u_2 \\
    \half \sigma^u_1 & \ldots & \ldots \\
    \half \sigma^u_2 & \ldots & \ldots
  \end{pmatrix}
  \begin{pmatrix}
    h^{\mu\nu} u_\nu \\ u^\lambda{}_{\!\!\lambda} h_{\mu\nu} u^\nu \\ u^{\mu\nu} u_\nu
  \end{pmatrix}
  + \mathcal{O}(u^2)~~.
\end{align}
The ellipsis denote terms quadratic or higher order in strain. For positive
semi-definiteness, the leading order transport coefficients $\eta(T)$,
$\zeta(T)$, and $\sigma(T)$ must be all non-negative, while the remaining ones
are unconstrained. It should be noted that the transport coefficient
$\bar \zeta^u$ does not cause any dissipation, and is an example of
non-dissipative non-hydrostatic transport in hydrodynamics.\footnote{We have not
  investigated constraints arising from Onsager's relations but it is expected
  that the non-dissipative non-hydrostatic coefficient $\bar\zeta^u$ is required
  to vanish.}  In the end, the complete set of constitutive relations for an
isotropic viscoelastic fluid up to first order in derivatives and linear in
strain is given by
\begin{align}\label{eq:linear-isotropic-consti}
  T^{\mu\nu}
  &= T \dow_T P_{\text{f}}\, u^\mu u^\nu
    + P_{\text{f}}\, g^{\mu\nu}
    - \eta\, \sigma^{\mu\nu}
    - \zeta\, \Theta P^{\mu\nu} \nn\\
  &\qquad
    + T \dow_T\fP\,u^\lambda{}_{\!\!\lambda}\, u^\mu u^\nu
    + \fP\, \lb h^{\mu\nu} + u^\lambda{}_{\!\!\lambda}\, g^{\mu\nu} \rb
    - 2 \fG\, \lb u^{\mu\nu}
    - \frac{1}{k} h^{\mu\nu} u^\lambda{}_{\!\!\lambda} \rb
    - \fB\, u^\lambda{}_{\!\!\lambda}\, h^{\mu\nu}\nn\\
  &\qquad
    - \eta^u_1\, u^\lambda{}_{\!\!\lambda} \sigma^{\mu\nu}
    - \eta^u_2\lb u^{(\mu}{}_{\!\sigma} \sigma^{\nu)\sigma}
    - \frac{1}{d} P^{\mu\nu} u_{\rho\sigma} \sigma^{\rho\sigma} \rb
    - 2\lb \zeta^u_2 - \bar\zeta^u \rb u^{\langle \mu\nu\rangle} \Theta \nn\\
  &\qquad
    - \lb \zeta^u_1\, u^\lambda{}_{\!\!\lambda} \Theta
    + \lb \zeta^u_2 + \bar\zeta^u \rb u_{\rho\sigma} \sigma^{\rho\sigma} \rb  P^{\mu\nu}
    + \mathcal{O}(u^2)~~,
\end{align}
where we have defined the expansion and shear of the fluid according to
\begin{equation}
  \Theta = \nabla_\mu u^\mu~~,~~
  \sigma^{\mu\nu}
  = 2 P^{\mu\rho} P^{\nu\sigma} \lb \nabla_{(\rho} u_{\sigma)} - \frac{1}{d}
  P_{\rho\sigma} \Theta \rb~~,
\end{equation}
and used \cref{eq:refchoice}.  Using \cref{dBphiI-map}, the $\phi^I$ equation of
motion takes the form
\begin{align} \label{eq:jopimp}
  u^\mu \dow_\mu \phi^I
  &= \frac{1}{\sigma} h^{IJ}\lB
    K_J^\ext
    + \nabla_\mu \lb \fP\, e^{\mu}_J
    - \fB\, u^\lambda{}_{\!\!\lambda}\, e^{\mu}_J
    - 2 \fG \lb u_{JK}
    - \frac{1}{k} h_{JK} u^\lambda{}_{\!\!\lambda} \rb e^{K\mu} \rb \rB \nn\\
  &\qquad
    - \frac{1}{\sigma^2} \lb
    \sigma^u_1\, u^\lambda{}_{\!\!\lambda} h^{IJ}
    + \sigma^u_2\, u^{IJ} \rb
    \lB
    K_J^\ext
    + \nabla_\mu \lb \fP\, e^{\mu}_J \rb \rB + \mathcal{O}(u^2)~~.
\end{align}
The first line in \cref{eq:linear-isotropic-consti} contains the usual
constitutive relations of an isotropic fluid, with $\eta(T)$ being the shear
viscosity and $\zeta(T)$ being the bulk viscosity. The terms in the second line
correspond to lattice pressure $\fP(T)$, shear modulus $\fG(T)$, and bulk
modulus $\fB(T)$, decoupled from the fluid except for the temperature dependence
of the coefficients, which are present at zero temperature as well (see
\cref{sec:zeroTemperature}). When $\fP=0$, then the second line describes the
well-known stresses of Hookean materials. The terms in the third and fourth
lines denote one-derivative corrections that are linear in strain and correspond
to the true coupling between fluid and elastic degrees of freedom. Such terms
have not been explicitly considered in traditional treatments
\cite{PhysRevA.6.2401, JAHNIG1972129, chaikin_lubensky_1995} neither in recent
ones \cite{Azeyanagi:2009zd, Azeyanagi:2010ab, Fukuma:2011kp, Fukuma:2011pr} and represent types of sliding frictional elements in rheology
analyses.\footnote{Some of these terms appear in the work of \cite{Fukuma:2012ws} but in the context of the specific conformal limit taken in \cite{Fukuma:2012ws}.}

\subsubsection{Rheology and phenomenological models}

Rheology is the study of stress/strain relations in flowing viscoelastic matter and is traditionally based on phenomenological models composed of mechanical building blocks designed for the purpose of describing observed properties of matter. The dynamics of viscoelastic materials studied in this paper is governed by energy-momentum conservation \eqref{eq:elastic} and the Goldstone equations \eqref{eq:phiI-EOM-formal}. In order to recast the equations in a more suitable form for comparison with rheology studies, it is useful to consider the implications of the Josephson condition \eqref{eq:jopimp}, namely
\begin{equation} \label{eq:rheology}
  \lie_\beta u_{\mu\nu}
  = \frac{1}{T} \left( \frac{1}{2}\sigma_{\mu\nu} + \frac{\Theta}{d} P_{\mu\nu} \right)
  + \mathcal{O}(\dow^2)~~,~~
  \lie_\beta \mathbb{h}_{\mu\nu} = \mathcal{O}(\dow^2)~~,
\end{equation}
where $\lie_\beta$ denotes the Lie derivative along $\beta^\mu$. These are the
\emph{rheology equations}. The first equation in \eqref{eq:rheology} expresses
the relation between the time-evolution of strains and viscous stresses while
the second is a consequence of one of the basic assumptions in this work,
namely, that the reference crystal metric is non-dynamical (i.e. absence of
plastic deformations). This corresponds to the elastic limit in the language of
\cite{Fukuma:2011pr} (see also app.~\ref{app:sakatani}).
\begin{figure}[h!]
  \begin{center}
    \includegraphics[width=0.6\textwidth]{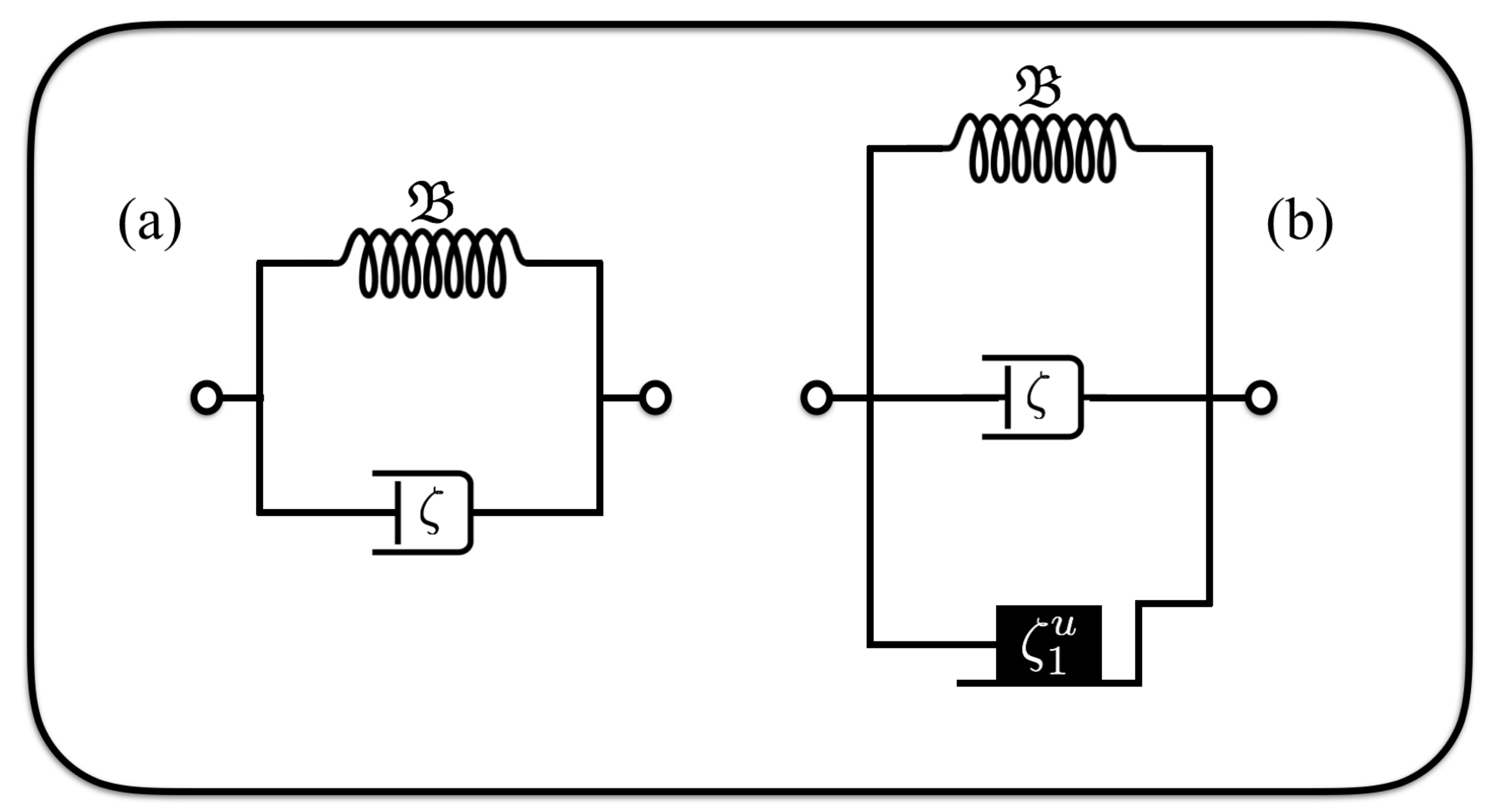} 
  \end{center}
  \caption{Material diagrams for the bulk stresses sector of (a) Kelvin-Voigt model and (b) Bingham-Kelvin model. As all elements are connected in parallel (spring, dashpot and sliding frictional element) the total stress is given by the sum of each of the individual contributions.} \label{fig:dashpot}
\end{figure}

Given the rheology equations \eqref{eq:rheology}, one can compare the
constitutive relations found here with existent viscoelastic models. First of
all, it should be noted that the last two lines in
\eqref{eq:linear-isotropic-consti} describe several couplings between fluid and
elastic degrees of freedom and a proper account of them in material models has
not been considered in generality. Doing so requires introducing many new
mechanical building blocks of the sliding frictional type. For simplicity, we
consider the case in which
$\fP=\eta^u_1=\eta_2^u=\zeta_1^u=\zeta_2^u=\bar\zeta^u=0$ which leads to the
energy-momentum tensor
\begin{equation}
  T^{\mu\nu}=
  \epsilon\, u^\mu u^\nu
    + P\, P^{\mu\nu}
    - \eta\, \sigma^{\mu\nu}
    - \zeta\, \Theta P^{\mu\nu} 
    - 2 \fG\, \lb u^{\mu\nu}
    - \frac{1}{d} h^{\mu\nu} u^\lambda{}_{\!\!\lambda} \rb
    - \fB\, u^\lambda{}_{\!\!\lambda}\, h^{\mu\nu}~~.
\end{equation}
This form of the stress tensor, together with \eqref{eq:rheology}, is known as
the Kelvin-Voigt model and usually represented as in \cref{fig:dashpot}(a),
where we have focused on the bulk stresses sector (i.e. we have depicted the
effect of bulk viscosity and bulk elastic modulus) and ignored the ideal fluid
part.

Another model that illustrates the use of coupling terms between elastic and
fluid degrees of freedom is the Bingham-Kelvin model for which
$\fP=\eta^u_1=\eta_2^u=\zeta_2^u=\bar\zeta^u=0$ and the energy-momentum tensor
becomes
\begin{equation} \label{eq:stressBV}
  T^{\mu\nu}=
  \epsilon\, u^\mu u^\nu
    + P\, P^{\mu\nu}
    - \zeta\, \Theta P^{\mu\nu} 
    - \fB\, u^\lambda{}_{\!\!\lambda}\, h^{\mu\nu}- \zeta^u_1\, u^\lambda{}_{\!\!\lambda} \Theta P^{\mu\nu}~~,
\end{equation}
where we have ignored the shear contribution to the stresses. The last term in
\eqref{eq:stressBV} is the term responsible for the frictional slide element
(black box) depicted in \cref{fig:dashpot}(b).

The entire possibility of linear responses \eqref{eq:linear-isotropic-consti}
allows for more intricate and rich material diagrams. The full nonlinear theory
of \cref{sec:oned} also allows for nonlinear responses to strain and hence for
the description of non-Newtonian fluids. However, it is not capable of
describing Maxwell-type models or Zener models as these violate the second
condition in \eqref{eq:rheology}. Such models allow for plastic deformations and
require that we consider a dynamical reference crystal metric $\mathbb{h}_{IJ}$
as in \cite{Fukuma:2011pr}. We intend to pursue this generalisation in the
future.


\subsection{Linearised fluctuations}
\label{sec:scalarperturbed}

In this section we study linearised fluctuations of equilibrium states of
isotropic crystals. We consider crystals coupled to a flat background
$g_{\mu\nu} = \eta_{\mu\nu}$ and vanishing external sources
$K^I_{\text{ext}}=0$, with static equilibrium configurations given by
\begin{equation}
  u^\mu = \delta^\mu_t~~,~~
  T = T_0~~,~~
  \phi^I = x^I~~.
\end{equation}
Note that in equilibrium we have $h^{IJ} = \bbdelta^{IJ}$, corresponding to a
crystal subjected to no strain. In general the system also admits solutions of
the type $\phi^I = \alpha\, x^I$ corresponding to a uniform strain. Such
configurations are allowed for a space filling crystal, but will need to be
supplied with appropriate boundary conditions if the crystal was finite in
extent. Since such configurations can be obtained by a trivial rescaling of
$\phi^I$'s, we do not consider them here.

\subsubsection{Modes}

Let us consider small perturbation of the equilibrium state parametrised by
\begin{equation}
  u^\mu = \delta^\mu_t + \delta u^\mu \quad\text{with}\quad \delta u^t = 0~~, \qquad
  T = T_0 + \delta T~~, \qquad
  \phi^I = x^I + \delta \phi^I~~.
\end{equation}
Plugging in a plane wave ansatz and solving the equations of motion
\eqref{eq:first-order-eom} linearly in the perturbations, we can find the
solutions
\begin{align}
  \delta u^I
  &= \lb 1 + \frac{(\omega^2 + k^2)\fP
    - k^2\lb \fB + 2\frac{d-1}{d}\fG + \fP'^2/s'\rb}
    {i\omega\sigma} \rb k^I  A_\parallel \E{i(k_I x^I - \omega t)} \nn\\
  &\qquad
    + \lb 1 + \frac{\omega^2\fP - k^2\fG}{i\omega\sigma}
    \rb A_\perp^I \E{i(k_I x^I - \omega t)}~~, \nn\\
  \delta\phi^I
  &= \frac{i}{\omega} \lb 1 + \frac{k^2}{i\omega\sigma}
    \frac{Ts\,\fP'}{s'} \rb  k^I A_\parallel \E{i(k_I x^I - \omega t)}
    + \frac{i}{\omega} A^I_\perp \E{i(k_I x^I - \omega t)}~~, \nn\\
  \delta T
  &= \frac{k^2}{\omega} \lb \frac{s + \fP'}{s'}
    + \frac{(\omega^2 + k^2)\fP
    - \lb \fB + 2\frac{d-1}{d}\fG \rb k^2}{i\omega\sigma s'/s}
    \rb A_\parallel \E{i(k_I x^I - \omega t)}~~.
\end{align}
We have suppressed the arguments of various transport coefficients but they are
understood to be evaluated on the equilibrium configuration. In addition, we have omitted 
the effect of viscosities but we consider it explicitly below. Primes denote a
derivative with respect to temperature. We have also used the isotropy of the
system to decompose
\begin{align}
  \frac{\dow P}{\dow h^{IJ}}\bigg|_{h=\alpha^2\bbdelta}
  &= - \frac{1}{2} \fP(T)\, \bbdelta_{IJ}~~, \nn\\
  \frac{\dow P}{\dow h^{IJ}\dow h^{KL}}\bigg|_{h=\alpha^2\bbdelta}
  &= \frac{1}{2} \lb\fP(T) -\fG(T) \rb \bbdelta_{I(L} \bbdelta_{K)J}
    -  \frac{1}{4} \lb \fB(T) - \frac{2}{d} \fG(T) \rb \bbdelta_{IJ}
    \bbdelta_{KL}~~, \nn\\
  \sigma_{IJ}\big|_{h=\alpha^2\bbdelta}
  &= \sigma(T)\, \bbdelta_{IJ}~~,
\end{align}
using the same transport coefficients introduced in
sec.~\ref{sec:isotropic}.\footnote{If we were to work around an equilibrium state
  with $\phi^I = \alpha x^I$, we would get the same expressions, except that
  these coefficients will be defined around the new equilibrium state and will not
  have an interpretation in terms of linear transport coefficients.}
The symbols $A_\parallel$ and $A_\perp^I$ (with $A_\perp^I k_I = 0$) denote
arbitrary amplitudes corresponding to ``longitudinal'' and ``transverse'' modes
respectively. The respective dispersion relations, in small momentum and
frequency regime, are given by
\begin{align}
  A_\parallel:\quad
  &\omega \lb  \omega^2  T s
    -  k^2 \lb \frac{(s + \fP')^2}{s'}
    + \omega \frac{Ts}{i\sigma} \frac{\fP'^2}{s'}
    \rb \rb
    + i \omega^2 k^2 \lb \zeta + 2 \frac{d-1}{d}\eta \rb\nn\\
  &\qquad
    + \lb \omega + \frac{Ts}{i\sigma} \lb  \omega^2 - \frac{s}{Ts'} k^2 \rb \rb
    \lb (\omega^2 + k^2) \fP - k^2  \lb \fB + 2\frac{d-1}{d} \fG \rb \rb
    = 0~~, \nn\\
  A_\perp:\quad
  & \omega^2 sT +  \lb 1 + \frac{\omega Ts}{i\sigma}  \rb
    \lb \omega^2\fP - k^2  \fG \rb
    + i \omega k^2  \eta =0~~.
\end{align}
Solving these equations, we find that in the longitudinal sector we have the
usual sound mode along with a new diffusion mode characteristic of a
lattice\footnote{This diffusion mode was identified in \cite{Delacretaz:2017zxd}
  and in holographic setups in \cite{Ammon:2019apj, Baggioli:2019aqf,
    Baggioli:2019abx}.}
\begin{equation}
  \omega(k) = \pm v_\parallel k - i \frac{\Gamma_\parallel}{2} k^2 +
  \mathcal{O}(k^3)~~, \qquad
  \omega(k) = -i D_\parallel k^2 + \mathcal{O}(k^3)~~,
\end{equation}
where
\begin{gather}
  v_\parallel^2
  = \frac{(s + \fP')^2/s' - \fP + \fB + 2 \frac{d-1}{d}\fG}{Ts + \fP}~~, \qquad
  \Gamma_\parallel
  =  \frac{T^2 s^2 v_\parallel^2}{\sigma (Ts + \fP)}
  \left(1-\frac{s+\fP'}{Ts' v_\parallel^2}\right)^2
  + \frac{\zeta + 2\frac{d-1}{d}\eta}{Ts+\fP}~~, \nn\\
  D_\parallel
  = \frac{s^2}{\sigma s'}\frac{- \fP + \fB + 2\frac{d-1}{d}\fG}{(Ts + \fP) v_\parallel^2}~~.
\end{gather}
On the other hand, in the transverse sector we have another sound mode
\begin{equation}
  \omega(k) = \pm v_\perp k - i \frac{\Gamma_\perp}{2} k^2 +
  \mathcal{O}(k^3)~~,
\end{equation}
where
\begin{equation}
  v_\perp^2 = \frac{\fG}{Ts + \fP}~~, \qquad
  \Gamma_\perp
  =  \frac{T^2 s^2 \fG}{\sigma  (Ts + \fP)^2}
  + \frac{\eta}{Ts+\fP}~~.
\end{equation}
We see that the transverse sound mode is controlled by the shear modulus
$\fG$. On the other hand, the new diffusive mode is controlled by the transport
coefficient $\sigma$ and $\eta$. In the absence of the lattice pressure $\fP$, these
expressions simplify
\begin{gather}
  v_\parallel^2
  = \frac{s}{Ts'} + \frac{\fB + 2 \frac{d-1}{d}\fG}{Ts}~,~
  \Gamma_\parallel
  =  \frac{1}{\sigma Ts}
  \left(\fB + 2 \frac{d-1}{d}\fG\right)^2+ \frac{\zeta + 2\frac{d-1}{d}\eta}{Ts}~,~
  D_\parallel
  = \frac{s}{\sigma Ts'}\frac{\fB + 2\frac{d-1}{d}\fG}{v_\parallel^2}~~, \nn\\
  v_\perp^2 = \frac{\fG}{Ts}~~, \qquad
  \Gamma_\perp
  =  \frac{\fG}{\sigma}+ \frac{\eta}{Ts}~~,
\end{gather}
which might be more familiar to some readers.

For linear stability of the system, the imaginary part of $\omega(k)$ must be
non-negative. This leads to the constraints
$v_\parallel^2, v_\perp^2, \Gamma_\parallel, \Gamma_\perp, D_\parallel > 0$. In
terms of coefficients, assuming that $Ts + \fP > 0$ and the second law
constraints $\eta,\zeta,\sigma\geq0$, we land on the parameter space
\begin{equation}
  \frac{(s + \fP')^2}{s'} - \fP + \fB + 2 \frac{d-1}{d}\fG > 0~~, \qquad
  \fG > 0~~, \qquad
  \frac{- \fP + \fB + 2\frac{d-1}{d}\fG}{s'} > 0~~.
\end{equation}
On the other hand, for causality, we require $v_\parallel^2, v_\perp^2 <
1$. This leads to
\begin{equation}
  \frac{(s + \fP')^2}{s'} - \fP + \fB + 2 \frac{d-1}{d}\fG < Ts + \fP~~, \qquad
  \fG < Ts + \fP~~.
\end{equation}
This gives the allowed range of parameters for a sensible evolution of the
dynamical equations.

\subsubsection{Linear response functions and Kubo formulas}

We can extend the analysis above to read out the linear response functions of
the theory by switching on plane wave background fluctuations. Let us start by
setting $\sigma\to\infty$ and $\eta,\zeta\to 0$ turning off the dissipative
corrections for simplicity. Let us take a perturbation of the background sources
\begin{equation}
  g_{\mu\nu} = \eta_{\mu\nu}
  + \delta g_{\mu\nu}~~,\qquad
  K^{\ext}_I = \delta K^\ext_I.
\end{equation}
We can read out the solution of the equations of motion
\bref{eq:first-order-eom} by a straightforward computation
\begin{align}
  \delta T
  &= \frac{s + \fP'}{s' \lb \omega^2 - v_\parallel^2  k^2 \rb}
    \lb \lb 2 k^2 v_\perp^2 k^{JK}
    - \omega^2 \bbdelta^{JK}
    \rb \half  \delta g_{JK}
    - \omega k^J \delta g_{tJ}
    - \half k^2 \delta g_{tt}
    - \frac{ik^J \delta K_J^\ext}{Ts+\fP}
    \rb~~, \nn\\
  \delta u^I
  &= \frac{\omega k^I}{\omega^2 - v_\parallel^2  k^2} \lb
    v_\perp^2 k^{JK} \delta g_{JK}
    - \frac{\omega k^J}{k^2} \delta g_{tJ}
    - \half \delta g_{tt}
    - v_\parallel^2 \half \bbdelta^{JK} \delta g_{JK}
    - \frac{ik^J/k^2 \delta K_J^\ext}{Ts+\fP}
    \rb \nn\\
  &\qquad
    - \omega \frac{\bbdelta^{IJ} - k^I k^J/k^2}{\omega^2 - v_\perp^2 k^2} \lb
    v_\perp^2 k^K \delta g_{JK}
    + \omega \delta g_{tJ}
    + \frac{i\delta K^\ext_J}{Ts+\fP} \rb~~, \nn\\
  \delta \phi^I
  &= \frac{i k^I}{\omega^2 - v_\parallel^2  k^2} \lb
  v_\perp^2 k^{JK} \delta g_{JK}
  - \frac{\omega k^J}{k^2} \delta g_{tJ}
  - \half \delta g_{tt}
    - v_\parallel^2 \half \bbdelta^{JK} \delta g_{JK}
    - \frac{ik^J/k^2 \delta K^\ext_J}{Ts+\fP}\rb \nn\\
  &\qquad
    - i \frac{\bbdelta^{IJ} - k^I k^J/k^2}{\omega^2 - v_\perp^2 k^2} \lb
    v_\perp^2 k^K \delta g_{JK}
    + \omega \delta g_{tJ}
    + \frac{i\delta K^\ext_J}{Ts+\fP}\rb~~.
\end{align}
The two point retarded Green's functions are defined as
\begin{gather}
  \mathcal{G}^{\mu\nu,\rho\sigma}_{TT,R}
  = - 2 \frac{\delta(\sqrt{-g}\, T^{\mu\nu})}{\delta
    g_{\rho\sigma}}\bigg|_{\delta g = 0,\delta K = 0}, \qquad
  \mathcal{G}^{I,J}_{\phi\phi,R}
  = \frac{\delta \phi^I}{\delta K_J^\ext}\bigg|_{\delta g = 0,\delta K = 0}, \nn\\
  \mathcal{G}^{I,\mu\nu}_{\phi T,R}
  = - 2 \frac{\delta \phi^I}{\delta
    g_{\mu\nu}}\bigg|_{\delta g = 0,\delta K = 0}
  =  \frac{\delta(\sqrt{-g}\, T^{\mu\nu})}{\delta K^\ext_I}\bigg|_{\delta g = 0,\delta K = 0}.
\end{gather}
Without loss of generality, we can choose the momentum to be in
$k^I = \delta^I_x$ and denote the remaining spatial indices by
$a,b,\ldots$. Defining
\begin{equation}
  \Delta_\parallel = \frac{\omega^2 - v_\parallel^2 k^2}{Ts + \fP}~~, \qquad
  \Delta_\perp = \frac{\omega^2 - v_\perp^2 k^2}{Ts + \fP}~~,
\end{equation}
we read out the respective two point functions; in the longitudinal sector we
have
\begin{gather}
  \mathcal{G}^{tt,tt}_{TT,R}
  = \frac{k^2}{\Delta_\parallel} 
  - \langle T^{tt} \rangle ~~, \qquad
  \mathcal{G}^{tt,tx}_{TT,R}
  = \frac{\omega k}{\Delta_\parallel}~~, \qquad
  \mathcal{G}^{tt,xx}_{TT,R}
  = \frac{v_\parallel^2 k^2}{\Delta_\parallel}
  + \langle T^{xx} \rangle~~, \qquad
  \mathcal{G}^{tx,tx}_{TT,R}
  = \frac{v_\parallel^2 k^2}{\Delta_\parallel}
  + \langle T^{tt} \rangle~~, \nn\\
  \mathcal{G}^{tx,xx}_{TT,R}
  = 
  \frac{\omega k v_\parallel^2}{\Delta_\parallel}~~, \qquad
  \mathcal{G}^{xx,xx}_{TT,R}
  = 
  \frac{v_\parallel^2\omega^2}{\Delta_\parallel}
  + \langle T^{xx} \rangle~~, \nn\\
  \mathcal{G}^{x,xx}_{\phi T,R} = \frac{-ik}{(Ts + \fP)\Delta_\parallel}~~,\qquad
  \mathcal{G}^{x,tx}_{\phi T,R} = \frac{-i\omega}{(Ts + \fP)\Delta_\parallel}~~,\qquad
  \mathcal{G}^{x,tx}_{\phi T,R} = \frac{-i\omega^2/k}{(Ts +
    \fP)\Delta_\parallel}
  + \frac{i}{k}~~, \nn\\
  \mathcal{G}^{x,x}_{\phi T,R} = \frac{-1}{(Ts + \fP)^2\Delta_\parallel}~~.
\end{gather}
Similarly, in the transverse sector
\begin{gather}
  \mathcal{G}^{tt,ab}_{TT,R}
  = \frac{k^2\lb v_\parallel^2 - 2 v_\perp^2\rb}{\Delta_\parallel}
  \bbdelta^{ab}
  + \langle T^{ab} \rangle ~~, \qquad
  \mathcal{G}^{ta,tb}_{TT,R}
  = \lb \frac{v_\perp^2 k^2}{\Delta_\perp} + \langle T^{tt} \rangle \rb \bbdelta^{ab}~~, \nn\\
  \mathcal{G}^{ab,cd}_{TT,R}
  = \lb 
  \frac{k^2 \lb v_\parallel^2 - 2v_\perp^2\rb^2}{\Delta_\parallel}
  + \frac{\lb s + \fP' \rb^2}{s'}
  + \fB -\fP \rb \bbdelta^{ab} \bbdelta^{cd} \nn\\
  + \lb P + \fP \rb
  \lb 2\delta^{c(a} \delta^{b)d} - \bbdelta^{ab} \bbdelta^{cd} \rb
  + 2 \fG \lb \delta^{c(a} \delta^{b)d} - \frac1d \bbdelta^{ab} \bbdelta^{cd}
  \rb~~, \nn\\
  \mathcal{G}^{a,tb}_{\phi T,R} = \frac{-i\omega}{(Ts + \fP)\Delta_\perp}
  \bbdelta^{ab}~~, \qquad
  \mathcal{G}^{a,b}_{\phi\phi,R} = \frac{-1}{(Ts + \fP)^2\Delta_\perp}
  \bbdelta^{ab}~~.
\end{gather}
Finally, we have non-zero contributions in the cross sector
\begin{gather}
  \mathcal{G}^{tx,ab}_{TT,R}
  = \frac{\omega k \lb v_\parallel^2 -  2 v_\perp^2 \rb}{\Delta_\parallel}
  \bbdelta^{ab}~~, \qquad
  \mathcal{G}^{ta,xb}_{TT,R}
  = \frac{\omega k v_\perp^2}{\Delta_\perp} \bbdelta^{ab}~~, \nn\\
  \mathcal{G}^{xx,ab}_{TT,R}
  = \frac{\omega^2\lb v_\parallel^2 - 2v_\perp^2 \rb}{\Delta_\parallel}
  \bbdelta^{ab}
  - \langle T^{ab} \rangle~~, \qquad
  \mathcal{G}^{xa,xb}_{TT,R}
  = 
  \frac{\omega^2 v_\perp^2}{\Delta_\perp} \bbdelta^{ab}
  + \langle T^{ab} \rangle~~, \nn\\
  \mathcal{G}^{x,ab}_{\phi T,R} = \frac{i k\lb \frac{2\fG}{Ts+\fP} -
    v_\parallel^2 \rb}{(Ts + \fP)\Delta_\parallel} \bbdelta^{ab}~~. \qquad
  \mathcal{G}^{a,xb}_{\phi T,R} = \frac{i}{k} + \frac{-i\omega^2/k}{(Ts + \fP)\Delta_\perp}~~.
\end{gather}
Note that all the correlators with odd number of transverse indices vanish due
to isotropy
\begin{gather}
  \mathcal{G}^{tt,ta}_{TT,R}
  = \mathcal{G}^{tx,ta}_{TT,R}
  = \mathcal{G}^{tt,xa}_{TT,R}
  = \mathcal{G}^{tx,xa}_{TT,R}
  = \mathcal{G}^{xx,xa}_{TT,R}
  = \mathcal{G}^{xx,ta}_{TT,R}
  = \mathcal{G}^{ta,bc}_{TT,R}
  = \mathcal{G}^{xa,bc}_{TT,R} = 0~~, \nn\\
  \mathcal{G}^{x,ta}_{\phi T,R}
  = \mathcal{G}^{x,xa}_{\phi T,R}
  = \mathcal{G}^{a,tt}_{\phi T,R}
  = \mathcal{G}^{a,tx}_{\phi T,R}
  = \mathcal{G}^{a,xx}_{\phi T,R}
  = \mathcal{G}^{a,bc}_{\phi T,R} = 0~~, \qquad
  \mathcal{G}^{x,a}_{\phi\phi,R} = 0~~.
\end{gather}

Upon turning on the dissipative transport coefficients, the two-point functions
become much more involved. However, we report the respective Kubo formulas
\begin{gather}
  \zeta + 2\frac{d-1}{d} \eta
  = - \lim_{\omega\to0}\lim_{k\to0}\frac{\omega^3}{k^4}\mathrm{Im}\,
  G^R_{T^{tt}T^{tt}}~~, \qquad
  \eta
  = - \lim_{\omega\to0}\lim_{k\to0}\frac{\omega}{k^2}\mathrm{Im}\,
  G^R_{T^{tx}T^{tx}}~~, \nn\\
  \frac{T^2s^2}{\sigma (Ts+\fP)^2}
  = \lim_{\omega\to0}\lim_{k\to0}\omega\,\mathrm{Im}\,
  G^R_{\phi^x\phi^x}.
\end{gather}
These can be used to read out the transport coefficients in terms of linear
responses functions.\footnote{We would like to note that these Kubo formulas are
  different from the ones being used in~\cite{Ammon:2019apj} due to the presence
  of lattice pressure. The authors find a discrepancy between their numerical
  results from holography and those predicted by hydrodynamics. It seems quite
  plausible that this mismatch will be resolved upon taking into account the
  lattice pressure in the constitutive relations and Kubo formulas.}

This finishes our quite detailed discussion of viscoelastic fluids. We have
written down the most generic constitutive relations determining the dynamics of
a viscoelastic fluid up to first order in the derivative expansion. In
particular, we specialised to linear isotropic materials and obtained the
respective constitutive relations, modes, and linear response functions. In the
next section we present an equivalent formulation of viscoelasticity in terms of
a fluid with partially broken higher-form symmetries.


\section{Viscoelastic fluids as higher-form superfluids}
\label{sec:higher-form}

In this section we present a formulation of hydrodynamics with partially broken
generalised global symmetries and show their relation to the theory of
viscoelastic fluids formulated in the previous section. Generalised global
symmetries are an extension of ordinary global symmetries with one-form (vector)
conserved currents and point-like conserved charges to higher-form conserved
currents and higher dimensional conserved charges such as strings and
branes~\cite{Gaiotto:2014kfa}. It has been observed that when a fluid with a
one-form symmetry\footnote{A conserved $(k+1)$-rank current
  $J^{\mu_1\ldots\mu_{k+1}}$ is said to be associated with a $k$-form
  symmetry. Consequently, the ordinary global symmetries are zero-form in this
  language.} has its symmetry partially broken along the direction of the fluid
flow, it implements a symmetry-based reformulation of
magnetohydrodynamics~\cite{Grozdanov:2016tdf, Hernandez:2017mch, Armas:2018atq,
  Armas:2018zbe}. In this section, we extend this partial symmetry breaking to
hydrodynamics with multiple higher-form symmetries and show that the resultant
theory is a dual description of viscoelastic fluids with translation broken
symmetries in arbitrary dimensions.  In $d$ spatial dimensions, one requires $d$
number of partially broken $(d-1)$-form symmetries in order to describe
viscoelasticity. The case of $d=2$ involving two one-form symmetries was
considered in~\cite{Grozdanov:2018ewh}, albeit in a very restrictive case and
ignoring the issues that require partial symmetry breaking.  The understanding
of partial symmetry breaking is essential for consistency of higher-form
hydrodynamics with thermal equilibrium partition functions, as has been
previously observed in~\cite{Armas:2018atq,Armas:2018ibg}.

\subsection{A dual formulation}
\label{sec:dual-formulation}

In viscoelastic fluids with translation broken symmetries, all the dependence on
the crystal field $\phi^I$ comes via its derivatives $e^I_\mu$. As we have
argued in \cref{sec:elastic-fluids}, the $\phi^I$ equations of motion can be
used to eliminate $u^\mu e^I_\mu$ in favour of $P^{I\mu} = P^{\mu\nu} e^I_\nu$
and other constituent fields in the theory. Thus, it should be possible to
reformulate the physics of viscoelastic fluids purely in terms of $P^{I\mu}$ and
the hydrodynamic fields $u^\mu$ and $T$, without referring to the microscopic
fields $\phi^I$.

To make this precise, we formally define a set of $d$-form currents associated
with the viscoelastic fluid by Hodge-dualising the derivatives of $\phi^I$
as
\begin{equation}\label{dForm-current-identification}
  J^{I\mu_1\ldots\mu_d} [u^\mu,T,P^{I\mu}; g_{\mu\nu}, K_I^\ext]
  = \epsilon^{\mu\mu_1\ldots\mu_d} \dow_\mu \phi^I
  = \epsilon^{\mu\mu_1\ldots\mu_d} \lb P^I_{\mu} -
  T u_\mu \delta_\scB \phi^I \rb~~.
\end{equation}
It is understood here that the $\phi^I$ equations of motion have been taken
onshell to eliminate $\delta_\scB \phi^I$ in terms of the remaining fields and background sources. 
Due to the symmetry of partial
derivatives, these currents are conserved by construction
\begin{equation} \label{eq:curc}
  \nabla_{\mu_1} J^{I\mu_1\ldots\mu_d} = 0~~.
\end{equation}
A priori, these conservation equations have $d(d+1)/2$ independent components
for every value of $I$ but only $d$ of these contain a time-derivative and
hence govern dynamical evolution, while the remaining $d(d-1)/2$ components are
constraints on an initial Cauchy slice. Conspicuously, these are the exact
number of dynamical equations required to evolve the $d$ physical components in
$P^{I\mu}$ for every value of $I$ (note that $u_\mu P^{I\mu} = 0$).

The conservation equation \eqref{eq:curc} implies that there is a set of $d$ topological conserved charges $Q^I$ of the form
\begin{equation}
Q^I[\Sigma_1]=\int_{\Sigma_1}\star J^I~~,
\end{equation}
where $\Sigma_1$ is a given one-dimensional surface and $\star$ is the Hodge operator in $d+1$ dimensional spacetime. The charges $Q^I[\Sigma_1]$ count the number of lattice hyperplanes that intersect the one-dimensional surface $\Sigma_1$.

The current \eqref{dForm-current-identification} couples to the field strength of a higher-form gauge field.
More precisely, we can replace the external currents $K_I^\ext$ by the field strength such that
\begin{equation}
  K_I^\ext = - \frac{1}{(d+1)!} \epsilon^{\mu_1\ldots\mu_{d+1}} H_{I\mu_1\ldots\mu_{d+1}}~~.
\end{equation}
Since $H_{I\mu_1\ldots\mu_{d+1}}$ is a full-rank form, locally it can be
re-expressed as an exact form
\begin{equation}
  H_{I\mu_1\ldots\mu_{d+1}} = (d+1) \dow_{[\mu_1}b_{I\mu_2\ldots\mu_{d+1}]}~~,
\end{equation}
where $b_{I\mu_1\ldots\mu_d}$ is a $d$-form gauge field defined up to a
$(d-1)$-form gauge transformation
\begin{equation}
  b_{I\mu_1\ldots\mu_d} \to b_{I\mu_1\ldots\mu_d}
  + d\, \dow_{[\mu_1}\Lambda_{I\mu_2\ldots\mu_{d}]}~~.
\end{equation}
Using this definition, the energy-momentum conservation equation \eqref{eq:elastic} takes the form
\begin{equation} \label{eq:stc}
  \nabla_\mu T^{\mu\nu}
  = - K_I^\ext \nabla^\nu \phi^I
  = \frac{1}{d!} H_I^{\nu\mu_1\ldots\mu_{d}} J^I_{\mu_1\ldots\mu_d}~~.
\end{equation}
In essence, we have reformulated viscoelastic fluids in terms of a fluid with
multiple $(d-1)$-form global symmetries. The background $d$-form gauge fields
$b_{I\mu_1\ldots\mu_d}$ couple to the $d$-form currents
$J^{I\mu_1\ldots\mu_d}$. The constitutive relations of viscoelastic fluids can be
equivalently re-expressed as
\begin{equation}
  T^{\mu\nu}[u^\mu, T, P^{I\mu}; g_{\mu\nu}, b_{I\mu_1\ldots\mu_d}]~~,~~
  J^{\mu_1\ldots\mu_d}_I [u^\mu, T, P^{I\mu}; g_{\mu\nu}, b_{I\mu_1\ldots\mu_d}]~~.
\end{equation}
The dynamics of the hydrodynamic fields $u^\mu$ and $T$, and $P^{I\mu}$ is governed
by  energy-momentum conservation \eqref{eq:stc} and $d$-form conservation equations \eqref{eq:curc}.\footnote{Note that, by the definition of the $d$-form currents, we have the following
relation
\begin{equation}\nonumber
  u_{\mu_1} J^{I\mu_1\ldots\mu_d}
  = \epsilon^{\mu\mu_1\ldots\mu_d}  P^I_{\mu} u_{\mu_1}~~,
\end{equation}
which can be understood as a frame choice from the higher-form hydrodynamic
perspective. In general, we can choose a different set of fields in the
hydrodynamic description which are aligned with $P_{I}^\mu$ in this particular
frame, but can be arbitrarily redefined otherwise.}

\subsection{Formalities of higher-form hydrodynamics}

\subsubsection{Ordinary higher-form hydrodynamics}

Having motivated a dual formulation of viscelastic fluids in terms of higher-form
symmetries, we consider higher-form hydrodynamics in its own right, following \cite{Armas:2018atq, Armas:2018ibg}. 
Consider a fluid living in $(d+1)$-dimensions that carries a conserved energy-momentum
tensor $T^{\mu\nu}$ and a $k$ number of conserved $d$-form currents
$J^{I\mu_1\ldots\mu_d}$ where $I=1,2,\ldots,k$. When coupled to a background
metric $g_{\mu\nu}$ and background $d$-form gauge fields
$b_{I\mu_1\ldots\mu_d}$, the associated conservation equations are given as
\begin{equation} \label{eq:higherform}
  \nabla_\mu T^{\mu\nu}
  = \frac{1}{d!} H_I^{\nu\mu_1\ldots\mu_d} J^I_{\mu_1\ldots\mu_d}, \qquad
  \nabla_{\mu_1} J^{I\mu_1\ldots\mu_d} = 0~~.
\end{equation}
In a generic number of dimensions, the conservation equations lead to
$(d + 1 + k d)$ dynamical equations and $kd (d-1)/2$ constraints. From this
counting procedure, it can be checked that eqs.~\eqref{eq:higherform} can provide dynamics for a set of symmetry
parameters
\begin{equation}
  \scB = \lb \beta^\mu~,~ \Lambda^{\beta}_{I\mu_1\ldots\mu_{d-1}} \rb~~.
\end{equation}
Under an infinitesimal symmetry transformation parametrised by $\scX =
(\chi^\mu, \Lambda^{\chi}_{I\mu_1\ldots\mu_{d-1}})$, they transform according to
\begin{equation}
  \delta_\scX \beta^\mu = \lie_\chi \beta^\mu~~,~~
  \delta_\scX \Lambda^{\beta}_{I\mu_1\ldots\mu_{d-1}}
  = \lie_\chi \Lambda^{\beta}_{I\mu_1\ldots\mu_{d-1}}
  - \lie_\beta \Lambda^{\chi}_{I\mu_1\ldots\mu_{d-1}}~~,
\end{equation}
where $\lie_\chi$ denotes the Lie derivative with respect to $\chi^\mu$. Let us
repackage these fields into the fluid velocity $u^\mu$, temperature $T$, and
$(d-1)$-form chemical potentials $\mu_{I\mu_1\ldots\mu_{d-1}}$ according to
\begin{equation}
  \frac{u^\mu}{T} = \beta^\mu~~,~~
  \frac{\mu_{I\mu_1\ldots\mu_{d-1}}}{T}
  = \Lambda^{\beta}_{I\mu_1\ldots\mu_{d-1}} + \beta^\nu b_{I\nu\mu_1\ldots\mu_{d-1}}~~,
\end{equation}
such that $u^\mu u_\mu = -1$. Interestingly, while the fields $u^\mu$ and $T$
are gauge invariant, $\mu_{I\mu_1\ldots\mu_{d-1}}$ transform akin to
$(d-1)$-form gauge fields
\begin{equation}
  \mu_{I\mu_1\ldots\mu_{d-1}} \to
  \mu_{I\mu_1\ldots\mu_{d-1}}
  - (d-1) T \dow_{\mu_1} \lb \beta^\nu \Lambda_{I\nu\mu_1\ldots\mu_{d-1}} \rb~~.
\end{equation}
Hence, they have the required $kd$ physical degrees of freedom, which, along with
 $u^\mu$ and $T$, match the number of dynamical components of the
conservation equations. 

Similar to our discussion in \cref{sec:conventional-setup}, higher-form fluids
need to obey a version of the second law of thermodynamics. In the current
context, this statement translates into the existence of an entropy
current $S^\mu$ that satisfies
\begin{equation}\label{eq:second-law}
  \nabla_\mu S^\mu
  + \frac{u_\nu}{T} \lb \nabla_\mu T^{\mu\nu}
  - \frac{1}{d!} H_I^{\nu\mu_1\ldots\mu_{d}} J^I_{\mu_1\ldots\mu_d} \rb
  - \frac{1}{(d-1)!}\frac{\mu_{I\mu_1\ldots\mu_{d-1}}}{T}
  \lb \nabla_{\mu} J^{I\mu\mu_1\ldots\mu_{d-1}} \rb = \Delta \geq 0~~.
\end{equation}
Compared to \cref{eq:elastic2ndLaw}, here we have also taken into account the
higher-form conservation equation and the respective multiplier has been chosen
to be $\mu_{I\mu_1\ldots\mu_{d-1}}/T$ using the inherent redefinition freedom in
the higher-form chemical potential. It is straightforward to formulate a higher-form
analogue of the adiabaticity equation to ease the derivative of the
constitutive relations. Defining the free energy current
\begin{equation}\label{FreeE-identification}
  N^\mu = S^\mu + \frac1T T^{\mu\nu}  u_\nu
  - \frac{1}{(d-1)!} \frac1T J^{I\mu\mu_1\ldots\mu_{d-1}}
  \mu_{I\mu_1\ldots\mu_{d-1}}~~,
\end{equation}
the adiabaticity equation for higher-form fluids reads
\begin{equation}\label{eq:adiabaticity}
  \nabla_\mu N^\mu
  = \half T^{\mu\nu}  \delta_\scB g_{\mu\nu}
  + \frac{1}{d!} J^{I\mu_1\ldots\mu_d} \delta_\scB b_{I\mu_1\ldots\mu_d}
  + \Delta~~,~~
  \Delta \geq 0~~.
\end{equation}
We have identified the variations of the various background fields according to
\begin{equation}
  \delta_\scB g_{\mu\nu} = 2 \nabla_{(\mu} \beta_{\nu)}~~,~~
  \delta_\scB b_{I\mu_1\ldots\mu_d}
  = - d\, \dow_{[\mu_1} \lb \frac1T \mu_{I\mu_2\ldots\mu_{d}]}  \rb
  + \beta^\nu H_{I\nu\mu_1\ldots\mu_d}~~.
\end{equation}
To obtain the constitutive relations of a higher-form fluid, it is required to
find the most generic expressions for $T^{\mu\nu}$ and $J^{I\mu_1\ldots\mu_d}$
in terms of the dynamical fields $u^\mu$, $T$, and
$\mu_{I\mu_1\ldots\mu_{d-1}}$, as well as background fields $g_{\mu\nu}$ and
$b_{I\mu_1\ldots\mu_d}$, arranged in a derivative expansion, that satisfy
\cref{eq:adiabaticity} for some $N^\mu$ and $\Delta$.

\subsubsection{Partial symmetry breaking}

When a higher-form symmetry is partially broken in its ground state, the
hydrodynamic description should include the associated Goldstone modes
$\varphi_{I\mu_1\ldots\mu_{d-2}}$ with
$u^{\mu_1} \varphi_{I\mu_1\ldots\mu_{d-2}} = 0$, that transform according
to\footnote{If the $(d-1)$-form symmetries were completely broken, we would
  instead introduce the $(d-1)$-form Goldstone fields
  $\phi_{I\mu_1\ldots\mu_{d-1}}$ that shift under a background gauge
  transformation according to
  \begin{equation} \nonumber
    \delta_\scX \phi_{I\mu_1\ldots\mu_{d-1}}
    = \lie_\chi \phi_{I\mu_1\ldots\mu_{d-1}}
    - \Lambda_{I\mu_1\ldots\mu_{d-1}}~~.
  \end{equation}
  The $(d-2)$-form Goldstones of partial symmetry breaking are essentially the
  components of the full Goldstones along the direction of the fluid flow, that is
  $\varphi_{I\mu_1\ldots\mu_{d-2}} =
  \beta^{\mu}\phi_{I\mu\mu_1\ldots\mu_{d-2}}$.  }
\begin{equation}
  \delta_\scX \varphi_{I\mu_1\ldots\mu_{d-2}}
  = \lie_\chi \varphi_{I\mu_1\ldots\mu_{d-2}}
  - \beta^\mu \Lambda_{I\mu\mu_1\ldots\mu_{d-2}}.
\end{equation}
This allows us to define a gauge-invariant version of the $(d-1)$-form chemical
potentials
\begin{equation}
  \zeta_{I\mu_1\ldots\mu_{d-1}} = (d-1)\, T \dow_{[\mu_1}
  \varphi_{I\mu_2\ldots\mu_{d-1}]}
  - \mu_{I\mu_1\ldots\mu_{d-1}}~~.
\end{equation}

The Goldstone fields are reminiscent of the crystal fields $\phi^I$ from
sec.~\ref{sec:conventional-setup} and are accompanied by their own equations of
motion, which can be schematically represented as
\begin{equation}
  K^{I\mu_1\ldots\mu_{d-2}} = 0~~.
\end{equation}
As in the previous formulation, the operator $K^{I\mu_1\ldots\mu_{d-2}}$ is not predetermined without the
knowledge of the microscopics but we can fix its form up to certain transport
coefficients by imposing the second law of thermodynamics. In this context,
this translates into the requirement that the fluid must admit an entropy current $S^\mu$
whose divergence must be positive-semi-definite in any arbitrary
$\varphi_I$-offshell configuration. Fixing the Lagrange multipliers associated
with the respective conservation equations to be $\beta^\mu$ and
$\zeta_{I\mu_1\ldots\mu_{d-1}}$, the statement of the second
can be written as
\begin{multline}\label{eq:second-law-K}
  \nabla_\mu S^\mu
  + \frac{u_\nu}{T} \lb \nabla_\mu T^{\mu\nu}
  - \frac{1}{d!} H_I^{\nu\mu_1\ldots\mu_{d}} J^I_{\mu_1\ldots\mu_d} \rb \\
  - \frac{1}{(d-1)!}\frac{\zeta_{I\mu_1\ldots\mu_{d-1}}}{T}
  \lb \nabla_{\mu} J^{I\mu\mu_1\ldots\mu_{d-1}}
  - (d-1)\, \beta^{\mu_1} K^{I\mu_2\ldots\mu_{d-1}} \rb
  = \Delta \geq 0~~,
\end{multline}
which is different than the corresponding statement in the symmetry unbroken phase
given in \eqref{eq:second-law}.  Defining the free energy current
\begin{equation}
  N^\mu = S^\mu + \frac1T T^{\mu\nu}  u_\nu
  - \frac{1}{(d-1)!} \frac1T J^{I\mu\mu_1\ldots\mu_{d-1}}
  \zeta_{I\mu_1\ldots\mu_{d-1}}~~,
\end{equation}
the associated adiabaticity equation becomes
\begin{equation}\label{eq:adiabaticity-K}
  \nabla_\mu N^\mu
  = \half T^{\mu\nu}  \delta_\scB g_{\mu\nu}
  + \frac{1}{d!} J^{I\mu_1\ldots\mu_d} \delta_\scB b_{I\mu_1\ldots\mu_d}
  + \frac{1}{(d-2)!} K^{I\mu_1\ldots\mu_{d-2}}
  \delta_\scB \varphi_{I\mu_1\ldots\mu_{d-2}}
  + \Delta~~,
\end{equation}
with $ \Delta\geq 0$ and where
\begin{equation}
  \delta_\scB \varphi_{I\mu_1\ldots\mu_{d-2}}
  = \frac{1}{T^2} u^\mu \zeta_{I\mu\mu_1\ldots\mu_{d-2}}~~.
\end{equation}

Similar to \cref{sec:conventional-setup}, we can use the leading order
adiabaticity equation \bref{eq:adiabaticity-K} to obtain the leading order
version of the $\varphi_{I\mu_1\ldots\mu_{d-2}}$ equation of motion, namely
\begin{equation}
  \delta_\scB \varphi_{I\mu_1\ldots\mu_{d-2}} = \mathcal{O}(\dow)
  \qquad\implies\qquad
  u^{\mu_1} \zeta_{I\mu_1\ldots\mu_{d-1}} = \mathcal{O}(\dow)~~.
\end{equation}
We can use some of the inherent redefinition freedom in
$\mu_{I\mu_1\ldots\mu_{d-1}}$ to convert this into an exact all order
statement.  Consequently, it is possible to take
$\varphi_{I\mu_1\ldots\mu_{d-2}}$ formally on-shell setting
$\delta_\scB \varphi_{I\mu_1\ldots\mu_{d-2}} = 0$, following which, the
adiabaticity equation \bref{eq:adiabaticity-K} turns into its
``symmetry-unbroken'' version \bref{eq:adiabaticity}. The constitutive relations
of a higher-form fluid with partially broken symmetry are given by the most
generic expressions for $T^{\mu\nu}$ and $J^{I\mu_1\ldots\mu_d}$ in terms of the
dynamical fields $u^\mu$, $T$, and $\zeta_{I\mu_1\ldots\mu_{d-1}}$ (with
$u^\mu u_\mu = -1$ and $u^{\mu_1}\zeta_{I\mu_1\ldots\mu_{d-1}} = 0$), and
background fields $g_{\mu\nu}$ and $b_{I\mu_1\ldots\mu_d}$, arranged in a
derivative expansion, that satisfy \cref{eq:adiabaticity} for some $N^\mu$ and
$\Delta$.

While $\zeta_{I\mu_1\ldots\mu_{d-1}}$ is fundamentally more transparent, for
most of the explicit computations, it will be helpful to work with a
Hodge-dualised version
\begin{equation}
  \psi^{\mu}_I = \frac{1}{(d-1)!}\epsilon^{\mu\mu_1\ldots\mu_d} u_{\mu_1}
  \zeta_{I\mu_2\ldots\mu_d}~~,~
  \zeta_{I\mu_1\ldots\mu_{d-1}}
  = \epsilon_{\mu\nu\mu_1\ldots\mu_{d-1}} \psi_I^{\mu} u^{\nu}~~,
\end{equation}
where we note that $u_\mu \psi_I^\mu = 0$. This representation makes it clear that
$\psi_I^\mu$, and hence $\zeta_{I\mu_1\ldots\mu_{d-1}}$, have the same degrees
of freedom as $P^{I\mu}$ and can be used as a fundamental hydrodynamic field for
viscoelastic fluids instead. The relation between the two is typically non-trivial
and needs to be obtained order-by-order in the derivative expansion.

Compared to \cref{eq:adiabaticity-elastic}, the adiabaticity equation
\bref{eq:adiabaticity} in the dual formulation does not exhibit order mixing,
as both $\delta_\scB g_{\mu\nu}$ and $\delta_\scB b_{I\mu_1\ldots\mu_d}$ are
$\mathcal{O}(\dow)$. This allows for a more transparent analysis of the
constitutive relations, as we shall see in the next section. Another benefit of
working in the dual formulation is that we directly obtain the constitutive
relations for (the Hodge dual of) the physically observable crystal momenta,
rather than for the equations of motion of the crystal fields. This can considerably
simplify the computation of the respective correlation functions and Kubo
formulae, as in the case of magnetohydrodynamics \cite{Armas:2018ibg}.

\subsection{Revisiting ideal viscoelastic fluids}

The constitutive relations of an ideal viscoelastic fluid in higher-form language are
characterised by an ideal order free energy current
\begin{equation} \label{eq:freehigher}
  N^\mu = p(T,\gamma_{IJ})\, \beta^\mu + \mathcal{O}(\dow)~~.
\end{equation}
Here $p(T,\gamma_{IJ})$ is an arbitrary function of all the available ideal
order scalars in the theory, namely the temperature $T$ and matrix
\begin{equation}
  \gamma_{IJ}
  = \frac{1}{(d-1)!}\zeta_{I\mu_1\ldots\mu_{d-1}} \zeta_{I\nu_1\ldots\nu_{d-1}}
  g^{\mu_1\nu_1}\ldots g^{\mu_{d-1}\nu_{d-1}}
  = \psi^{\mu}_I \psi^{\nu}_J g_{\mu\nu}~~.
\end{equation}
Introducing this into \cref{eq:adiabaticity} and noting that\footnote{Note that
  $\zeta_I{}^{\mu}{}_{\mu_2\ldots\mu_{d-1}} \zeta_J^{\nu\mu_2\ldots\mu_{d-1}}
  =(d-2)! (\gamma_{IJ} P^{\mu\nu} - \psi_I^{\nu} \psi_J^{\mu})$.}
\begin{equation}
  \delta_\scB \gamma_{IJ}
  = - \frac{2}{(d-1)!}  u^{\mu_1}
  \zeta_{(I}^{\mu_2\ldots\mu_d} \delta_\scB b_{J)\mu_1\ldots\mu_{d}}
  + \lb \psi^{\mu}_I \psi_J^{\nu} - \gamma_{IJ} g^{\mu\nu} \rb \delta_\scB g_{\mu\nu}~,
\end{equation}
we obtain the ideal viscoelastic fluid constitutive relations\footnote{In \cite{Armas:2018ibg}, we have given a formulation of higher-form hydrodynamics with a single conserved current. In app.~\ref{app:hfh} we provide the comparison between the ideal order higher-form hydrodynamics of this section with that of \cite{Armas:2018ibg}.}
\begin{align} \label{eq:tcSF}
  T^{\mu\nu}
  &= \lb \epsilon + p \rb u^\mu u^\nu
    + p g^{\mu\nu}
    - \frac{q^{IJ}}{(d-2)!}
    \zeta_I^{\mu}{}_{\mu_2\ldots\mu_{d-1}} \zeta_J^{\nu\mu_2\ldots\mu_{d-1}}
    + \mathcal{O}(\dow)~~, \nn\\
  J^{I\mu_1\ldots\mu_d}
  &= - d\, q^{IJ} u^{[\mu_1} \zeta_J^{\mu_2\ldots\mu_{d}]}
    + \mathcal{O}(\dow)~~.
\end{align}
Here we have defined
\begin{equation} \label{eq:thermoSF}
  \df p = s\df T + \half q^{IJ} \df \gamma_{IJ}~~,~~
  \epsilon + p = sT + q^{IJ}\gamma_{IJ}~~.
\end{equation}
These are the most generic constitutive relations of an ideal viscoelastic fluid in
higher-form language. As promised earlier, there is no order-mixing across
derivative orders in this formulation. It is useful to explicitly work out the equations of motion for ideal
viscoelastic fluids, which result in
\begin{gather}
  \nabla_\mu (s u^\mu) = 0~~, \nn\\
  \lb s T P^{\nu\mu} + q^{IJ} \psi^\nu_{J} \psi^\mu_I \rb
  u^\lambda \delta_\scB g_{\mu\lambda}
  + \frac{1}{d!} q^{IJ} \psi^\nu_{J} \epsilon^{\mu\mu_1\ldots\mu_{d}} u_{\mu}
  \delta_\scB b_{I\mu_1\ldots\mu_d} = 0~~, \nn\\
  \nabla_{[\mu} \lb q^{IJ} \psi_{J\nu]} \rb = 0~~.
\end{gather}

In order to find the relation between the two formalisms, we need to perform the
identification according to \cref{dForm-current-identification}, which results
in the following map between formulations
\begin{equation}
  e^{I}_\mu = q^{IJ} \psi_{J\mu}
  + \mathcal{O}(\dow)
  \qquad\implies\qquad
  h^{IJ} = q^{IK} q^{JL} \gamma_{KL} + \mathcal{O}(\dow)~~.
\end{equation}
Additionally, introducing this into the energy-momentum tensor, we find
\begin{gather} \label{eq:map} p = P - r_{IJ} h^{IJ} + \mathcal{O}(\dow)~~,~~
 \nn\\
  q^{IJ} = - q^{IK} q^{JK} r_{KL} \implies q^{IJ} = - (r^{-1})^{IJ}~~,
\end{gather}
while temperature, entropy density and energy density 
agree in both formulations. 

\subsection{One derivative corrections}
\label{sec:dual-one-der-corrections}

Following the same arguments as the previous subsection, we consider the
hydrostatic free energy density in the dual picture
\begin{equation} \label{eq:hydroparthigher}
  \mathcal{N}
  = p + \tilde f_1^I \frac{1}{T} \psi_I^{\mu} \dow_\mu T
  + 2T \tilde f_2^{[IJ]} \psi_I^{\mu} \psi_J^{\nu} \dow_{[\mu} u_{\nu]}
  + \tilde f_3^{I(JK)} \psi_{I\mu} \dow_\mu \gamma_{JK}~~.
\end{equation}
The hydrostatic constitutive relations can be obtained using the variations
given above and refer the reader to \cref{app:hs-details} for details. In turn,
in the non-hydrostatic sector, we can expand
\begin{equation}
  T^{\mu\nu}_{\text{nhs}} = \psi_I^{\mu}\psi_J^{\nu} \mathcal{T}^{IJ}~~,~~
  J^{I\mu_1\ldots\mu_{d}}_{\text{nhs}}
  = \epsilon^{\mu\mu_1\ldots\mu_d} u_\mu \mathcal{J}^{I}~.
\end{equation}
The most general non-hydrostatic constitutive relations are correspondingly 
\begin{equation}
  \begin{pmatrix}
    \mathcal{T}^{IJ} \\ \mathcal{J}^I
  \end{pmatrix}
  =
  - T \begin{pmatrix}
    \tilde\eta^{IJKL} &
    \tilde\chi^{IJK} \\
    \tilde\chi'^{IKL} &
    \tilde\sigma^{IK}
  \end{pmatrix}
  \begin{pmatrix}
    \half \psi_K^{\mu} \psi_L^{\nu} \delta_\scB g_{\mu\nu} \\
    \frac{1}{d!} \epsilon^{\mu\mu_1\dots\mu_d}u_\mu \delta_\scB b_{K\mu_1\ldots\mu_d}
  \end{pmatrix}~~.
\end{equation}
The transport coefficient matrices have necessary symmetry properties. The positivity constraint requires that the symmetric part of the $5\times 5$  transport
coefficient matrix is positive semi-definite.

In order to provide the map of transport coefficients at first order between the
two formulations, we first use the identification in
\cref{dForm-current-identification} in order to obtain
\begin{equation}
  P^I_{\mu}
  =
  \frac{1}{(d-1)!} \epsilon_{\mu\mu_1\ldots\mu_d} u^{\mu_1} u_\nu J^{I\nu\mu_2\ldots\mu_d}~~,~~
  u^\mu \dow_\mu \phi^I
  = - \frac{1}{d!} \epsilon_{\mu\mu_1\ldots\mu_d} u^\mu J^{I\mu_1\ldots\mu_d}~~.
\end{equation}
This provides a definition of $P_{I}^\mu$ and $\delta_\scB \phi^I$ in the
conventional formulation in terms of the dual formulation variables
\begin{align}\label{eq:variable_map_1der}
  P^{I\mu}
  &= \lB q^{IJ} 
    + \frac{2}{T} \frac{\dow \tilde f_1^K }{\dow \gamma_{IJ}}
    \psi_K^{\lambda} \dow_\lambda T
    + 4T \frac{\dow \tilde f_2^{[KL]}}{\dow \gamma_{IJ}} \psi^{\rho}_K
    \psi^{\sigma}_L \dow_{[\rho} u_{\sigma]}
    + 2 \frac{\dow \tilde f_3^{L(MK)}}{\dow \gamma_{IJ}}
    \psi^{\lambda}_L \dow_\lambda \gamma_{MK}
    - 2 \nabla_\rho \lb \tilde f_3^{K(IJ)} \psi^{\rho}_K \rb
     \rB \psi^{\mu}_J \nn\\
  &\qquad
    + \tilde f_1^I  \frac{1}{T} P^{\mu\nu}\dow_\nu T
    + 4T \tilde f_2^{[IJ]} P^{\mu\nu} \psi^{\rho}_J \dow_{[\nu} u_{\rho]} 
    + \tilde f_3^{I(JK)} P^{\mu\nu} \dow_\nu \gamma_{JK}~~, \nn\\
  \delta_\scB \phi^I
  &= \tilde\chi'^{IKL} \half \psi_K^{\mu} \psi_L^{\nu} \delta_\scB g_{\mu\nu}
    + \tilde\sigma^{IK} \frac{1}{d!} \epsilon^{\mu\mu_1\dots\mu_d}u_\mu \delta_\scB b_{K\mu_1\ldots\mu_d}.
\end{align}

We wish to begin the comparison with the free energy currents in the two
formulations. Using \cref{FreeE-identification}, we know that
\begin{equation}
  N^\mu
  = N^\mu_{\text{elastic}}
  + \frac{u^\mu}{T}
   \psi^{I\nu} P_{I\nu}
  - \psi^{I\mu} \delta_\scB \phi_I~~.
\end{equation}
Using the map \bref{eq:variable_map_1der} and the results from
\cref{app:hs-details}, we infer the map between the hydrostatic transport
coefficients
\begin{gather}
  p(T,\gamma_{IJ}) = P(T, q^{IK} q^{JL} \gamma_{KL}) + q^{IJ} \gamma_{IJ}~~, \nn\\
  \tilde f_1^{I}
  = f^1_J q^{JI} 
  + 2T q^{IM} q^{JN} \gamma_{NL} \frac{\dow q^{KL}}{\dow T} f^3_{M(JK)}~~, \nn\\
  \tilde f_2^{[IJ]} = f^2_{[MN]} q^{MI} q^{NJ}~~, \nn\\
  \tilde f_3^{I(JK)}
  = f^3_{N(LM)} q^{IN} q^{JL} q^{KM}
  + 2 q^{IM} q^{AN} \gamma_{NL} \frac{\dow q^{BL}}{\dow \gamma_{JK}} f^3_{M(AB)}~~.
\end{gather}
To obtain the map in the non-hydrostatic sector, we need to compare the
energy-momentum tensors and $\delta_\scB \phi^I$ in the two formulations. In
hindsight, we allow for a relative field redefinition of the fluid velocity
between the two formulations, i.e
$u^\mu_{\text{elastic}} = u^\mu + \delta u^\mu$ with $u_\mu \delta u^\mu =
0$. The $\varphi^I$-equation of motion \bref{dBphiI-map}, upon using the said
field redefinition, implies that
\begin{align}\label{eq:dphi-map}
  \sigma_{IJ} \delta_\scB \phi^J
  + \frac{1}{T} \sigma_{IJ} e^J_\lambda \delta u^\lambda
  &= \frac{1}{T}\bigg(
  K_J^\ext
  - \nabla_\mu \lb r_{JK} e^{K\mu} \rb
  - \chi'_{JKL} P^{K\mu} P^{L\nu} \nabla_{(\mu} u_{\nu)} \bigg) \nn\\
  &= \lb \frac{1}{d!} \epsilon^{\lambda\mu_1\ldots\mu_d} u_\lambda
  \delta_\scB b_{I\mu_1\ldots\mu_d} 
  - \frac{1}{2} \chi'_{IMN} q^{MK} q^{NL} \psi^{\mu}_K \psi^{\nu}_K
  \delta_\scB g_{\mu\nu} \rb.
\end{align}
On the other hand, using the results from \cref{app:hs-details}, it is
straight-forward, albeit cumbersome, to obtain that
\begin{equation}
  T^{\mu\nu}_{\text{elastic,hs}}
  = T^{\mu\nu}_{\text{hs}}
  + 2T u^{(\mu} \lb s \delta u^{\nu)}
  - \psi^{\nu)}_I \delta_\scB \phi^I \rb.
\end{equation}
Given that all the non-hydrostatic corrections are in the Landau frame, we must
choose the velocity field-redefinition to be
$\delta u^\mu = \psi^{\mu}_I\delta_\scB \phi^I/s$, mapping the hydrostatic
sectors of the two formulations to each other. Comparing $\delta_\scB \phi^J$
from \cref{eq:dphi-map} to \cref{eq:variable_map_1der} and comparing the
non-hydrostatic energy-momentum tensors in the two formulations, we obtain the
map
\begin{gather}
  \tilde\eta^{IJKL}
  = q^{IA}q^{JB}q^{KC}q^{LD}
  \lb \eta_{ABCD} - \chi_{ABR} \tilde\sigma^{RS} \chi'_{SCD} \rb~~,
  \nn\\
  \tilde\chi^{IJK}
  = q^{IA}q^{JB} \chi_{ABC} \tilde\sigma^{CK}~~, \nn\\
  \tilde\chi'^{IJK} = - \tilde\sigma^{IL} 
  \chi'_{LMN} q^{JM} q^{KN}~~, \nn\\
  (\tilde\sigma^{-1})_{IJ} = \sigma_{IJ} 
  + \frac{\sigma_{IK}}{Ts} q^{KL} \gamma_{LJ}~~.
\end{gather}

This completes the formulation of viscoelastic hydrodynamics in terms of
generalised global symmetries and shows that it can exactly accommodate
viscoelastic hydrodynamics with broken translation invariance.  In the next
section we look at particular realisations of both these formulations in the
context of holography.

\section{Conformal viscoelastic fluids and holography}
\label{sec:holography}

In this section we provide, and study the properties of, holographic models in
$D=4,5$ bulk dimensions (i.e. $d=2,3$ spatial dimensional fluids). The models we
consider in general break conformal symmetry due to double trace deformations
but conformal symmetry can be recovered in a specific case.  Thus, in the
beginning of this section we consider conformal fluids. In connection with
viscoelastic holography, we consider two classes of models that have been
considered in the literature. The first class of models has translation broken
symmetries involving a set of $(D-2)$ scalar fields $\Phi_I$ minimally coupled
to gravity. The second class is formulated in the context of generalised global
symmetries and involve a set of $(D-2)$ gauge fields $B_{Ia_1...a_{D-2}}$
minimally coupled to gravity \cite{Grozdanov:2018ewh}. The latter class
describes particular equilibrium states of the higher-form hydrodynamics
described in sec.~\ref{sec:higher-form}, which we explicitly show by
generalising the work of \cite{Grozdanov:2018ewh} to $D=5$. Given that in this
case the dual fluid is governed by conservation equations alone (i.e. no
dynamical fields), which has been one of the motivations in the holographic
digressions of \cite{Hofman:2017vwr, Grozdanov:2017kyl, Grozdanov:2018ewh}, we
consider it first. Later, generalising aspects of \cite{Faulkner:2012gt}, we
``dualise'' the model with higher-form symmetries and obtain the class of
viscoelastic models with translation broken symmetries, which consist of the
model of \cite{Andrade:2013gsa} but with an alternative quantisation of the
scalar fields and a double trace deformation of the boundary theory. We show
that this process results in the dual fluid given in \cref{sec:elastic-fluids}.

\subsection{Conformal viscoelastic fluids}
\label{sec:conformalfluids}

A viscoelastic fluid is said to be conformal if it is invariant under the
conformal rescaling of the background metric
$g_{\mu\nu} \to \Omega^2 g_{\mu\nu}$ for some arbitrary function $\Omega(x)$. In
practice, it implies that the energy-momentum tensor of the theory is traceless (modulo conformal anomalies)
and the constitutive relations are only constructed out of the conformal
covariants.

\subsubsection{Constitutive relations}

\begin{subequations}  \label{eq:conformal-constraint}
Focusing on the non-anomalous case, setting the trace of the energy-momentum tensor \eqref{eq:net-EM} to zero, we
get certain constraints on the respective transport coefficients, namely
  \begin{gather}\label{eq:conformal-constraint-1}
    \epsilon = d\, P - r_{IJ} h^{IJ}, \qquad f^1_I = f^3_{I(JK)} = 0~~, \qquad
    (d-3) f^2_{[IJ]} - T \frac{\dow f^2_{[IJ]}}{\dow T}
    - 2 h^{KL} \frac{\dow f^2_{[IJ]}}{\dow h^{KL}}  = 0~~, \nn\\
    h^{IJ} \eta_{IJKL} = h^{IJ}\chi_{IJK} = 0~~.
  \end{gather}
  Furthermore, requiring $T^{\mu\nu}$ and $K_I$ to only involve conformal
  covariants requires
  \begin{equation}
    \eta_{IJKL} h^{KL} = \chi'_{IKL} h^{KL} = 0~~.
  \end{equation}
\end{subequations}
The first equation in \cref{eq:conformal-constraint-1} determines the energy
density in conformal fluids as expected.
At the one-derivative hydrostatic order, we see that we are only left with
$f^2_{[IJ]}$ which is the only conformal covariant term in the free-energy
current. In the non-hydrostatic sector, we essentially just eliminate the
conformal-non-invariant $\nabla_\mu u^\mu$ term from the constitutive
relations. Consequently we get
\begin{align}\label{eq:consti-conformal}
  T^{\mu\nu}_{\text{conformal}}
  &= (\epsilon + P) u^\mu u^\nu + P g^{\mu\nu}
    - r_{IJ} e^{I\mu} e^{J\nu}
    + T^{\langle\mu\nu\rangle}_{f_2} \nn\\
  &\qquad
    - P^{I\langle\mu} P^{J\nu\rangle} \eta_{IJKL} P^{K\langle\rho} P^{L\sigma\rangle}
    \nabla_{\rho} u_{\sigma}
    - P^{I\langle\mu} P^{J\nu\rangle} \chi_{IJK} u^\rho \dow_\rho \phi^K
    + \mathcal{O}(\dow^2)~~,
\end{align}
and
\begin{equation}
  u^\mu \dow_\mu \phi^I\big|_{\text{conformal}}
  = (\sigma^{-1})^{IJ}\lB
  K_J^\ext
  - \nabla_\mu \lb r_{JK} e^{K\mu} \rb
  - \chi'_{JKL} P^{K\langle\mu} P^{L\nu\rangle} \nabla_{\mu} u_{\nu} \rB
  + \mathcal{O}(\dow^2)~~.
\end{equation}
We will now focus on a special case of these constitutive relations.

\subsubsection{Linear conformal isotropic materials}
\label{sec:linear-conformal}

For conformal viscoelastic fluids truncated to linear order in strain, we need
to additionally impose the constraints \eqref{eq:conformal-constraint} on top of
the constitutive relations \cref{eq:linear-isotropic-consti}, leading to
\begin{equation}\label{eq:conformal-linear-constraints}
  T \dow_T P_{\text{f}}
  = (d+1)\, P_{\text{f}} + d\,\fP~~,~~
  T \dow_T\fP
  = (d+1)\,\fP - d\,\fB~~,~~
  \zeta = \zeta^u_1 = \zeta^u_2 = \bar \zeta^u = 0~~.
\end{equation}
This gives the following constitutive relations of a linear conformal viscoelastic fluid
\begin{align}\label{eq:linear-isotropic-consti-conformal}
  T^{\mu\nu}
  &= (d+1)\, P_{\text{f}}\, u^\mu u^\nu
    + P_{\text{f}}\, g^{\mu\nu}
    - \eta\, \sigma^{\mu\nu} \nn\\
  &\qquad
    + T \dow_T\fP\, u^\lambda{}_{\!\!\lambda}
    \lb u^\mu u^\nu + \frac{1}{d} h^{\mu\nu} \rb
    + \fP\, \lb
    d\, u^\mu u^\nu
    +  h^{\mu\nu}
    + u^\lambda{}_{\!\!\lambda}\lb g^{\mu\nu}
    - \frac{d+1}{d} h^{\mu\nu} \rb \rb \nn\\
  &\qquad
    - 2 \fG\, \lb u^{\mu\nu}
    - \frac{1}{k} h^{\mu\nu} u^\lambda{}_{\!\!\lambda} \rb 
    - \eta^u_1\, u^\lambda{}_{\!\!\lambda} \sigma^{\mu\nu}
    - \eta^u_2\lb u^{(\mu}{}_{\!\sigma} \sigma^{\nu)\sigma}
    - \frac{1}{d} P^{\mu\nu} u_{\rho\sigma} \sigma^{\rho\sigma} \rb
    + \mathcal{O}(u^2)~~,
\end{align}
while the $\phi^I$ equation of motion is still given by \cref{eq:jopimp}. In the
special case that the internal pressure of the lattice $\fP=0$, we infer that
the bulk modulus $\fB = 0$ and these constitutive relations, along with the
$\phi^I$ equations of motion \eqref{eq:jopimp}, simplify to
\begin{align}\label{eq:linear-isotropic-consti-conformal-noP}
  T^{\mu\nu}_{\text{conformal}}
  &= (d+1)\, P_{\text{f}}\, u^\mu u^\nu
    + P_{\text{f}}\, g^{\mu\nu}
    - \eta\, \sigma^{\mu\nu}  \nn\\
  &\qquad
    - 2 \fG\, \lb u^{\mu\nu}
    - \frac{1}{k} h^{\mu\nu} u^\lambda{}_{\!\!\lambda} \rb 
    - \eta^u_1\, u^\lambda{}_{\!\!\lambda} \sigma^{\mu\nu}
    - \eta^u_2\lb u^{(\mu}{}_{\!\sigma} \sigma^{\nu)\sigma}
    - \frac{1}{d} P^{\mu\nu} u_{\rho\sigma} \sigma^{\rho\sigma} \rb
    + \mathcal{O}(u^2)~~, \nn\\
  u^\mu \dow_\mu \phi^I\big|_{\text{conformal}}
  &= \frac{1}{\sigma} h^{IJ}\lB
    K_J^\ext
    - \nabla_\mu \lb 2 \fG \lb u_{JK}
    - \frac{1}{k} h_{JK} u^\lambda{}_{\!\!\lambda} \rb e^{K\mu} \rb \rB \nn\\
  &\qquad
    - \frac{1}{\sigma^2} \lb
    \sigma^u_1\, u^\lambda{}_{\!\!\lambda} h^{IJ}
    + \sigma^u_2\, u^{IJ} \rb
    K_J^\ext + \mathcal{O}(u^2)~~.
\end{align}
The first line of the energy-momentum tensor represents the constitutive
relations for an ordinary uncharged conformal fluid. The second line has the
expected shear modulus term along with the variants of shear viscosities
$\eta^u_1$, $\eta^u_2$ representing the coupling of the conformal fluid to the
strain of the crystal.

We would like to note that the constitutive relations
\eqref{eq:linear-isotropic-consti-conformal} are, in principle, different from
the ones obtained in~\cite{Fukuma:2012ws}.  The strain $u_{\mu\nu}$ defined in
\cref{eq:defn-strain} transforms inhomogeneously under a conformal
transformation, i.e.
$u_{\mu\nu} \to \Omega^2 u_{\mu\nu} + \half\lb \Omega^2 - 1 \rb
\mathbb{h}_{\mu\nu}$, because conformal rescaling only acts on the physical
distances and not on the reference distances between the crystal cores. It
follows that the transport coefficients $\fP(T)$, $\fG(T)$, and
$\eta^u_{1,2}(T)$ appearing in \cref{eq:linear-isotropic-consti-conformal} do
not have a homogeneous conformal scaling. This is in contrast
to~\cite{Fukuma:2012ws}, which chooses the conformal transformations to scale
the reference metric as well, i.e.
$\mathbb h_{IJ} \to \Omega^2 \mathbb h_{IJ}$, leading to a homogeneous scaling
of the strain tensor $u_{\mu\nu} \to \Omega^2 u_{\mu\nu}$. In turn, the
coefficients $\fP(T)$, $\fG(T)$, and $\eta^u_{1,2}(T)$ will all scale
homogeneously. With this alternative choice, however, invariance of the
partition function under conformal transformations does not agree with a
traceless energy-momentum tensor. To wit, 
\begin{equation}
  - \delta_\Omega\ln\mathcal{Z}
  = \half \Omega^2 g_{\mu\nu} \langle T^{\mu\nu}
  \rangle - \Omega^2 \mathbb h_{IJ} (\delta \ln\mathcal{Z}/\delta \mathbb h_{IJ})
  = 0
  \qquad\implies\qquad
  g_{\mu\nu} \langle T^{\mu\nu} \rangle \neq 0~~,
\end{equation}
up to anomalies. Furthermore, we find that the conformal viscoelastic fluids
obtained from holographic models below lead to inhomogenously scaling transport
coefficients, thus the scaling proposed in~\cite{Fukuma:2012ws} describing the case in which the 
reference metric also transforms under conformal transformations, does not describe the conformal
fluids that appear in holographic models.



\subsubsection{Modes}

Specialising to the conformal case, we can revisit the modes of linear
fluctuations obtained in \cref{sec:scalarperturbed}. Using
\cref{eq:conformal-linear-constraints} we find that the speed of
longitudinal/transverse sound and diffusive constant are determined to be
\begin{gather}
  v_\parallel^2
  = \frac1d
  + 2 \frac{d-1}{d} \frac{\fG}{Ts+\fP}~~, \qquad
  \Gamma_\parallel
  =  \frac{T^2 s^2}{\sigma (Ts + \fP)^2}
  \frac{4\frac{(d-1)^2}{d} \fG^2}{Ts+\fP+ 2 (d-1) \fG}
  + \frac{2\frac{d-1}{d}\eta}{Ts+\fP}~~, \nn\\
  D_\parallel
  = \frac{s^2T}{\sigma d(s+\fP')}\frac{\fP - T \fP' + 2 (d-1)\fG}{Ts+\fP+ 2
    (d-1) \fG}~~, \nn\\
  v_\perp^2 = \frac{\fG}{Ts+\fP}~~, \qquad
  \Gamma_\perp
  = \frac{\fG}{\sigma}\frac{T^2s^2}{(Ts+\fP)^2}
  + \frac{\eta}{Ts+\fP}~~.
\end{gather}
Most of these expressions are not particularly illuminating, except that the
transverse and  longitudinal sound modes satisfy the simple identity
\begin{equation} \label{eq:vconformal}
  v_\parallel^2 =  \frac{1}{d}  + 2 \frac{d-1}{d} v_\perp^2~~,
\end{equation}
while the diffusion coefficients satisfy
\begin{equation} \label{eq:newconformal}
  \frac{D_\parallel}{\Gamma_\parallel - 2\frac{d-1}{d}\Gamma_\perp }
  =
  - 1
  + \frac{T\fP'-\fP}{T(s+\fP')}
  \frac{\frac{1}{d}  + 2 \frac{d-1}{d} v_\perp^2}{2\frac{d-1}{d}v_\perp^2 }~~.
\end{equation}
The relation \eqref{eq:vconformal} is well known for conformal lattices, as obtained in \cite{Esposito:2017qpj}, however we have generalised it to finite temperature and included the presence of lattice pressure. On the other hand, the relation \eqref{eq:newconformal} is novel, also holding for conformal lattices and was not identified in \cite{Esposito:2017qpj}.
In the special case that $\fP= 0$, we
simplify these to
\begin{gather}
  v_\parallel^2
  = \frac1d
  + 2 \frac{d-1}{d} \frac{\fG}{Ts}~~, \qquad
  \Gamma_\parallel
  =  \frac{1}{\sigma}
  \frac{4\frac{(d-1)^2}{d} \fG^2}{Ts+ 2 (d-1) \fG}+ \frac{2\frac{d-1}{d}\eta}{Ts}~~, \qquad
  D_\parallel
  = \frac{sT}{\sigma d}\frac{2 (d-1)\fG}{Ts+ 2 (d-1) \fG}~~, \nn\\
  v_\perp^2 = \frac{\fG}{Ts}~~, \qquad
  \Gamma_\perp
  = \frac{\fG}{\sigma}+ \frac{\eta}{Ts}~~.
\end{gather}
However, we find that the holographic models that we consider below generically
lead to a non-zero $\fP$ coefficient.\footnote{We believe that the mismatch between holographic and hydrodynamical approaches reported in \cite{Ammon:2019apj} is due to the fact that the authors of \cite{Ammon:2019apj} have not taken into account the presence of lattice pressure.}

\subsection{Models with higher-form symmetries}
\label{sec:holohigher}

This section deals with models whose bulk gravity metric describes fluids with
higher-form symmetries living on the AdS boundary. In $D=4$, this model was
considered in \cite{Grozdanov:2018ewh} and here we generalise it to include
$D=5$ as well. It should be noted that this model does not encompass the full
description of higher-form fluids as discussed in section \ref{sec:higher-form} but only
the hydrodynamics of fluids whose equilibrium states have
$\varphi^I = \text{constant}$.\footnote{The complete model should involve at
  least an extra set of scalar fields whose equations of motion admit the
  solution $\varphi^I=\text{constant}$, in which case reduces to the model
  studied here.}

\subsubsection{The model}

Denoting the bulk metric by $G_{ab}$ where $a,b,\ldots$ are
spacetime indices in the bulk, the bulk action takes the form (with
$\ell_{\text{AdS}} = 1$)
\begin{equation} \label{eq:blkact}
  S_{\text{bulk}}
  = \frac{M_p^2}{2}
  \int  \sqrt{-G}\, \df x^{D} \left(R+(D-1)(D-2)
    - \frac{1}{2(D-1)!} \bbdelta^{IJ}
    \widetilde H^{a_1\ldots a_{D-1}}_I \widetilde H_{J a_1\ldots a_{D-1}} \right)~~,
\end{equation}
where $\widetilde H_I = \df B_I$ and $B_{I a_1...a_{D-2}}$ are the $(D-2)$-form
gauge fields. The bulk action must also be supplemented by an appropriate
boundary action at some cutoff surface $r=\Lambda_c$ where $r$ is the
holographic direction and $\Lambda_c\to\infty$ the boundary. The boundary action
has the form
\begin{equation} \label{eq:bdyact}
  S_{\text{bdy}}
  = M_p^2 \int_{r=\Lambda_c} \sqrt{-\gamma}\, \df x^{D-1}\,
  \left(K-(D-2)
    + \frac{1}{4\kappa(\Lambda_c)(D-2)!}
    \bbdelta^{IJ}
    \mathcal{H}^{\mu_1\ldots\mu_{D-2}}_I \mathcal{H}_{J\mu_1\ldots\mu_{D-2}}
  \right)~~,
\end{equation}
where
$\mathcal{H}_{I\mu_1...\mu_{D-2}} = n^a \widetilde H_{Ia\mu_1\ldots\mu_{D-2}}$,
$K=G^{ab}\Df_a n_b$ is mean extrinsic curvature of the cutoff surface, $\Df_{a}$
the bulk covariant derivative compatible with $G_{ab}$, $n^a$ is a unit
normalised outward-pointing normal vector to the surface, and $\mu,\nu,\ldots$
label indices along the surface.\footnote{In \eqref{eq:bdyact} we have assumed
  that the boundary metric is flat. It is straightforward to add the usual
  boundary terms that render the on-shell action finite for non-flat boundary
  metrics \cite{Balasubramanian:1999re}.}  In \eqref{eq:bdyact}, we have
introduced the induced metric on the cutoff surface $\gamma_{\mu\nu}$, in turn
related to the boundary metric $g_{\mu\nu}$ by a conformal factor
$g_{\mu\nu}=\lim_{\Lambda_c\to\infty}\gamma_{\mu\nu}\Lambda_c^{-2}$. Additionally,
$\kappa(\Lambda_c)$ is a function of the cutoff and in particular
$\Lambda^{D-3}/\kappa(\Lambda)$ is a coupling constant of double trace
deformations of the boundary field theory, which can be fixed by demanding the
sources to be physical (i.e. independent of $\Lambda_c$) as we shall now
explain.

\subsubsection{Holographic renormalisation with higher-form symmetries}

The procedure employed here follows closely that of \cite{Hofman:2017vwr, Grozdanov:2017kyl, Grozdanov:2018ewh}. We focus on asymptotically AdS solutions which have metric of the form
\begin{equation} \label{eq:metric}
  \df s^2
  = \frac{1}{r^2f(r)} \df r^2
  + r^2 \left(-f(r)\df t^2
    + \bbdelta_{IJ} \df x^I \df x^J \right)~~,~~
  f(r\to\infty)\to 1~~.
\end{equation}
The equations of motion for the set of gauge fields take the form
$\Df_{a_1} \widetilde H_I^{a_1...a_{D-2}} = 0$, leading to the near boundary behaviour of
the gauge fields
\begin{equation}
  B_{I\mu_1...\mu_{D-2}}
  = \frac{r^{D-3}}{D-3} \mathcal{J}_{I\mu_1\ldots\mu_{D-2}}(x)
  + \hat B_{I\mu_1\ldots\mu_{D-2}}(x)
  + \mathcal{O}(1/r)~~,
\end{equation}
for $D=4,5$ and where $x^\mu$ are boundary coordinates. Performing a variation
of the total on-shell action with respect to the $B_I$ fields, one obtains 
\begin{equation} \label{eq:bvar}
  \delta_{B} S
  = - \int_{r=\Lambda_c} \sqrt{-\gamma}\, \df x^{D-1}\,
  \frac{1}{(D-2)!} J^{I\mu_1...\mu_{D-2}} \delta b_{I\mu_1\ldots\mu_{D-2}}~~,
\end{equation}
where the boundary current $J^{I\mu_1\ldots\mu_{D-2}}$ and the boundary gauge field
source $b_{I\mu_1...\mu_{D-2}}$ are, respectively, given by
\begin{equation} \label{eq:holocur}
\begin{split}
  J^{I\mu_1...\mu_{D-2}}
  &= M_p^{2}\, \bbdelta^{IJ} \lim_{\Lambda_c\to\infty} \Lambda_c^{-(D-3)}
  n_{a} {\widetilde H}_{J}^{a\mu_1\ldots\mu_{D-2}}~~,\\
  b_{I\mu_1...\mu_{D-2}}
  &= \half \lb \hat B_{I\mu_1\ldots\mu_{D-2}}
    + \Lambda_c^{D-3} \left(\frac{1}{D-3} - \frac{1}{\kappa(\Lambda_c)}\right)
    \mathcal{J}_{I\mu_1...\mu_{D-2}} \rb~~.
\end{split}
\end{equation}
We have chosen the pre-factor in $J^{I\mu_1\ldots\mu_{D-2}}$ in such a way as to
have a unit pre-factor in \eqref{eq:bvar}. The requirement that the source is
independent of the cutoff $\Lambda_c$, that is
$\df b_{I\mu_1\ldots\mu_{D-2}}/\df\Lambda_c=0$ implies that
\begin{equation} \label{eq:renorm}
  \frac{\Lambda_c^{D-3}}{\kappa(\Lambda_c)}
  = \frac{\Lambda_c^{D-3}}{D-3} - \mathcal{M}^{D-3}~~,
\end{equation}
agreeing with \cite{Grozdanov:2018ewh} for $D=4$. This condition not only
renders the source physical but also guarantees that the on-shell action is
finite. The constant $\mathcal{M}$ is the renormalisation group scale, which can
only be fixed by experiments. Given the well-posed formulation of the
variational problem for this class of models, it is possible to extract the
on-shell boundary stress tensor, which takes the form
\begin{multline} \label{eq:holostressH}
  T^{\mu\nu}
  = M_p^{2} \lim_{\Lambda_c\to\infty} \Lambda_c^{D+1} \bigg[
  K\gamma^{\mu\nu} - K^{\mu\nu}
  - (D-2)\gamma^{\mu\nu} \\
  + \frac{1}{8(D-3)!} \bbdelta^{IJ} \left(
    \mathcal{H}_I^{\mu\mu_2\ldots\mu_{D-2}} \mathcal{H}_J{}^{\nu}{}_{\mu_2\ldots\mu_{D-2}}
    - \frac{1}{2(D-2)}\gamma^{\mu\nu}
    \mathcal{H}_I^{\mu_1\ldots\mu_{D-2}} \mathcal{H}_{J\mu_1\ldots\mu_{D-2}}
  \right)\bigg]~~.
\end{multline}
Following the same footsteps as in \cite{Grozdanov:2018ewh}, it is straightforward to show that the following Ward identities are satisfied
\begin{equation} \label{eq:Ward1}
  \nabla_\mu T^{\mu\nu}
  = \frac{1}{(D-2)!}
  H_{I}^{\nu\mu_1\ldots\mu_{D-2}} J^I_{\mu_1\ldots\mu_{D-2}}~~,~~
  \nabla_{\mu_1} J^{I\mu_1\ldots\mu_{D-2}} = 0~~,
\end{equation}
in agreement with the hydrodynamic expectations of \cref{sec:higher-form}. The
Ward identities \eqref{eq:Ward1} also follow directly from the on-shell action,
given that the sources $b_I$ inherit the gauge and diffeomorphism transformation
properties of the bulk fields $B_I$. We will now look at specific examples of
thermal states that describe equilibrium viscoelastic fluids. 

\subsubsection{Thermal state in $D=4$}
\label{sec:bbhi4}

This case was studied in \cite{Grozdanov:2018ewh} and here we simply review
it. The bulk black hole geometry has metric function $f(r)$ and field strengths
given by\footnote{We have rescaled $m\to\sqrt{2}m$ compared to
  \cite{Grozdanov:2018ewh}.\label{note:rescaling}}
\begin{equation} \label{eq:t1}
  f(r)
  = 1 - \frac{m^2}{r^2}
  - \left(1-\frac{m^2}{r_h^2} \right) \frac{r_h^3}{r^3}~~,~~
  H_{1,txr} = H_{2,tyr} = -\sqrt{2}m~~,
\end{equation}
where $r=r_h$ denotes the location of the black hole horizon and $m$
parametrises the dipole charge. The goal is to identify the dual thermodynamics
as those corresponding to a fluid with partially broken higher-form symmetries
as in \cref{sec:higher-form}. To that aim, we note that according to
\eqref{eq:freehigher}, the free energy density of the fluid is equal to (minus)
the pressure. In turn, the free energy density of the black brane geometry
\eqref{eq:t1} can be obtained by evaluating the Euclidean on-shell action. The
total action $S_{T}$ is the sum of the bulk \eqref{eq:blkact} and surface
\eqref{eq:bdyact} contributions. Thus the pressure is given by (upon setting
$M_p^2 = 2$)
\begin{equation}
  p = - T S_{T}^{E}
  = r_h^3 \left[
    1 + \left(\frac{2\mathcal{M}}{r_h} - 3\right)\frac{m^2}{r_h^2}
  \right]~~,
\end{equation}
where $S_{T}^{E}$ is the Wick rotated version of $S_{T}$ after integration over
the time circle with period $1/T_0$ set to 1. The temperature and entropy of the
black hole are easily computed while the components of the stress tensor and
current are evaluated using \eqref{eq:holostressH} and \eqref{eq:holocur}
yielding
\begin{gather}
  T = \frac{r_h}{4\pi}\left(3-\frac{m^2}{r_h^2}\right)~~,~~
  s = 4\pi r_h^2~~,~~\nn\\
  T^{tt} = 2r_h^3
  \left[1 + \left(\frac{\mathcal{M}}{r_h} - 1\right)\frac{m^2}{r_h^2}\right]~~,~~
  T^{xx} = T^{yy}
  = r_h^3\left(1-\frac{m^2}{r_h^2}\right)~~,~~
  J^{1,tx} = J^{2,ty} = \sqrt{2}m~~,
\end{gather}
where we have set $M_p=1$ for simplicity. We wish to match these results with
the constitutive relations and thermodynamics of a viscoelastic fluid in
$d=2$. From \eqref{eq:tcSF}, we can read out the quantities appearing in
energy-momentum and charge currents 
\begin{gather}
  u^\mu = \delta^\mu_t~~,~~
  \zeta^\mu_I = - \sqrt{2}m\left(\mathcal{M}-r_h\right) \delta^\mu_I~~,\nn\\
  \epsilon = 2r_h^3
  \left[1 + \left(\frac{\mathcal{M}}{r_h} -
      1\right)\frac{m^2}{r_h^2}\right]~~,~~
  q^{IJ} = \frac{\bbdelta^{IJ}}{\mathcal{M}-r_h}~~.
\end{gather}
Note that the boundary metric is $g_{\mu\nu} = \eta_{\mu\nu}$. With these
identifications, the black brane geometry \eqref{eq:t1} describes a dual
viscoelastic fluid obeying the thermodynamic relations \eqref{eq:thermoSF}. The
one-form $\zeta_{I\mu}$ can be thought of as a chemical potential associated
with the background sources $b_{I\mu\nu}$, in particular
$\zeta_{I\mu} = b_{t\mu}$. This fixes
$\hat B_{It\mu} = 2\sqrt 2 m r_{h} \delta_{I\mu}$ using \cref{eq:holocur}. It is
worth mentioning that the renormalisation group scale $\mathcal{M}$ is not a
thermodynamic quantity but simply a constant that parametrises a family of
solutions. The pressure and energy density are only functions of $T$ and
$\gamma_{IJ} = \zeta^\mu_I \zeta_{J\mu} = 2 m^2(\mathcal{M} - r_h)^2
\bbdelta_{IJ}$.



\subsubsection{Thermal state in $D=5$} \label{sec:bbhi5}

In $D=5$, the metric \eqref{eq:metric} solves the bulk Einstein equations given
the following metric function and field strengths
\begin{equation}
  f(r) = 1 - \frac{m^2}{r^2} - \left(1-\frac{m^2}{r_h^2}\right)
  \frac{r_h^4}{r^4}~~,~~~
  H_{1,tyzr} = H_{2,tzxr} = H_{3,txyr} = -2mr~~.
\end{equation}
Using the renormalisation procedure of \eqref{eq:renorm} we evaluate the
on-shell Euclidean action in order to find the pressure and extract the
temperature and entropy from the black brane geometry
\begin{equation} \label{eq:pb5}
  p = r_h^4 \left(1 - \frac{5m^2}{r_h^2} + \frac{6m^2\mathcal{M}^2}{r_h^4}
    + \frac{9}{4} \frac{m^4}{r_h^4}\right)~~,~~
  T = \frac{r_h}{\pi}\left(1-\frac{m^2}{2r_h^2}\right)~~,~~s=4\pi r_0^3~~.
\end{equation}
The boundary stress tensor \eqref{eq:holostress} and charge currents \eqref{eq:holocur} have the following non-vanishing components
\begin{gather}
  T^{tt} = 3 \left(r_h^4+m^2 \left(2
      \mathcal{M}^2-r_h^2\right)+\frac{m^4}{4}\right)~~,~~
  T^{xx}=T^{yy}=T^{zz}
  = r_h^4-m^2 \left(2 \mathcal{M}^2+r_h^2\right)+\frac{m^4}{4}~~, \nn\\
  J^{1,tyz} = J^{2,tzx} = J^{3,txy} = 2m~~.
\end{gather}
We wish to interpret the stress tensor and currents as a higher-form fluid with
three global currents. From \eqref{eq:tcSF}, we get
\begin{gather}
  u^\mu = \delta^\mu_t~~,~~ 
  \zeta_{I\mu\nu} = - \frac12 m\lb m^2+4 \mathcal{M}^2-2 r_h^2\rb
  \epsilon_{t\mu\nu I}~~, \nn\\
  \epsilon = 3 \left(r_h^4+m^2 \left(2
      \mathcal{M}^2-r_h^2\right)+\frac{m^4}{4}\right)~~,~~
  q^{IJ} = \frac{4\bbdelta^{IJ}}{m^2+4 \mathcal{M}^2-2 r_h^2}~~.
\end{gather}
These quantities satisfy the expected thermodynamic relations
\eqref{eq:thermoSF}. Demanding the interpretation of $\zeta_{I\mu\nu}$ as a
chemical potential associated with $b_{I\mu\nu\rho}$, we get that
$\hat B_{It\mu\nu} = m \left(2r_h^2-m^2\right) \epsilon_{t\mu\nu I}$. These
thermodynamic properties provide a non-trivial example of a fluid with
generalised global symmetries.


\subsection{Models with translational broken symmetries}
\label{sec:holotrans}

\subsubsection{The model}

In this section we propose a model of viscoelasticity based on that of \cite{Andrade:2013gsa} usually studied in the
context of momentum dissipation. The model takes the form of 
gravity in AdS space minimally coupled to a set of $(D-2)$ scalar fields. In
units where the AdS radius is set to $\ell_{\text{AdS}}=1$, the bulk action
takes the form \cite{Andrade:2013gsa}
\begin{equation} \label{eq:X}
  S_{\text{bulk}} = \frac{M_p^2}{2}\int
  \sqrt{-G}\, \df x^{D}
  \left(R+(D-1)(D-2)-2\mathfrak{m}^2 X\right)~~,~~
  X = \frac{1}{2} \bbdelta^{IJ} G^{ab} \partial_a \Phi_I \partial_b \Phi_J~~,
\end{equation}
where $\mathfrak{m}$ is a free parameter. Varying the bulk action with respect to the metric and scalar field yields the equations of motion
\begin{gather}
  R_{ab}-\frac{R}{2}G_{ab}-\frac{(D-1)(D-2)}{2}G_{ab}-\frac{1}{M_p^2}T_{ab}^{M}=0~~, \\
  \nabla_a \nabla^a \Phi_I=0~~,~~T^{ab}_M
  = M_p^2\, \mathfrak{m}^2\,
  \bbdelta^{IJ} \left(\partial^{a}\Phi_I \partial^b\Phi_J
    - \frac{1}{2}G^{ab}\partial_{c}\Phi_I\partial^{c}\Phi_J\right)~~.
  \label{eq:blkeom}
\end{gather}
For the total action to be well-defined for asymptotically AdS solutions one needs to perform holographic renormalisation as to determine the boundary action under appropriate boundary conditions. Works that study momentum dissipation treat the massless scalar fields $\Phi_I$ as sources in the boundary field theory. Assuming a flat boundary metric, this choice leads to the boundary action
\begin{equation} \label{eq:bound1}
  S = S_{\text{bulk}} + S_{\text{bdy}}~~,~~
  S_{\text{bdy}} = M_p^2 \int_{r=\Lambda_c} \sqrt{-\gamma}\, \df x^{D-1}
  \left(K-(D-2)
    + \frac{\mathfrak{m}^2}{(D-3)}\bar X\right)~~,
\end{equation}
where $\bar X$ given by \eqref{eq:X} but where the contraction is performed with $\gamma^{\mu\nu}$. Variation of the total action with respect to $\Phi_I$, upon using the bulk equations \eqref{eq:blkeom} leads to the boundary term
\begin{equation}
  \delta_{\Phi} S
  = \int_{r=r_h}\sqrt{-\gamma}\, \df x^{D-1}\, \mathcal{O}^I \delta \Phi_I~~,
\end{equation}
for some scalar operator $\mathcal{O}_I$. It is clear from here that if the
boundary action \eqref{eq:bound1} is considered then the set of $\Phi_I$ are
taken as sources in the boundary theory. In the hydrodynamic limit, this is the
setting of forced fluid dynamics~\cite{Bhattacharyya:2008ji} and of momentum
relaxation~\cite{Blake:2015epa} in which case the scalar fields $\Phi_I$ are
background sources to which the fluid couples to but not dynamical fields as in
viscoelasticity.\footnote{As we will see below, the relation between $\Phi_I$
  and $\phi^I$ is given by $\phi^I = \sqrt{2}\mathfrak{m}\, \bbdelta^{IJ}
  \Phi_J$. } In order to use the holographic model \eqref{eq:X} for describing viscoelastic materials, another type of boundary conditions is necessary.

\subsubsection{Dualising the holographic model}

It is well known that in $D=3$, the dynamics of a $U(1)$ gauge field $A_\mu$ is
equivalent to the dynamics of a scalar field $\Phi$ since
$\df A\sim \star \df\Phi$. It is also known that a massless scalar field in
$D=3$ AdS admits two possible quantisations corresponding to different
dimensions of the boundary theory operator ($\Delta=1\pm 1$)
\cite{Klebanov:1999tb}. This was exploited in \cite{Faulkner:2012gt} to show
that the correct boundary conditions that describe the dynamics of the gauge
field $A_\mu$ are not those that fix the scalar $\Phi$ to be the source but
instead those that fix its conjugate momentum. This corresponds to the
quantisation with $\Delta = 0$. In this section we generalise the analysis of
\cite{Faulkner:2012gt} to higher-form fields in order to dualise the model of
sec.~\ref{sec:holohigher}. The difference between the cases considered here and
that of \cite{Faulkner:2012gt} is that the dualisation takes a theory with mixed
boundary conditions (in the sense of \cite{Witten:2001ua}) and maps to another
theory with mixed boundary conditions.

Naturally, in $D$ bulk dimensions, the dynamics of the bulk gauge field
$B_{I\mu_1...\mu_{D-2}}$ is equivalent to the dynamics of a scalar field $\Phi_I$
since $\df B_I \sim\star \df\Phi_I$.
This can be seen directly at the level of the path integral. Consider the action
for the $B_I$ field as in \eqref{eq:blkact} but integrate over the field
strength $\widetilde H_I = \df B_I$ instead of over $B$. In such case, one needs
to enforce the Bianchi identity $\df\widetilde H_I = 0$ by introducing a
Lagrange multiplier $\Phi_I$ such that
\begin{equation}
  \mathcal{Z}_{\text{blk}}
  = \int\mathcal{D}\widetilde H_I\,
  \mathcal{D} \Phi_I\,
  \exp \left(-S_{\text B}
    + \frac{M_p^2}{(D-1)!}\int_{\mathcal{W}} \sqrt{-G}\,\df x^{D}\,
    \bbdelta^{IJ}\,\Phi_I\, \epsilon^{a_1...a_{D}}\partial_{a_1}\widetilde H_{J a_2...a_{D}}\right)~~,
\end{equation}
where $S_{\text B} = M_p^2\widetilde H^{2}/4(D-1)!$. The quadratic action in
$\widetilde H_I$ can be integrated out by imposing the equations of motion for
$\widetilde H_I$, namely
\begin{equation}
  \epsilon^{a_1...a_D}\partial_{a_1}\Phi_I
  =- \half \widetilde H^{a_2...a_{D}}_I ~~,
\end{equation}
such that the path integral becomes
\begin{equation}
  \mathcal{Z}_{\text{blk}}
  = \int\mathcal{D}\Phi_I \,
  \exp \left(- M_p^2 \int_{\mathcal{W}} \sqrt{-G}\, \df x^D\,
    \bbdelta^{IJ} \partial_a \Phi_I \partial^{a}\Phi_J \right)~~,
\end{equation}
which is that of a massless scalar field in $D$ dimensions. Having established
the duality at the level of the bulk path integrals, one may include the
boundary action. Focusing just on the boundary term in \eqref{eq:bdyact}
involving the $B_I$ field one readily finds that
\begin{equation} \label{eq:ztotal}
  \mathcal{Z}_{\text{bdy}}
  = \int\mathcal{D}\Phi_I\, \exp
  \left(-\frac{1}{2(D-2)!}\int_{r=\Lambda_c}\sqrt{-\gamma}\,\df x^{D-1}\,
    J^{I\mu_1...\mu_{D-2}} b_{I\mu_1...\mu_{D-2}}\right)~~,
\end{equation}
where $J^{\mu_1...\mu_{D-2}}$ is a conserved current and hence can be expressed
in terms of a scalar operator
$J^{\mu_1...\mu_{D-2}} = \epsilon^{\mu_1...\mu_{D-2}\mu_{D-1}}
\partial_{\mu_{D-1}}\mathcal{O}$. Inserting this into \eqref{eq:ztotal} and
integrating by parts yields
\begin{equation}
  \mathcal{Z}_{\text{bdy}}
  = \int\mathcal{D}\Phi_I\, \exp
  \left(\frac{1}{2(D-2)!}\int_{r=\Lambda_c}\sqrt{-\gamma}\,\df x^{D-1}\,
    \epsilon^{\mu_1...\mu_{D-1}}
    \nabla_{\mu_1}b_{I\mu_2...\mu_{D-1}}\, \mathcal{O}^I\right)~~.
\end{equation}
Thus the operator $\mathcal{O}^I$ couples to the source
$\epsilon^{\mu_1...\mu_{D-1}} \nabla_{\mu_1}b_{I\mu_2...\mu_{D-1}}$, which is
proportional to the field strength $H_I=\df b_I$. Using \eqref{eq:holocur} one
derives
\begin{equation} \label{eq:pi}
\begin{split}
  \lim_{\Lambda_c\to\infty}
  \frac{\epsilon^{\mu_1...\mu_{D-1}}}{(D-1)!}H_{I\mu_1...\mu_{D-1}}
  &= \lim_{\Lambda_c\to\infty}\frac{\epsilon^{\mu_1...\mu_{D-1}}}{2(D-1)!}
  \left(\widetilde
    H_{I\mu_1...\mu_{D-1}}-\frac{1}{\kappa(\Lambda_c)}
    \nabla_{\mu_1}(n^{a}\widetilde H_{Ia\mu_2...\mu_{D-1}})\right) \\
  &= \lim_{\Lambda_c\to\infty}
  \left[n^a\partial_a\Phi_I
    + \frac{1}{\kappa(\Lambda_c)}\nabla_\mu\nabla^\mu \Phi_I
  \right]~~.
\end{split}
\end{equation}
Hence, in order to describe viscoelastic fluids, the sources in the model \eqref{eq:X} must be taken to be the conjugate momenta to the scalars $\Phi_I$, and naturally involve some coupling constant $\Lambda_c^{D-3}/\kappa(\Lambda_c)$.

These realisations lead us to consider the following boundary action 
\begin{equation} \label{eq:bdynew}
  S_{\text{bdy}}
  = M_p^2 \int_{r=\Lambda_c}
  \sqrt{-\gamma}\,\df x^{D-1}\,
  \left(K-(D-2)
    + \mathfrak{m}^2\, \bbdelta^{IJ} \Phi_I n^{a} \partial_a \Phi_J
    + \frac{\mathfrak{m}^2}{\kappa(\Lambda_c)}\bar X\right)~~,
\end{equation}
for some function $\kappa(\Lambda_c)$ of the cutoff surface $r=\Lambda_c$. Under
a variation of the total action $S$ with respect to $\Phi_I$ we obtain
\begin{equation} \label{eq:scalarop}
  \delta_{\Phi}S
  = \int_{r=\Lambda_c} \mathcal{O}^I \delta\Pi_I~~,
\end{equation}
where
\begin{equation}
  \mathcal{O}^I = \sqrt{2}\mathfrak{m}\,\bbdelta^{IJ} \Phi_J~~,~~
  \Pi_I =
  \frac{\mathfrak{m}}{\sqrt{2}} M_p^2 \lim_{\Lambda_c\to\infty}
  \sqrt{-\gamma}\left(n^{a}\partial_a\Phi_I
    + \frac{1}{\kappa(\Lambda_c)}\nabla_\mu\nabla^\mu\Phi_I\right)~~.
\end{equation}
We observe that with this specification of boundary action, the sources are the conjugate momentum $\Pi_I$, as in \eqref{eq:pi}, and the currents are proportional to the scalar fields $\Phi_I$. As in the case of holographic renormalisation for higher-form fields, we demand the sources $\Pi_I$ to be independent of the cutoff. The near boundary expansion of the fields $\Phi_I$ is \cite{Skenderis:2002wp}
\begin{equation} \label{eq:phinear}
  \Phi_I
  = \Phi_I^{(0)}(x)+\frac{\nabla_\mu\nabla^\mu \Phi_I^{(0)}(x)}{2(D-3)r^2}+\mathcal{O}\left(\frac{1}{r^4}\right)~~,
\end{equation}
for some function $\Phi_I^{(0)}(x)$ of the boundary coordinates. Eq.~\eqref{eq:phinear} is consistent with \eqref{eq:pi} and again implies \eqref{eq:renorm} for some renormalisation group scale $\mathcal{M}$, rendering the onshell action finite. This choice of boundary conditions corresponds to dimension $\Delta=0$ of the operators $\mathcal{O}^I$ \cite{Klebanov:1999tb}.

Given the total action we can obtain the Ward identities. Varying the onshell
action with respect to $\gamma_{\mu\nu}$ yields the boundary stress tensor
\begin{equation} \label{eq:holostress}
  M_p^{-2} T^{\mu\nu}
  = \lim_{\Lambda_c\to\infty} \Lambda_c^{D+1} \left[
    K \gamma^{\mu\nu}
    - K^{\mu\nu}
    - (D-2) \gamma^{\mu\nu}
    + \frac{\mathfrak{m}^2}{\kappa(\Lambda_c)}
    \left(\gamma^{\mu\nu} \bar X
      - \bbdelta^{IJ} \partial^{\mu}\Phi_I \partial^\nu\Phi_J\right)
  \right]~~.
\end{equation}
Acting with the covariant derivative on the boundary stress tensor, using the
Codazzi-Mainardi equation $n^{a}R_{a\mu}= - \nabla_\mu K + \nabla_\nu {K^\nu}_\mu$ (see e.g. \cite{Armas:2017pvj}) and the bulk equations \eqref{eq:blkeom}, one obtains the Ward identity
\begin{equation}
  \nabla_\mu T^{\mu\nu}
  = - \Pi_I \partial^\nu \mathcal{O}^I~~.
\end{equation}
Comparing this with \eqref{eq:elastic} we identify $K_{I}^{\text{ext}} = \Pi_I$
and $\mathcal{O}^I = \phi^I$. Thus fixing the boundary value of the source
$\Pi_I$ provides dynamics for the Goldstone scalars $\phi^I$ and has the
interpretation of applying external forces to the crystal lattice. We will now
study thermal states within the model \eqref{eq:X} with boundary action
\eqref{eq:bdynew}.

\subsubsection{Thermal state in $D=4$}

The bulk metric in $D=4$ was considered in \cite{Baggioli:2018vfc} but the
thermodynamic properties have not been properly evaluated. The metric takes the
form \eqref{eq:metric} but with metric function and scalar fields
\begin{equation} \label{eq:bb4}
  f(r) = 1 - \frac{\mathfrak{m}^2}{r^2}
  - \left(1-\frac{\mathfrak{m}^2}{r_0^2}\right)
  \frac{r_0^3}{r^3}~~,~~
  \Phi_1 = x~~,~~
  \Phi_2 = y~~,
\end{equation}
where $r=r_0$ is the location of the horizon. We now wish to determine the
thermodynamics of this black brane and the holographic stress tensor and scalar
currents for later interpretation in terms of a viscoelastic fluid. Noting that
the free energy of the viscoelastic fluid is (minus) the pressure $P$ as in
\eqref{eq:free_elastic}, we obtain the pressure from the onshell action while
the entropy and temperature are extracted from the black brane
\begin{equation} \label{eq:thermod4}
  P = r_0^3 \left(1+\frac{\mathfrak{m}^2}{r_0^2}
    - \frac{2\mathfrak{m}^2\mathcal{M}}{r_0^3}\right)~~,~~
  T = \frac{r_0}{4\pi}\left(3-\frac{\mathfrak{m}^2}{r_0^2}\right)~~,~~
  s = 4\pi r_0^2~~,
\end{equation}
where we have set $M_p=1$. In order to obtain the stress tensor we use \eqref{eq:holostress} and for the scalar operators we use \eqref{eq:scalarop}, finding
\begin{gather}
  T^{tt} = 2r_0^3 \left(1-\frac{\mathfrak{m}^2}{r_0^2}
    + \frac{\mathfrak{m}^2\mathcal{M}}{r_0^3}\right)~~,~~
  T^{xx} = T^{yy} = r_0^3\left(1-\frac{\mathfrak{m}^2}{r_0^2}\right)~~,~~ \nn\\
  \phi^1 = \sqrt{2}\mathfrak{m}x~~,~~
  \phi^2 = \sqrt{2}\mathfrak{m}y~~,
\end{gather}
while the sources in this case vanish, i.e. $\Pi_I=0$.\footnote{The trace of the energy-momentum tensor is non-vanishing except if $\mathcal{M}=0$, in which case the theory is conformal.} We now identify the
thermodynamic properties of the viscoelastic fluid by comparison with
\eqref{eq:const_elastic}. We find
\begin{gather}
  u^\mu = \delta^\mu_t~~,~~
  h^{IJ} = 2\mathfrak{m}^2\bbdelta^{IJ}~~,\nn\\
  \epsilon = 2r_0^3 \left(1-\frac{\mathfrak{m}^2}{r_0^2}
    + \frac{\mathfrak{m}^2\mathcal{M}}{r_0^3}\right)~~,~~
  r_{IJ} = \lb r_0 - \mathcal{M}\rb \bbdelta_{IJ} ~~.
\end{gather}
These quantities satisfy the thermodynamic relations \eqref{eq:elastic_thermo}.
Introducing these quantities in the map \eqref{eq:map}, we obtain exactly the
same thermodynamic properties as in \cref{sec:bbhi4} provided we identify
$r_h = r_0$ and $m=\mathfrak{m}$. In the case $\mathcal{M}=0$, these thermodynamic quantities
describe a conformal fluid as in sec.~\ref{sec:linear-conformal}.

We would like to note that the strain of the viscoelastic fluid is given by
\begin{equation}
  u_{IJ} = \half (h_{IJ} - \bbdelta_{IJ})
  = \half \lb \frac{1}{2\mathfrak m^2} - 1\rb \bbdelta_{IJ}~~.
\end{equation}
Therefore $\mathfrak{m}$, in some sense, controls the strength of the elastic
strain and hence holography can provide models of viscoelasticity with arbitrary
strains.\footnote{In previous considerations of viscoelastic holography (see
  e.g. \cite{Ammon:2019apj}), the bulk field $\Phi_I$ has been related to the
  crystal displacement field $\delta\phi^I = \phi^I - x^I$ at the boundary and
  not with $\phi^I$ itself. Unlike our model, where the strainless limit is
  given by $\fm = 1/\sqrt 2$, this alternative choice places the strainless
  limit at $\fm= 0$. Realising that the theory in the bulk becomes an ordinary
  charged black brane at $\fm = 0$ that is known to be dual to a pure fluid at
  the boundary and not an unstrained crystal, we do not make this
  choice. Furthermore, it is unclear if this choice can be implemented at a
  non-linear level in strain. A similar choice has been made in the higher-form
  setup of~\cite{Grozdanov:2017kyl}, but the authors there approached it as
  fluctuations around a state without ``dynamical defects'' (no crystal cores),
  distinct from a crystal phase where such defects are obviously present.} Let
us focus on the linear regime to make contact with
\cref{sec:linear-conformal}. Expanding
\begin{align}
  r_0(T,h^{IJ})
  &=  \frac16 \lb 4 \pi T + \sqrt{3\, \bbdelta_{IJ} h^{IJ} + 16 \pi^2 T^2} \rb
    \nn\\
  &= \frac16 \lb 4 \pi T + \sqrt{6 + 16 \pi^2 T^2} \rb
    - \frac{1}{2\sqrt{6 + 16 \pi^2 T^2}} u_{IJ}h^{IJ}
    + \mathcal{O}(u^2)~~,
\end{align}
and noting \cref{eq:thermo-expansions-linear}, we can read out the lattice pressure and bulk modulus
\begin{equation}
  \fP = \mathcal{M} - \frac16 \lb 4 \pi T + \sqrt{6 + 16
    \pi^2 T^2} \rb~~,~~
  \fP - \fB = \frac{1}{2\sqrt{6 + 16 \pi^2 T^2}}~~.
\end{equation}
The fluid pressure $P_{\text f}$ can be read out trivially from $P$ by setting
strain to zero. We cannot comment on the shear modulus $\fG$ because we are
working in a state with diagonal strain. We find that the coefficients $\fP$ and
$\fB$ do not scale homogeneously under conformal transformations. This is in
accordance with the comments made on inhomogeneous conformal scaling of strain in
\cref{sec:linear-conformal}.

Since the thermodynamic properties derived in \eqref{eq:thermod4} describe an
equilibrium state with vanishing elastic shear tensor, it is not possible to
extract from it the shear modulus. Thus, we do not have a complete knowledge of
the transverse phonon and the longitudinal sound dispersion relations.  However,
\cite{Grozdanov:2018ewh} showed that for small $\mathfrak{m}$ the dispersion
relations under the assumption of vanishing shear modulus agree with numerical
results.\footnote{We believe that the discrepancy between hydrodynamic and
  numerical results identified in \cite{Grozdanov:2018ewh} is due to the fact
  that \cite{Grozdanov:2018ewh} assumed a vanishing shear modulus in their
  hydrodynamic calculations.} According to the analysis of
\cite{Grozdanov:2018ewh}, it is expected that a Gregory-Laflamme like
instability \cite{Gregory:1993vy} is present for specific values of the
parameters, including when $\mathcal{M}=0$. However, a more in-depth analysis is
necessary in order to make definite statements.

\subsubsection{Thermal state in $D=5$}

Bulk metrics dual to viscoelastic fluids in $D=5$ have not been studied in depth but they straightforwardly generalise their lower dimensional counterpart. The bulk metric function and scalar fields that solve \eqref{eq:blkeom} are given by
\begin{equation} 
f(r)=1-\frac{\mathfrak{m}^2}{r^2}-\left(1-\frac{\mathfrak{m}^2}{r_0^2}\right)\frac{r_0^{4}}{r^{4}}~~,~~\Phi_1=\sqrt{2}x~~,~~\Phi_2=\sqrt{2}y~~,~~\Phi_3=\sqrt{2}z~~.
\end{equation}
The onshell action (i.e. pressure), the temperature and entropy are given by
\begin{equation}
P=r_0^4\left(1+\frac{\mathfrak{m^2}}{r_0^2}-\frac{3}{4}\frac{\mathfrak{m}^4}{r_0^4}-6\frac{\mathfrak{m}^2\mathcal{M}^2}{r_0^4}\right)~~,~~T=\frac{r_0}{2\pi}\left(2-\frac{\mathfrak{m}^2}{r_0^2}\right)~~,~~s=4\pi r_0^3~~,
\end{equation}
while the boundary stress tensor takes the form
\begin{gather}
  T^{tt} = \frac{3}{4}r_0^4\left(\left(2-\frac{\mathfrak{m}^2}{r_0^2}\right)^2
    + 8\frac{\mathfrak{m}^2\mathcal{M}^2}{r_0^4}\right)~~,~~
  T^{xx} = T^{yy} = T^{zz}
  =
  r_0^4\left(1-\frac{\mathfrak{m}^2}{r_0^2}+\frac{\mathfrak{m}^4}{4r_0^4}-2\frac{\mathfrak{m}^2\mathcal{M}^2}{r_0^4}\right)~~, \nn\\
  \phi^1 = 2x~~,~~
  \phi^2 = 2y~~,~~
  \phi^3 = 2z~~.
\end{gather}
Comparing with the constitutive relations \eqref{eq:elastic_thermo}, we
identify
\begin{gather}
  u^\mu = \delta^\mu_t~~,~~
  h^{IJ} = 4\mathfrak{m}^2 \bbdelta^{IJ}~~, \nn\\
  \epsilon = \frac{3}{4}r_0^4\left[\left(2-\frac{\mathfrak{m}^2}{r_0^2}\right)^2
    + 8\frac{\mathfrak{m}^2\mathcal{M}^2}{r_0^4}\right]~~,~~ r_{IJ} =
  \frac{\bbdelta_{IJ}}{2} \left(r_0^2 -
    \frac{\mathfrak{m}^2}{2}-2\mathcal{M}^2\right)~~.
\end{gather}
Again, these quantities satisfy the thermodynamic relations
\eqref{eq:const_elastic} and using the map \eqref{eq:map} they lead to the
constitutive relations and thermodynamic relations of sec.~\ref{sec:bbhi5}
provided $r_h=r_0$ and $m=\mathfrak{m}$.

Similar to $D=4$, we can perform a small strain expansion. We find that
\begin{equation}
  \fP = \mathcal{M}^2
  - \frac{\pi T}{8}  \lb 2 \pi T + \sqrt{2 + 4 \pi^2 T^2} \rb~~,~~
  \frac23 \fP - \fB = \frac{\pi T}{12 \sqrt{2 + 4 \pi^2 T^2}}~~,
\end{equation}
and corresponds to a conformal fluid when $\mathcal{M}=0$.  Once again, the
transport coefficients appearing here are not homogeneous under conformal
scalings.

\section{Outlook}
\label{sec:outlook}

In this paper we introduced two novel formulations of relativistic viscoelastic
hydrodynamics capable of dealing with elastic and smectic crystals phases. The
first formulation of sec.~\ref{sec:elastic-fluids} follows traditional
treatments \cite{PhysRevA.6.2401, JAHNIG1972129, chaikin_lubensky_1995} where
the elastic degrees of freedom are described by the dynamics of Goldstones of
translational broken symmetries.  However, it generalises earlier literature by
considering the effect of external currents, anisotropy, and
nonlinearities. When applied to the case of linear isotropic materials we
uncovered 6 new transport coefficients (5 dissipative and 1 non-dissipative) in
\cref{sec:isotropic} that characterise the coupling between elastic and fluid
degrees of freedom, constituting sliding frictional forces in viscoelastic
material diagrams. The second formulation of \cref{sec:higher-form} uses the
framework of generalised global symmetries in order to recast traditional
viscoelastic treatments as higher-form superfluidity. This provides a fully
symmetry-based approach to viscoelastic hydrodynamics and we show how the two
formulations map one-to-one.

In \cref{sec:holography} we studied holographic models to both these
formulations and proposed a new and simple model for viscoelasticity, consisting
of the model of \cite{Andrade:2013gsa} with an alternative quantisation for the
scalar fields and a double trace deformation. We also classified conformal
linear isotropic viscoelastic materials in \cref{sec:conformalfluids} and shown
that they correctly reproduce holographic results when there is no double trace
deformation. We also identified new holographic transport coefficients, which
have not been considered in earlier holographic works \cite{Alberte:2017cch,
  Alberte:2017oqx, Esposito:2017qpj, Baggioli:2018bfa, Grozdanov:2018ewh,
  Andrade:2019zey, Ammon:2019wci, Ammon:2019apj, Baggioli:2019abx}. Namely, by
expanding the equation of state in a small strain expansion, we notice that
there is a linear term in strain in the free energy, which is the lattice
pressure and usually ignored in classical elasticity treatments (see
sec.~\ref{sec:holotrans}).

The work presented here naturally opens up the possibility for various
extensions and generalisations which we now describe.

\noindent
\textbf{Non-homogeneity and dynamical reference metric:} The hydrodynamic
formulations considered here assumed homogeneity of the crystal lattice and a
non-dynamical reference metric. As such, it was shown that phenomenological
viscoelastic models such as the Kelvin-Voigt and Bingham-Voigt models are
special cases of the general constitutive relations obtained here. However,
other existent models such as the Maxwell model are not captured within this
approach. In order to do so, it is required to consider dynamical reference
metrics, allowing for the possibility of plastic deformations. This has been
implemented in \cite{Fukuma:2011pr} but only for linear strains.\footnote{One can also
study non-homogeneous models by introducing a potential for the scalar crystal
fields. In the context of holography, this is the premise of ``massive Goldstone'' models studied in, for
example,~\cite{Musso:2018wbv,Musso:2019kii,Amoretti:2018tzw,Amoretti:2019cef}.}

\noindent
\textbf{Disclinations and dislocations:} We have assumed that the fields
$\phi^I$ are surface forming, that is, the Bianchi-type condition in
\eqref{eq:elastic} is satisfied. This implies that no defects
(e.g. disclinations or dislocations) are present. It would be interesting to
understand viscoelastic hydrodynamics in the presence of defects. In terms of
generalised global symmetries this implies that the higher-form currents
$J^{I\mu_1...\mu_{d-1}}$ are not conserved, which makes it harder to understand
from this dual point of view. However, it may be the case that in some cases,
the violation of current conservation can be understood as an anomaly in quantum
field theories with generalised global symmetries.

\noindent
\textbf{Other crystal phases:} This paper was mostly concerned with elastic
(solid) crystal phases, although some of the results are valid for smectic-A
phases. However, liquid crystals can exhibit many other types of phases such as
other smectic, nematic, and hexatic phases. The hydrodynamics of these phases
have been consider in traditional treatments \cite{PhysRevA.6.2401,
  JAHNIG1972129, chaikin_lubensky_1995, Sonin_1998}, though without a careful
analysis of the constitutive relations. It would be interesting to revisit these
works using modern hydrodynamics and to develop equivalent models in terms of
higher-form symmetries.

\noindent
\textbf{Charged lattices, charge density waves, and holography:} Charged Wigner
crystals and charge density waves are charged generalisations of elastic and
nematic phases of neutral crystals. It would be interesting to consider such
extensions as it can aid in the understanding of recent holographic studies
\cite{Delacretaz:2017zxd, Amoretti:2017frz, Amoretti:2017axe}. It is natural to
consider the works \cite{Alberte:2017cch, Baggioli:2018bfa, Baggioli:2018vfc,
  Andrade:2019zey, Baggioli:2019jcm} and charged generalisations
\cite{Delacretaz:2017zxd, Amoretti:2017frz, Amoretti:2017axe} within the
framework of generalised global symmetries. We would also like to understand
whether sec.~\ref{sec:elastic-fluids} is sufficient for modelling the
viscoelastic fluids encountered in \cite{Alberte:2017cch, Alberte:2017oqx,
  Esposito:2017qpj, Baggioli:2018bfa, Andrade:2019zey, Ammon:2019wci,
  Ammon:2019apj, Baggioli:2019abx}).

\noindent
\textbf{Finite relaxation time:} As mentioned in the introduction to this work,
Maxwell's original idea of viscoelasticity consisted of materials that exhibited
elasticity at short time scales and fluidity at long time scales. In this paper
we focused on situations in which elasticity and fluidity coexist at long time
scales and long wavelengths by assuming the strain relaxation times to be very
large. It would be interesting to consider the case of arbitrary finite
relaxation times as in \cite{Fukuma:2011pr} in such a way that deviations away
from the hydrodynamic regime are under control. In these situations, the
framework of quasi-hydrodynamics will most likely be useful, as in
\cite{Grozdanov:2018fic}.

\noindent
\textbf{Fluid/gravity of viscoelastic fluids:} We explored holographic models to
viscoelastic hydrodynamics in \cref{sec:holography} but only at ideal order in a
long wavelength expansion. It is clear from the analysis of
\cref{sec:holography} that the holographic models describe constitutive
relations nonlinear in strain. As it is non-trivial to categorise all possible
materials nonlinearly in strain, it would be interesting to continue the
expansion one order higher and to uncover viscoelastic transport coefficients
that are present in gravity, ultimately obtaining a possible phenomenological
model of viscoelasticity. In this context, it will be possible to uncover the
conformal fluid structure of \cref{sec:holography} for small strains.

\noindent
\textbf{Shear elastic modulus and instabilities:} In \cref{sec:scalarperturbed},
we have performed a thorough analysis of the dispersion relations of
viscoelastic fluids, identified a longitudinal sound, transverse sound and a
diffusive mode as well as generalised the relation between longitudinal and
sound modes in a conformal solid \cite{Esposito:2017qpj} to the case of finite
temperature and in the presence of lattice pressure. In order to apply these
results to the holographic models of \cref{sec:holography}, it is necessary to
obtain the shear elastic modulus. This transport coefficient does not follow
from the equation of state in equilibrium since the equilibrium state that we
have considered has vanishing elastic shear tensor. The Kubo formulae of
\cref{sec:scalarperturbed} can be used in the holographic models studied here to
obtain the shear modulus. Nevertheless, the results of \cite{Grozdanov:2018ewh}
suggest the existence of an instability for certain values of parameters in the
model of \cite{Andrade:2013gsa} with alternative boundary conditions. It would
be interesting to obtain the shear modulus precisely and study instabilities in
these models more thoroughly.

\noindent
Finally, the work presented here shows that formulating hydrodynamics in terms
of generalised global symmetries can be extremely useful, not only because it
allows to rewrite hydrodynamic theories with dynamical fields just based on
symmetries (and their spontaneous breaking), but also because it allows for a
cleaner understanding of potential holographic models and their boundary
conditions. It should be noted that, as in the case of magnetohydrodynamics,
viscoelasticity when written in the language of generalised global symmetries
has global symmetries partially spontaneously broken (along the fluid flows). In
the context of condensed matter systems, broken global symmetries are only
natural \cite{Wen:2018zux}. At this point, we are not aware of a physical
hydrodynamic system with unbroken generalised global symmetries at the boundary as described in
\cite{Armas:2018zbe}. This fact has repercussions to several other constructions
studied in \cite{Grozdanov:2018fic}, where the Goldstone modes of spontaneous
broken global symmetries have not been taken into account. We wish to study
these constructions more carefully in the near future.

\acknowledgements

We would like to thank M. Baggioli, M. Fukuma, L. Giomi, R. S. Green, P. Kovtun, N. Poovuttikul and Y. Sakatani for various helpful discussions and comments on earlier drafts of this work. JA is partly
supported by the Netherlands Organization for Scientific Research (NWO). AJ is
supported by the NSERC Discovery Grant program of Canada.


\appendix

\section{Geometry of crystals}
\label{app:geometry}
In this appendix we give further details on the geometry of crystals. In order to characterise the crystals' response and symmetries, it is useful to
introduce auxiliary structures and give further details on how to describe the
crystals' geometry.  Given the objects described above, it is natural to
formulate the geometry of the worldsheets of $(d-k)$-dimensional crystal cores in
terms of the geometry of surface folitations (see e.g. \cite{Carter:1997pb,
  Armas:2017pvj, Speranza:2019hkr}).  The indices $I,J,K,...$ are indices on the
crystal-space (transverse to the crystal cores) and can be raised/lowered using
$h^{IJ}$ and $h_{IJ}$. In order to be able to define derivatives of tensor
structures that live on the crystal-space, we introduce a connection the acts on
the crystal indices as\footnote{In the language of surfaces, \eqref{eq:spincon}
  is usually referred to as the spin connection.}
\begin{equation}
\label{eq:spincon}
  C^I_{\mu J} =
  - e^\lambda_J \nabla_\mu e^I_\lambda
  = - e^\lambda_J \dow_\mu e^I_\lambda
  + e^\lambda_J \Gamma^{\sigma}_{\mu\lambda} e^I_\sigma~~,
\end{equation}
where $\nabla_\mu$ denotes the spacetime covariant derivative compatible with $g_{\mu\nu}$ and
associated with the Christoffel connection $\Gamma^{\lambda}_{\mu\nu}$. Additionally, we introduce the covariant derivative $\Df_\mu$, associated with $\Gamma^{\lambda}_{\mu\nu}$
and $C^I_{\mu J}$ and  compatible with both $g_{\mu\nu}$ and $h_{IJ}$. The covariant derivative $\Df_\mu$ transforms as a tensor under
$\GL(k)$ transformations of the normal one-forms. With this definition at hand, it is easy to check that 
\begin{equation}
\label{eq:ids}
  h^{\lambda\nu} \Df_\mu e^I_\nu
  = h_{\lambda\nu} \Df_\mu e^\nu_I
  = \Df_\mu h^{IJ}
  = \Df_\mu h_{IJ} = 0~~.
\end{equation}
The projection of the structures $\Df_\mu e^I_\lambda, \Df_\mu e^\lambda_I$  along the crystal directions vanishes as in \eqref{eq:ids} but in general 
\begin{equation}
  \Df_\mu e^I_\nu
  = \bar h_{\nu}{}^{\!\lambda}\, \nabla_\mu e^{I}_\lambda~~.
\end{equation}
Finally, the curvature associated with the connection $C^I_{\mu J}$ is not
independent and is related to projections of the spacetime Riemann curvature
tensor $R_{\mu\nu}{}^\rho{}_\sigma$, that is
\begin{equation}
 2\dow_{[\mu}C^I_{\nu] J} + 2 C^I_{[\mu K} C^K_{\nu] J} =
  R_{\mu\nu}{}^\rho{}_\sigma e^I_\rho e^\sigma_J - 2 h_{JK} \Df_\mu
  e^{[I}_{\lambda} \Df_\nu e^{K]\lambda}~~,
 \end{equation}
which is a generalised Ricci-Voss equation (see e.g. \cite{Carter:1996wr}).

The crystal metric $h_{IJ}$ and the reference metric $\mathbb{h}_{IJ}$ contain
all the information about the internal geometry of the crystal and, in
particular, about the deformations of the crystalline structure. The one-forms
$e^I_\mu$, however, contain additional information about the shape and
orientation of the crystal as embedded into the spacetime. If the spacetime does
not have broken rotational invariance except for the existence of the crystal
itself, this extra information can only be accessed via
derivatives.\footnote{This will no longer be the case when additional degrees of
  freedom are introduced in the theory, such as the hydrodynamic fluid velocity
  and temperature as we discuss in the next section.} Given the structures
introduced above, we see that all the information about the derivatives of
$e^I_\mu$ is stored in $\Df_\mu e^I_\nu$ and $C^I_{\mu J}$. In turn, the
derivatives of $h^{IJ}$ are all captured by
\begin{equation}
C^{(IJ)}_\mu = -\frac{1}{2} \nabla_\mu h^{IJ}~~.
\end{equation}
Hence, $\Df_\mu e^I_\nu$ and $C^{[IJ]}_\mu$ can be seen as containing the
additional information about the crystal embedding. $C^{[IJ]}_\mu$ captures the
differential of the local $\SO(k)$ orientation of the crystal with respect to
the reference coordinate system, while $\Df_\mu e^I_\nu$ captures the shape of
the crystal cores via the tangential extrinsic curvature tensor
$K^I_{\mu\nu} = \bar h_\mu{}^{\!\lambda} \Df_\lambda e^I_{\nu}$ and that of the
crystal via the normal extrinsic curvature tensor
$L^{IJ}_{\nu} = - e^{J\lambda} \nabla_\lambda e^I_{\nu}$. It is well known that
a generic set of one-forms $e^I_\mu$ does not have to be surface forming,
i.e. there might not exist a foliation of crystal core worldsheets normal to all
the $e^I_\mu$. For this to be the case, one needs to invoke the Frobenius
theorem, ensuring that there must exist a set of spacetime one-forms
$a^I_{\nu J}$ such that $\nabla_{[\mu} e^I_{\nu]} = - a^I_{[\mu J}
e^J_{\nu]}$. This is equivalent to the condition that the extrinsic curvature
tensor is symmetric $K^I_{\mu\nu} = K^I_{\nu\mu}$. Introducing the
normal one-forms as in sec.~\ref{sec:elasticity},  i.e.
$e^I_\mu(x) = \Lambda^I{}_{\!J}(x) \dow_\mu \phi^J(x)$ leads to the relations
\begin{equation}
  C^I{}_{\mu[J} e^\mu_{K]} = 0~~,~~
  C^I{}_{\mu J} \bar h^{\mu}{}_\nu = e^\lambda_J \Df_\lambda e^I_\nu~~,~~
  \bar h^{\mu[\rho} \bar h^{\sigma]\nu} \Df_\mu e^I_{\nu} = 0~~.
\end{equation}

\section{Details of hydrostatic constitutive relations} 
\label{app:hs-details} 

\subsection{Conventional formulation}

The hydrostatic free energy density in the conventional formulation is given by
\cref{eq:N-conventional}. Let us vary each of the one-derivative terms
independently. We find
\begin{align}
  \frac{1}{\sqrt{-g}}
  &\delta_\scB \lb \sqrt{-g}\frac{f^1_I}{T} e^{I\mu} \dow_\mu T \rb
    - \nabla_\mu \lb f^1_I \frac{1}{T} e^{I\mu} \delta_\scB T \rb \nn\\
  &= \frac{1}{T} e^{I\mu} \dow_\mu T
    \frac{1}{\sqrt{-g}} \delta_\scB \lb \sqrt{-g}\, f^1_I \rb
    + \lB - \nabla_\rho \lb f^1_I e^{I\rho} \rb u^\mu u^\nu
    - 2f^1_I \frac{1}{T} e^{I(\mu} \nabla^{\nu)} T
    \rB \half \delta_\scB g_{\mu\nu} + \mathcal{O}(\dow^2)~~, \nn\\[1ex]
  \frac{1}{\sqrt{-g}}
  &\delta_\scB \lb \sqrt{-g}\, 2 f^2_{[IJ]} e^{I\mu} e^{J\nu}
    \dow_{[\mu} (T u_{\nu]}) \rb
    - \nabla_\mu \lb f^2_{[IJ]} 2 T e^{I\mu} e^{J\nu} \delta_\scB u_{\nu} \rb \nn\\
  &= 2 T e^{I\mu} e^{J\nu} \dow_{[\mu} u_{\nu]} \frac{1}{\sqrt{-g}}
    \delta_\scB \lb \sqrt{-g}\, f^2_{[IJ]} \rb \nn\\
  &\qquad
    + \lB
    - 8 f^2_{[IJ]} e^{J\rho} \nabla_{[\lambda} (T u_{\rho]}) g^{\lambda(\mu} e^{I\nu)}
    - 2 \nabla_{\rho} \lb f^2_{[IJ]} 2 e^{I\rho} e^{J\lambda} \rb T u^{(\mu}
    P^{\nu)}_\lambda \rB \half \delta_\scB g_{\mu\nu}
    + \cO(\dow^2) \nn\\[1ex]
  \frac{1}{\sqrt{-g}}
  &\delta_\scB \lb \sqrt{-g} f^3_{I(JK)} e^{I\mu} \dow_\mu h^{JK} \rb
    - \nabla_\mu \lb f^3_{I(JK)} e^{I\mu} \delta_\scB h^{JK} \rb \nn\\
  &= e^{I\mu} \dow_\mu h^{JK} \frac{1}{\sqrt{-g}}
    \delta_\scB \lb \sqrt{-g} f^3_{I(JK)} \rb \nn\\
  &\qquad
    + \lB - 2 f^3_{I(JK)} e^{I(\mu} \nabla^{\nu)} h^{JK} 
    + 2 \nabla_\lambda \lb f^3_{I(JK)} e^{I\lambda} \rb e^{J\mu} e^{K\nu}
    \rB \half \delta_\scB g_{\mu\nu}
    + \cO(\dow^2)~~.
\end{align}
The first terms in the respective expressions can be expanded using the identity
\begin{align}
  S\frac{1}{\sqrt{-g}} \delta_\scB \lb \sqrt{-g}\, X(T,h^{IJ}) \rb
  &= S \lB
    X g^{\mu\nu}
    + T \frac{\dow X}{\dow T} u^\mu u^\nu
    - 2 \frac{\dow X}{\dow h^{IJ}} e^{I\mu} e^{J\nu}
    \rB \half \delta_\scB g_{\mu\nu} \nn\\
  &\qquad
    - \nabla_\mu \lb S \frac{\dow X}{\dow h^{IJ}} 2 e^{J\mu} \rb
    \delta_{\scB} \phi^{I}
    + \nabla_\mu \lb S \frac{\dow X}{\dow h^{IJ}} 2 e^{(I\mu} 
    \delta_\scB \phi^{J)}  \rb~~. 
\end{align}
The last line can be ignored to first order in the derivative
expansion. Consequently, we obtain the constitutive relations for the
energy-momentum tensor
\begingroup \allowdisplaybreaks
\begin{align}\label{eq:consti-hs-conventional}
  T^{\mu\nu}_{f_1}
  &= \frac{1}{T} e^{I\lambda} \dow_\lambda T \lB
    f^1_I g^{\mu\nu}
    + T \frac{\dow f^1_I}{\dow T} u^\mu u^\nu
    - 2 \frac{\dow f^1_I}{\dow h^{JK}} e^{J\mu} e^{K\nu}
    \rB \nn\\
  &\qquad
    - \nabla_\rho \lb f^1_I e^{I\rho} \rb u^\mu u^\nu
    - 2f^1_I \frac{1}{T} e^{I(\mu} \nabla^{\nu)} T
    + \mathcal{O}(\dow^2), \nn\\[1ex]
  T^{\mu\nu}_{f_2}
  &= 2 T e^{I\rho} e^{J\sigma} \dow_{[\rho} u_{\sigma]}  \lB
    f^2_{[IJ]} g^{\mu\nu}
    + T \frac{\dow f^2_{[IJ]}}{\dow T} u^\mu u^\nu
    - 2 \frac{\dow f^2_{[IJ]}}{\dow h^{KL}} e^{K\mu} e^{L\nu}
    \rB \nn\\
  &\qquad
    - 8 f^2_{[IJ]} e^{J\rho} \nabla_{[\lambda} (T u_{\rho]}) g^{\lambda(\mu} e^{I\nu)}
    - 2 \nabla_{\rho} \lb f^2_{[IJ]} 2 e^{I\rho} e^{J\lambda} \rb T u^{(\mu}
    P^{\nu)}_\lambda, \nn\\[1ex]
  T^{\mu\nu}_{f_3}
  &= e^{I\lambda} \dow_\lambda h^{JK} \lB
    f^3_{I(JK)} g^{\mu\nu}
    + T \frac{\dow f^3_{I(JK)}}{\dow T} u^\mu u^\nu
    - 2 \frac{\dow f^3_{I(JK)}}{\dow h^{IJ}} e^{L\mu} e^{M\nu}
    \rB \nn\\
  &\qquad
    - 2 f^3_{I(JK)} e^{I(\mu} \nabla^{\nu)} h^{JK} 
    + 2 \nabla_\lambda \lb f^3_{I(JK)} e^{I\lambda} \rb e^{J\mu} e^{K\nu}~~.
\end{align}
\endgroup
The one derivative terms in \cref{eq:N-conventional} only affect the $\phi^I$
equation of motion at two-derivative order and can be ignored for our purposes.

\subsection{Dual formulation} 

In the dual formulation, the variations of the ideal order fields are given by
\begin{gather}
  \delta_\scB T = \frac{T}{2} u^\mu u^\nu \delta_\scB g_{\mu\nu}~~, \nn\\
  \delta_\scB \gamma_{IJ}
  = - \frac{2}{(d-1)!}  u^{\mu_1}
  \zeta_{(I}^{\mu_2\ldots\mu_d} \delta_\scB b_{J)\mu_1\ldots\mu_{d}}
  + \lb
  \psi^{\mu}_I \psi^{\nu}_J
  - \gamma_{IJ} g^{\mu\nu} \rb \delta_\scB g_{\mu\nu}~~, \nn\\
  \delta_\scB \psi^{\lambda}_I
  = 
  - \frac{1}{(d-1)!}  u^{\mu_1}
  \epsilon^{\lambda\nu\mu_2\ldots\mu_d}u_{\nu} \delta_\scB b_{I\mu_1\ldots\mu_{d}}
  + \lb 2 u^\lambda \psi_I^{(\mu} u^{\nu)} - \psi_I^{\lambda} g^{\mu\nu} \rb
  \half \delta_\scB g_{\mu\nu}~~, \nn\\
  \delta_\scB (T u_\lambda)
  = T u^{(\mu} P^{\nu)}{}_\lambda \delta_\scB g_{\mu\nu}~~.
\end{gather}
In turn, the variations of the first order hydrostatic scalars take the form
\begingroup \allowdisplaybreaks
\begin{align}
  \frac{1}{\sqrt{-g}} \delta_\scB
  &\lb \sqrt{-g} \, \frac{\tilde f_1^I}{T} \psi_I^{\mu} \dow_\mu T \rb
    - \nabla_\mu \lb \tilde f_1^I \frac{1}{T} \psi_I^{\mu}
    \delta_\scB T \rb \nn\\
  &= \frac{1}{T} \psi_I^{\lambda} \dow_\lambda T \frac{1}{\sqrt{-g}} \delta_\scB
    \lb \sqrt{-g} \, \tilde f_1^I \rb
    + \lB  \frac{\tilde f_1^I}{T} \dow_\lambda T \lb 2 u^\lambda
    \psi_I^{(\mu} u^{\nu)} - \psi_I^{\lambda} g^{\mu\nu} \rb
    - \nabla_\lambda \lb \tilde f_1^I \psi_I^{\lambda} \rb u^\mu u^\nu
    \rB \half \delta_\scB g_{\mu\nu} \nn\\
  &\qquad
    + \lB - d\, \tilde f_1^I  \frac{1}{T} \dow_\lambda T  u^{[\mu_1}
  \epsilon^{\lambda\nu|\mu_2\ldots\mu_d]}u_{\nu} \rB \frac{1}{d!} \delta_\scB
    b_{I\mu_1\ldots\mu_{d}}~~, \nn\\[1ex]
  \frac{1}{\sqrt{-g}}
  &\delta_\scB \lb \sqrt{-g}\, 2T \tilde f_2^{[IJ]} \psi_I^{\mu}
    \psi_J^{\nu} \dow_{[\mu} u_{\nu]} \rb
    - \nabla_\mu \lb 2 \tilde f_2^{[IJ]} \psi_I^{\mu}
    \psi_J^{\sigma} \delta_\scB (T u_{\sigma}) \rb \nn\\
  &= 2T \psi_I^{\rho} \psi_J^{\sigma} \dow_{[\rho} u_{\sigma]}
    \frac{1}{\sqrt{-g}} \delta_\scB \lb \sqrt{-g}\, \tilde f_2^{[IJ]} \rb \nn\\
  &\qquad
    + \lB - 2 \nabla_\rho \lb 2 \tilde f_2^{[IJ]} \psi_I^{\rho}
    \psi_J^{\sigma}\rb T u^{(\mu} P^{\nu)}{}_\sigma
    - 4 \tilde f_2^{[IJ]} \psi^{J\rho}
    \dow_{[\rho} (T u_{\lambda]})
    \lb 2 u^\lambda \psi_I^{(\mu} u^{\nu)} - \psi_I^{\lambda} g^{\mu\nu} \rb
    \rB \half \delta_\scB g_{\mu\nu} \nn\\
  &\qquad
    + \lB 4 d \tilde f_2^{[IJ]} \psi_J^{\rho} \dow_{[\rho} (T u_{\lambda]})
    u^{[\mu_1}
    \epsilon^{\lambda\nu|\mu_2\ldots\mu_d]}u_{\nu} \rB
    \frac{1}{d!} \delta_\scB b_{I\mu_1\ldots\mu_{d}}~~, \nn\\[1ex]
  \frac{1}{\sqrt{-g}} \delta_\scB
  &\lb \sqrt{-g} \, \tilde f_3^{I(JK)} \psi_I^{\mu} \dow_\mu \gamma_{JK} \rb
    - \nabla_\mu \lb \tilde f_3^{I(JK)} \psi_I^{\mu} \delta_\scB \gamma_{JK} \rb \nn\\
  &= \psi_I^{\lambda} \dow_\lambda \gamma_{JK} \frac{1}{\sqrt{-g}} \delta_\scB
    \lb \sqrt{-g} \, \tilde f_3^{I(JK)} \rb \nn\\
  &\qquad
    + \lB  \tilde f_3^{I(JK)} \dow_\lambda \gamma_{JK} \lb 2 u^\lambda
    \psi_I^{(\mu} u^{\nu)} - \psi_I^{\lambda} g^{\mu\nu} \rb
    - 2 \nabla_\lambda \lb \tilde f_3^{I(JK)} \psi_I^{\lambda} \rb
    \lb
    \psi_J^{\mu} \psi_K^{\nu}
  - \gamma_{JK} g^{\mu\nu} \rb
    \rB \half \delta_\scB g_{\mu\nu} \nn\\
  &\qquad
    + d\, u^{[\mu_1} \epsilon^{\lambda\nu|\mu_2\ldots\mu_d]} u_\nu
    \lB - \tilde f_3^{I(JK)} \dow_\lambda \gamma_{JK}
    + 2 \nabla_\rho \lb \tilde f_3^{K(IJ)} \psi_K^{\rho} \rb
      \psi_{J\lambda}
    \rB \frac{1}{d!} \delta_\scB b_{I\mu_1\ldots\mu_{d}}~~.
\end{align}
\endgroup It is useful to consider the variation of an arbitrary function of $T$
and $\gamma_{IJ}$ is given by
\begin{align}
  \frac{1}{\sqrt{-g}} \delta_\scB
  &\lb \sqrt{-g} X(T,\gamma_{IJ}) \rb \nn\\
  &= \lb T \frac{\dow X}{\dow T} u^\mu u^\nu
    + \lb X - 2\frac{\dow X}{\dow \gamma_{IJ}} \gamma_{IJ} \rb g^{\mu\nu}
    + 2\frac{\dow X}{\dow \gamma_{IJ}}
  \psi_I^{\mu} \psi_J^{\nu}
    \rb \half \delta_\scB g_{\mu\nu} \nn\\
  &\qquad
    + \lB - 2d \frac{\dow X}{\dow \gamma_{IJ}}  u^{[\mu_1}
  \epsilon^{\mu\nu|\mu_2\ldots\mu_d]} \psi_{J\mu} u_\nu \rB \frac{1}{d!} \delta_\scB b_{I\mu_1\ldots\mu_{d}}~~.
\end{align}
We now record the contributions to the energy-momentum tensor, higher-form currents and free energy current that arise from each of the scalars in the hydrostatic effective action \eqref{eq:hydroparthigher}. In particular, we have 
\begingroup \allowdisplaybreaks
\begin{align}
  T^{\mu\nu}_{\tilde f_1}
  &= \frac{1}{T} \psi_I^{\lambda} \dow_\lambda T \lB
    T \frac{\dow \tilde f_1^I}{\dow T} u^\mu u^\nu
    - 2\frac{\dow \tilde f_1^I}{\dow \gamma_{KL}} \gamma_{KL} g^{\mu\nu}
    + 2\frac{\dow \tilde f_1^I}{\dow \gamma_{KL}}
  \psi_K^{\mu} \psi_L^{\nu}
    \rB \nn\\
  &\qquad
    + \frac{2\tilde f_1^I}{T} u^\lambda \dow_\lambda T \psi_I^{(\mu} u^{\nu)}
    - \nabla_\lambda \lb \tilde f_1^I \psi_I^{\lambda} \rb u^\mu u^\nu, \nn\\
  J^{I\mu_1\ldots\mu_d}_{\tilde f_1}
  &= - d\, u^{[\mu_1} \epsilon^{\mu\nu|\mu_2\ldots\mu_d]}u_{\nu}
    \lB \frac{1}{T} 2 \frac{\dow \tilde f_1^K }{\dow \gamma_{IJ}} 
    \psi_{J\mu} \psi_K^{\lambda} \dow_\lambda T
    + \tilde f_1^I  \frac{1}{T} \dow_\mu T \rB, \nn\\
  N^\mu_{\tilde f_1}
  &= \frac{2\tilde f_1^I}{T^2} u^{[\mu} \psi_I^{\nu]} \dow_\nu T~~,
  \nn\\[1ex]
  T^{\mu\nu}_{\tilde f_2}
  &= 2T \psi_I^{\rho} \psi_J^{\sigma} \dow_{[\rho} u_{\sigma]}
    \lB T \frac{\dow \tilde f_2^{[IJ]}}{\dow T} u^\mu u^\nu
    - \lb \tilde f_2^{[IJ]} + 2\frac{\dow \tilde f_2^{[IJ]}}{\dow \gamma_{KL}} \gamma_{KL} \rb g^{\mu\nu}
    + 2\frac{\dow \tilde f_2^{[IJ]}}{\dow \gamma_{KL}}
  \psi_K^{\mu} \psi_L^{\nu}
    \rB \nn\\
  &\qquad
    - 2 \nabla_\rho \lb 2 \tilde f_2^{[IJ]} \psi_I^{\rho}
    \psi_J^{\sigma}\rb T u^{(\mu} P^{\nu)}{}_\sigma
    - 8 \tilde f_2^{[IJ]} \psi_J^{\rho} u^\sigma \dow_{[\rho} (T u_{\sigma]})
    \psi_I^{(\mu} u^{\nu)}, \nn\\
  J^{I\mu_1\ldots\mu_d}_{\tilde f_2}
  &= - d\, u^{[\mu_1} \epsilon^{\mu\nu|\mu_2\ldots\mu_d]} u_\nu \lB
    4T \frac{\dow \tilde f_2^{[KL]}}{\dow \gamma_{IJ}}  \psi_{J\mu} \psi_K^{\rho} \psi_L^{\sigma} \dow_{[\rho} u_{\sigma]}
    - 4T \tilde f_2^{[IJ]} \psi_J^{\rho} \dow_{[\rho} u_{\mu]} \rB, \nn\\
  N^\mu_{\tilde f_2}
  &= \frac{6 \tilde f_2^{[IJ]}}{T} u^{[\mu} \psi_I^{\rho} \psi_J^{\sigma]}
    \nabla_\rho (T u_\sigma)~~,
  \nn\\[1ex]
  T^{\mu\nu}_{\tilde f_3}
  &= \psi_I^{\lambda} \dow_\lambda \gamma_{JK} \lB
    T \frac{\dow \tilde f_3^{I(JK)}}{\dow T} u^\mu u^\nu
    - 2\frac{\dow \tilde f_3^{I(JK)}}{\dow \gamma_{LM}} \gamma_{LM} g^{\mu\nu}
    + 2\frac{\dow \tilde f_3^{I(JK)}}{\dow \gamma_{LM}}
  \psi_L^{\mu} \psi_M^{\nu}
    \rB \nn\\
  &\qquad
    + 2 \tilde f_3^{I(JK)} u^\lambda \dow_\lambda \gamma_{JK} \psi_I^{(\mu} u^{\nu)} 
    - 2 \nabla_\lambda \lb \tilde f_3^{I(JK)} \psi_I^{\lambda} \rb
    \lb \psi_J^{\mu} \psi_K^{\nu} - \gamma_{JK} g^{\mu\nu} \rb, \nn\\
  J^{I\mu_1\ldots\mu_d}_{\tilde f_3}
  &= - d\, u^{[\mu_1} \epsilon^{\mu\nu|\mu_2\ldots\mu_d]} u_\nu
    \lB
    2 \frac{\dow \tilde f_3^{L(JK)}}{\dow \gamma_{IM}}
    \psi_{M\mu} \psi_L^{\lambda} \dow_\lambda \gamma_{JK}
    + \tilde f_3^{I(JK)} \dow_\mu \gamma_{JK}
    - 2 \nabla_\rho \lb \tilde f_3^{K(IJ)} \psi_K^{\rho} \rb
      \psi_{J\mu}
    \rB~~, \nn\\
  N^\mu_{\tilde f_3}
  &= 2 \tilde f_3^{I(JK)} u^{[\mu}\psi_I^{\nu]} \dow_\nu \gamma_{JK}~~.
\end{align}
\endgroup


\section{Comparison with previous works}
\label{app:comparison}

\subsection{Comparison with Fukuma-Sakatani formulation}
\label{app:sakatani}
In this section we compare our work with that of \cite{Fukuma:2011pr}. 
The work of \cite{Fukuma:2011pr} differs from ours in several ways: it assumes small (linear) strains; 
isotropy of the material; absence of external forces; 
introduces an extra scalar field related to breaking of time translations; it considers the possibility of 
a dynamical reference metric; and introduces a finite strain relaxation time. By contrast, the work
presented here works fully non-linear in strain; incorporates external forces; includes anisotropy; assumes a non-dynamical reference metric; does not
break time translations; and considers very large relaxation times. Within this regime, it is capable of capturing many
viscoelastic effects which \cite{Fukuma:2011pr} cannot.

As a result of not breaking time translations, the strain tensor in the work presented here is
purely spatial. The $(d+1)$-dimensional spacetime strain tensor
$u_{\mu\nu}$ is defined via the pullback of the $d$ dimensional crystal strain tensor
\begin{equation}
  u_{\mu\nu} = e^I_\mu e^J_\nu u_{IJ}~~,
\end{equation}
which implies that there exists a timelike vector $v^\mu$ such that
$v^\mu u_{\mu\nu} = 0$. Physically, the vector $v^\mu$ can be understood as a
reference frame in which the stress tensor is purely spatial. In equilibrium,
this frame is the same as the fluid frame of reference. We see that the
formulation of viscoelastic fluids given in this paper necessitates the strain
tensors to admit such a reference frame.\footnote{Although this property of the
  strain tensor seems to stand on firm physical grounds, this constraint has not
  been imposed in the generic definition in~\cite{Fukuma:2011pr}. It appears
  that the formulation presented here can be relaxed by including yet another scalar field
  $\phi_0$, whose derivative is timelike. This could potentially be understood as the
  time-translation symmetry also being spontaneously broken. The physical
  implications of this are unclear to us.} In the notation of~\cite{Fukuma:2011pr} we
identify $E^\FS_{\mu\nu} = u_{\mu\nu}$ where the upperscript FS denotes quantities defined in \cite{Fukuma:2011pr}. 
We can split the strain tensor into components according to
\begin{gather}
  \varepsilon_\FS^{\mu\nu}
  = P^{\mu\rho} P^{\nu\sigma} E^\FS_{\rho\sigma}
  = P^{I\mu} P^{J\nu} u_{IJ}~~,~~  %
  \tr\varepsilon_\FS = P^{\mu\nu} E^\FS_{\mu\nu}
  = h^{IJ} u_{IJ}
  + T^2 u_{IJ} \delta_\scB \phi^I \delta_\scB \phi^J~~, \nn\\
  \varepsilon^\mu_\FS = - 2 P^{\mu\rho} u^\nu E^\FS_{\rho\nu}
  = - 2T P^{I\mu} u_{IJ} \delta_\scB \phi^J~~,~~
  \theta_\FS = u^\mu u^\nu E^\FS_{\mu\nu}
  = T^2 u_{IJ} \delta_\scB \phi^I \delta_\scB \phi^J~~.
\end{gather}
The extrinsic curvature defined in~\cite{Fukuma:2011pr} is given as
\begin{equation}
  K_{\mu\nu}^\FS = \half \lie_u P_{\mu\nu}
  = \nabla_{(\mu} u_{\nu)}
  + u^\alpha u_{(\mu} \nabla_\alpha u_{\nu)}
  = \frac{T}{2} P_\mu{}^\rho P_\nu{}^\sigma \delta_\scB g_{\rho\sigma}~~.
\end{equation}
We first examine the $\phi^I$ equations of motion. 
In sec.~\ref{sec:elastic-fluids} we deduced that
\begin{equation}
  \delta_\scB \phi^I = \mathcal{O}(\dow)~~,
\end{equation}
and hence, in the formulation presented in this paper, $\varepsilon^\mu_\FS$ is pushed to
$\mathcal{O}(\dow)$ and $\theta_\FS$ to $\mathcal{O}(\dow^2)$ and are both
algebraically determined. On the other hand, noting that
\begin{equation}
  \delta_\scB u_{\mu\nu}
  = \half h_\mu^{\rho} h_\nu^{\sigma} \delta_\scB g_{\rho\sigma}
  + \lb 2 \delta^\lambda_{(\mu} e^J_{\nu)} u_{IJ} 
  - h^\lambda_{(\mu} e^J_{\nu)} h_{IJ} \rb \nabla_{\lambda} \delta_\scB \phi^{I}
  = \frac1T K_{\mu\nu}^\FS
  + \mathcal{O}(\dow^2)~~,
\end{equation}
we also have
\begin{equation}
  \delta_\scB \varepsilon^\FS_{\mu\nu}
  = \frac{1}{T} K_{\mu\nu}^\FS
  + \mathcal{O}(\dow^2)
  \qquad\implies\qquad
  \bar K_{\mu\nu}^\FS
  = - \lie_u \lb \varepsilon^\FS_{\mu\nu} - \half P_{\mu\nu} \rb = \mathcal{O}(\dow^2)~~.
\end{equation}
This is the elastic limit of the rheology equations given in equation (3.3) of~\cite{Fukuma:2011pr}.\footnote{This also agrees with the general rheology equations (2.50)-(2.52) of \cite{Fukuma:2011pr} when the relaxation times are taken to be very large, that is, $\tau_s,\tau_\sigma,\tau_\pm\to\infty$.}

We are now ready to compare the constitutive relations in~\cite{Fukuma:2011pr}
to the ones presented in the core of this paper. Noting that $\theta = \mathcal{O}(\dow^2)$, up to
one-derivative order and linear in strain, the energy-momentum tensor (2.41)
of~\cite{Fukuma:2011pr} can be expressed as
\begin{align}
  T^{\mu\nu}
  &= \lb \epsilon_{\text f} + P_{\text f} \rb u^\mu u^\nu
    + P_{\text f}\, g^{\mu\nu}
    +
    \lb \fB - \frac{2}{d} \fG \rb u^\lambda{}_{\!\lambda} h^{\rho\sigma}
    + 2 \fG u^{\rho\sigma} \nn\\
  &\qquad
    + 2 T u^{(\mu} P^{I\nu)}
    \lB 
    \lb \fB
    - \frac{2}{d} \fG \rb u^\lambda{}_{\!\lambda} h_{IJ}
    + 2 \fG u_{IJ} \rB \delta_\scB \phi^J \nn\\
  &\qquad
    - \lB 2\eta P^{\mu\rho} P^{\nu\sigma}
    + \lb \zeta - \frac{2\eta}{d} \rb P^{\mu\nu} P^{\rho\sigma} 
    \rB \frac{T}{2} \delta_\scB g_{\rho\sigma} + \mathcal{O}(\dow^2) + \mathcal{O}(u^2)~~,
\end{align}
where we have identified various transport coefficients as
\begin{gather}
  \text{Fluid pressure:}~ P_{\text f}(T) = P_\FS(T)~~,~~
  \text{Fluid energy density:}~ \epsilon_{\text f}(T) = e_\FS(T)
  = T P'_\FS(T) - P_\FS(T)~~, \nn\\
  \text{Shear viscosity:}~ \eta = \frac{\eta^\FS_3}{T}~~,~~
  \text{Bulk viscosity:}~ \zeta = \frac{\zeta^\FS_6}{T}~~, \nn\\
  \text{Shear modulus:}~ \fG = - (\mathcal{G}^\FS - \eta^\FS_2)~~,~~
  \text{Bulk modulus:}~ \fB = - (\mathcal{K}^\FS - \zeta^\FS_4)~~.
\end{gather}
To match these with \eqref{eq:linear-isotropic-consti}, we perform a hydrodynamic frame
transformation of the fluid velocity such that
\begin{equation}
  u^\mu \to u^\mu + P^{I\mu} \delta u_I,~~,~~
  \delta u_I
  = - \frac{T}{e_\FS + P_\FS}
  \lB 
  \lb \fB
  - \frac{2}{d} \fG \rb u^\lambda{}_{\!\lambda} h_{IJ}
  + 2 \fG u_{IJ} \rB \delta_\scB \phi^J~~,
\end{equation}
which also induces a shift in $\delta_\scB \phi^I$ such that
\begin{equation}
  \delta_\scB \phi^I \to \delta_\scB \phi^I + \frac{1}{T} P^I_\mu P^{J\mu} \delta u_J~~.
\end{equation}
Up to linear order in $u_{\mu\nu}$ and up to first order in derivatives, this gives
\begin{align}
  T^{\mu\nu}
  &= \lb \epsilon_{\text f} + P_{\text f} \rb u^\mu u^\nu
    + P_{\text f}\, g^{\mu\nu}
    + \lB \lb \fB - \frac{2}{d} \fG \rb u^\lambda{}_{\!\lambda} h^{\rho\sigma}
    + 2 \fG u^{\rho\sigma} \rB \nn\\
  &\qquad
    - \lB 2\eta P^{\mu\rho} P^{\nu\sigma}
    + \lb \zeta - \frac{2\eta}{d} \rb P^{\mu\nu} P^{\rho\sigma} 
    \rB \frac{T}{2} \delta_\scB g_{\rho\sigma} + \mathcal{O}(\dow^2) + \mathcal{O}(u^2)~~.
\end{align}
These directly match the constitutive relations presented in the core of this paper if we identify
\begin{gather}
  P(T,h^{IJ}) = P_{\text f}(T) + \half C^{IJKL} u_{IJ} u_{KL}~~, \nn\\
  f^1_I = f^2_{[IJ]} = f^3_{I(JK)} = 0~~, \nn\\
  \eta_{IJKL} = 2\eta\, h_{IK} h_{JL}
  + \lb \zeta - \frac{2\eta}{d} \rb h_{IJ} h_{KL}~~,~~
  \chi_{IJK} = 0~~,
\end{gather}
where the elastic modulus $C^{IJKL}$ is given in \cref{eq:modulii}.

\subsection{Comparison with Grozdanov-Poovuttikul formulation}

The authors of \cite{Grozdanov:2018ewh} considered the case $d=2$ in the language of generalised global 
symmetries. The work of \cite{Grozdanov:2018ewh} only investigated ideal order dynamics for very specific 
equilibrium states. In this work we have provided a different interpretation of the fundamental hydrodynamic degrees of freedom.
In particular, the formulation of \cite{Grozdanov:2018ewh} did not consider the existence of the Goldstones $\varphi^I$ of partial spontaneous breaking of 1-form symmetry, rendering it
incapable to characterise all equilibrium states by means of an equilibrium partition function or effective action.

In $d=2$ space dimensions, the conservation equations \eqref{eq:higherform} decompose into 7 dynamical equations
and 2 constraints. For each one-form symmetry $I$, one can reformulate hydrodynamics as a string fluid
with a partially broken symmetry~\cite{Armas:2018atq}.  In this case, there is a
 set of 7 hydrodynamic variables
\begin{equation}
  u^\mu~~,~~
  T~~,~~
  \zeta^I_\mu~~,
\end{equation}
with $\zeta^I_\mu u^\mu = 0$. In terms of the one-form chemical potentials
$\mu^I_\mu$ and scalar Goldstone $\varphi^I$~\cite{Armas:2018atq}, the one-forms
$\zeta^I_\mu$ can be written as
\begin{equation}
  \zeta^I_\mu = T \dow_\mu \varphi^I - \mu_\mu^I~~.
\end{equation}
When comparing with \cite{Grozdanov:2018ewh} we deduce that $\zeta^I_\mu = \varpi^I h^I_\mu$ with no sum over $I$, where $h^I_\mu$ are unit normalised vectors 
and $\varpi^I$ are chemical potentials introduced in \cite{Grozdanov:2018ewh}.\footnote{We have redefined $\mu^I$ as introduced in \cite{Grozdanov:2018ewh} such that $\mu^I\to\varpi^I$.}
The scalar Goldstones $\varphi^I$ are a consequence of partially broken one-form
symmetries and are distinct from the translation breaking Goldstones $\phi^I$
used in sec.~\ref{sec:elastic-fluids}. The one-form chemical potential
$\mu_\mu^I$ is a gauge field with $4$ independent degrees of freedom which is the same number 
of degrees of freedom contained in $\zeta^I_\mu$. Under a gauge transformation
$\mu^I_\mu \to \mu^I_\mu - T \dow_\mu(u^\mu\Lambda_\mu^I/T)$. The Goldstones
$\varphi^I$, on the other hand, are completely determined by
$\zeta^I_\mu u^\mu = 0$ leading to
\begin{equation}
  u^\mu \dow_\mu \varphi^I = \frac1T u^\mu \mu_\mu^I~~.
\end{equation}
In $d=2$, this gives the same number of degrees of freedom as in \cite{Grozdanov:2018ewh}, except that there the authors
have set $q_{12} = 0$. In order to recover the exact constitutive relations of \cite{Grozdanov:2018ewh} we set
$q_{12} = 0$ and define
\begin{equation}
  q_{11} = \frac{\rho_1}{\varpi_1}~~,~~
  q_{22} = \frac{\rho_2}{\varpi_2}~~,~~
  \zeta^\mu_1 = - \varpi_1 h^\mu_1~~,~~
  \zeta^\mu_2 = - \varpi_2 h^\mu_2~~,
\end{equation}
from which it follows that $\gamma^{11} = \varpi_1^2$, $\gamma^{22} = \varpi_2^2$ and
$\gamma^{12} = \varpi_1\varpi_2 h_1\cdot h_2$. Note however that $\dow p/\dow
\gamma^{12} = 0$ in \cite{Grozdanov:2018ewh}.

\subsection{Comparison with higher-form hydrodynamics}
\label{app:hfh}

In this section, we compare our construction of viscoelastic fluids in terms of
higher-form symmetries given in \cref{sec:higher-form} to the higher-form
hydrodynamics formulated in~\cite{Armas:2018ibg}. The work
of~\cite{Armas:2018ibg} only involves a single higher-form symmetry as opposed
to the multiple copies generically required in viscoelastic
hydrodynamics. Therefore, a precise comparison is only possible for $k=1$.

Let us take a special case of our discussion in \cref{sec:higher-form} with only
one higher-form symmetry. This corresponds to a smectic phase, as opposed to
elastic, where the lattice order is only present in one spatial direction. We
can define $(d-1)$- and $d$-dimensional volume forms
\begin{equation}
  \text{Vol}_{d-1} = -
  \frac{1}{\sqrt{\gamma_{11}}(d-1)!}\zeta_{1\mu_1\ldots\mu_{d-1}}
  \df x^{\mu_1}\wedge\ldots\wedge \df x^{\mu_{d-1}}~~, \qquad
  \text{Vol}_{d} = u \wedge \text{Vol}_{d-1}~~.
\end{equation}
We can also define a projector
\begin{equation}
  \Pi^{\mu\nu} = \frac{1}{\gamma_{11}(d-2!)}
  \zeta_1{}^{\mu}{}_{\mu_2\ldots\mu_{d-1}} \zeta_1^{\nu\mu_2\ldots\mu_{d-1}}~~.
\end{equation}
In terms of these, the ideal order constitutive relations \bref{eq:tcSF} reduce
to
\begin{align}
  T^{\mu\nu}
  &= \lb \epsilon + p \rb u^\mu u^\nu
    + p g^{\mu\nu}
    - q^{11} \gamma_{11} \Pi^{\mu\nu}
    + \mathcal{O}(\dow)~~, \nn\\
  J^{1}
  &= q^{11} \sqrt{\gamma_{11}}\, \text{Vol}_d
    + \mathcal{O}(\dow)~~.
\end{align}
This can be compared directly to section 2 of~\cite{Armas:2018ibg} with
$Q = q^{11} \sqrt{\gamma_{11}}$ and $\mu = \sqrt{\gamma_{11}}$. In principle, we
can extend this comparison to include one-derivative corrections as
well. However for technical simplicity, in this work we have only focused on
one-derivative corrections for elastic fluids ($k=d$), so such a comparison is
beyond the scope of the current analysis.

\providecommand{\href}[2]{#2}\begingroup\raggedright\endgroup

\end{document}